\documentclass[a4paper]{report}
\usepackage{./dissertation}
\usepackage{mathtext}
\usepackage{latexsym}
\usepackage[english]{babel}
\usepackage{graphicx}
\usepackage{euscript}
\begin{document}
\begin{titlepage}
\begin{center}
\large
UNIVERSTITE LYON I -- CLAUDE BERNARD\\
LOMONOSOV MOSCOW STATE UNIVERSITY\\
\normalsize
le type de doctorat : Arr\^et\'e du 06 janvier 2005

\vskip 25mm

\large {\bf Igor CHILINGARIAN}

\vskip 10mm

\normalsize
date de soutenance: 23/Nov/2006

\vskip 10mm

\large{\large{\bfseries{
FORMATION ET EVOLUTION DES GALAXIES ELLIPTIQUES NAINES
}}}
\vskip 10mm

directeurs de la th\`ese:\\ Philippe PRUGNIEL / Olga SIL'CHENKO
\end{center}

\vskip 20mm
\begin{flushleft}
membres du jury:\\
\vskip 5mm
Philippe PRUGNIEL (CRAL Observatoire de Lyon, FRANCE)\\
Georges PATUREL (CRAL Observatoire de Lyon, FRANCE)\\
prof. Jacques BERGEAT (CRAL Observatoire de Lyon, FRANCE)\\
Olga SIL'CHENKO (Sternberg Astronomical Institute, RUSSIA)\\
prof. Yuri EFREMOV (Sternberg Astronomical Institute, RUSSIA){\small -- rapporteur}\\
Dmitry MAKAROV (Special Astrophysical Observatory, RUSSIA){\small -- rapporteur}\\
prof. Boris SHUSTOV (Institute of Astronomy RAS, RUSSIA){\small -- pr\'esident}\\
prof. Ariane LANCON (Observatoire de Strasbourg, FRANCE){\small -- rapporteur}\\
\end{flushleft}
\end{titlepage}

\newpage
\begin{center}
{\bf ABSTRACT}
\end{center}

\begin{flushleft}
{\small Cette th\`ese pr\'esente des \'etudes d'observation de l'\'evolution
des galaxies elliptiques naines. Les dE sont population dominante dans les
amas des galaxies, mais leur origine et \'evolution est une question de
discussion. Plusieurs sc\'enarios d'enl\`evement de gaz des dE existent :
vents galactiques, pression dynamique d\'epouillant, harassment
gravitationnel. Nous pr\'esentons la nouvelle m\'ethode d'\'evaluation des
param\`etres de population stellaires et de la cin\'ematique interne,
bas\'es sur des spectres observ\'es par ajustage de pr\'ecision par les
populations synth\'etiques de PEGASE.HR. Nous nous appliquons cette
technique aux observations 3D-spectroscopic des galaxies dE dans l'ama Virgo
et les groupes proches et la spectroscopie multiobject de plusieurs
douzaines de dEs dans l'ama Abell 496. Nous pr\'esentons la d\'ecouverte de
jeunes noyaux dans les dE galaxies lumineuses dans l'ama Virgo. Bas\'e sur
l'analyse des donn\'ees de nos observations nous concluons cela : (1) il y a
une connection \'evolutionnaire entre les dEs et les dIrrs (2) le sc\'enario
le plus probable de l'enl\'evement de gaz est pression dynamique
d\'epouillant par le mati\`ere intergalactique.}

{\small {\bf Mots-cl\'es:} galaxies: dwarf -- galaxies: evolution --
galaxies: elliptical and lenticular, dE -- galaxies: stellar content}

\vskip 10mm

{FORMATION AND EVOLUTION OF DWARF ELLIPTICAL GALAXIES}

{\small This thesis presents observational studies of evolution of dwarf
elliptical galaxies. dE's are numerically dominant population in clusters of
galaxies, but their origin and evolution is a matter of debate. Several
scenarios of gas removal from dE's exist: galactic winds, ram pressure
stripping, gravitaional harassment. We present new method to estimate
stellar population parameters and internal kinematics, based on fitting
observed spectra in the pixel space by PEGASE.HR synthetic populations. We
apply this technique to 3D-spectroscopic observations of dE galaxies in the
Virgo cluster and nearby groups and multiobject spectroscopy of several
dozens of dE's in the Abell 496 cluster. We present discovery of young
nuclei in bright dE galaxies in the Virgo cluster. Based on the analysis of
observational data we conclude that: (1) there is an evolutionary connection
between dE's and dIrr's, (2) the most probable scenario of gas removal is
ram pressure stripping by the intergalactic medium.
}


\vskip 5mm
Centre de Recherche Astronomique de Lyon,
Observatoire de Lyon, 9 avenue Charles Andr\'e, 
F-69230 Saint-Genis Laval, France ; CNRS, UMR 5574

\vskip 5mm
Sternberg Astronomical Institute of the Moscow State University,
13 Universitetski prospect, Moscow, 119992, Russia

\end{flushleft}

\tableofcontents

\introduction{Introduction}

\section*{general characteristics of the work}
This work is devoted to studies of formation and evolution of dwarf
elliptical galaxies (diffuse elliptical galaxies, dE) -- the most common
type of galaxies in the present Universe. Though dwarf elliptical galaxies
represent over 70 percent of the population in dense regions of the Universe
(clusters and rich groups), their origin and evolution are not yet
clarified, especially questions of gas loss, and consequently interruption
of star formation. Presently considered possibilities include: (1) ram
pressure stripping, (2) gas removal by galactic winds due to supernovae, (3)
gas loss due to gravitational harassment.

Recent studies demonstrated great variety of observational appearances of
diffuse elliptical galaxies: many of them rotate, but some do not (they
might be supported by anisotropic velocity dispersions); many of them contain
embedded structures: discs, bars; some of them show evidence for a presence
of ISM; several objects exhibit kinematically-decoupled structures. All
these phenomena comfort the origin of dE galaxies with late-type disc dwarf
galaxies, experienced morphological transformation and lost their gas during
lifetime in clusters or groups.

For the argued choice of the scenario of dE galaxy evolution we decided
to investigate possible connections between stellar kinematics and
parameters of the stellar population (age, metallicity, [$\alpha$/Fe]
abundance ratios) exploiting integral field spectroscopy of nearby dE
galaxies and multiobject spectroscopy of a larger sample of more distant
objects.

Research work that has been conducted by the author during last 3 years
results in original technique for extraction of the stellar kinematical
parameters (radial velocity, velocity dispersion) and parameters of stellar
population (age, metallicity) from the spectra, integrated along a line of
sight. This method has been applied to the observations of dE galaxies
obtained with the MPFS IFU spectrograph at the Russian 6-m telescope and
FLAMES-Giraffe spectrograph at ESO VLT. Approach for data storage and access
mechanisms for 3D spectroscopic data in a frame of the International Virtual
Observatory has been developed.

\section*{actuality of the topic}
Presently, studies of the galaxy evolution is one of them most popular
topics of the modern astrophysics. While the mechanisms of evolution of
giant galaxies (both: elliptical and spiral) are investigated quite well,
the same cannot be said about dE galaxies, which are much more numerous, but
also more difficult for studies. Taking into account recent data on the
stellar populations in dE's: relatively high metallicities and intermediate
ages, the original idea considering dE's as building blocks for larger
systems is strongly criticized. Hence, question about formation and
evolution of dE galaxies is a corner-stone for understanding processes of
galaxy evolution in general.

Classical approach to determine stellar population parameters by measuring
parametrized spectral line strength (Lick indices) was proposed as empirical
over 20 years ago, but its first astrophysical justification was made in
1994 (Worthey et al., 1994). Since that time evolutionary synthesis
techniques evolved dramatically, and it became possible to synthesize
complete spectral energy distributions of stellar populations at high
spectral resolution, but not only parameters of selected spectral features.
Taking into account a progress in the instrumentation and observational
techniques, creation of the qualitatively new approach of estimating
parameters of stellar populations as a vital task for analysis of modern
spectral data.

\section*{immediate goal}
To create a new technique to analyse absorption-line spectra, including spectra
with low signal-to-noise ratios, and to apply it to the IFU data for dE
galaxies.

To analyse kinematics and stellar populations of dE galaxies in order to
make argued choice of the scenario of their formation and evolution.

\section*{novelty of the research}
\begin{enumerate}
\item Original technique for extracting stellar population and internal
kinematics by fitting integrated spectra in the pixel space has been developed
\item Parameters of stellar population, its chemical composition and
central velocity dispersion values are obtained for the
statistically-significant sample of dE galaxies in the Abell~496 cluster
\item Young nuclei discovered in dE galaxies in the Virgo cluster
\item Based on the results obtained in this work, a conclusion of the most
probable scenario of gas removal from dE, ram pressure stripping, is made
\end{enumerate}

\section*{practical value}
\begin{enumerate}
\item Spectral fitting technique proposed in this thesis gives the same
precision as existing approaches (e.g. Lick indices), but for the
signal-to-noise values 2-5 times lower, which allows to reduce significantly
exposure times during observations and makes possible to study low surface
brightness objects

\item Spectral fitting technique allowed to analyse observations of low
surface brightness dE galaxies, and in the future will provide a possibility
of re-processing existing absorption-line spectra on a qualitatively new
level

\item Method for storing and accessing 3D data in the Virtual Observatory
gives opportunity to build science-ready data archives containing 3D spectra

\end{enumerate}

\section*{approbation of results}
Results presented in this thesis have been presented by the author on the
seminars of INASAN (Russian Academy of Sciences), Sternberg Astronomical
Institute of MSU, CRAL Observatoire de Lyon, GEPI Observatoire de
Paris-Meudon, and international conferences listed below:
\begin{enumerate}
\item Lomonosov-2003, physics-astronomy section (Moscow, Russia, 14 March
2003)
\item ADASS-XIII (Strasbourg, France, 12-15 Oct 2003)
\item Russian National Astronomical Conference VAK-2004 (Moscow, 24-28 May
2004)
\item JENAM-2004 (Granada, Spain, 14-17 September 2004)
\item ADASS-XIV (Pasadena, USA, 24-27 October 2004)
\item International Astronomical Union Colloquium 198 (Les Diablerets,
Switzerland, 14-18 March 2005)
\item IVOA Interoperability Meeting (Kyoto, Japan, 14-18 May 2005)
\item ADASS-XV (San Lorenzo de El Escorial, Spain, 2-5 Oct 2005)
\item IVOA Interoperability Meeting (Villafranca del Castillo, Spain,
6-7 october 2005)
\item ESO Workshop: Science Perspectives for 3D Spectroscopy (Garching,
Germany 10-13 Oct 2005)
\item IVOA Interoperability Meeting (Victoria, Canada, 15-18 May 2006)
\item Mapping the Galaxy and Nearby Galaxies (Ishigaki, Japan, 26-30 Jun
2006) 
\item IVOA Interoperability Meeting (Moscow, Russia, 18-22 Sep 2006)
\end{enumerate}

\section*{publications and author's personal input}
Main results of the thesis are presented in 10 papers, published in the
refereed journals (2) and conference proceedings (8), and also in the
Standard, proposed by the IVOA to be used in the astronomical data archives.

In the mentioned publications author has:
\begin{itemize}
\item developed method for analysis of spectra, implemented it as a
software package, applied to the spectral data and interpreted the results
in [1-5,8,9,11];
\item provided method for analysis of spectra as a software package in [7];
\item applied Characterisation Data Model to the 3D data, created examples
of characterisation metadata for real 3D datasets obtained with MPFS in [6];
\item applied Characterisation Data Model to the 3D and longslit spectra,
scanning Fabry-Perot interferometer, and also edited the document in [10]
\end{itemize}

\section*{structure of dissertation}
Dissertation consists of: introduction, four chapters, conclusions, and
appendix. It includes 107 pages, 28 figures, 12 tables.
Bibliography includes 134 references.

{\bf Chapter~1} starts with brief review of existing methods of estimating
stellar population parameters. New technique for estimating stellar
population parameters is presented and discussed there. Stability,
precision, possible biases of the new approach are investigated.

{\bf Chapter~2} presents results of analysis of the integral field
spectroscopy for four Virgo cluster dwarfs: IC~783, IC~3468, IC~3509, and
IC~3653. Embedded rotating stellar disc is found in IC~3653. Young nuclei
are revealed in IC~783, IC~3468, and IC~3509.

{\bf Chapter~3} includes analysis of the IFU data for two rather unusual
low-luminosity galaxies in groups: NGC~770 (NGC~772 group), exhibiting
counter-rotating core, and NGC~127 (NGC~128 group), showing evidences for
ongoing cross-fueling from NGC~128.

{\bf Chapter~4} is devoted to the studies of a large sample of early-type
galaxies in the Abell~496 cluster, based on multi-object spectroscopy and
deep multicolour photometry. Fundamental properties of the objects are
discussed.

{\bf Conclusions} chapter includes major results of the thesis and brief
discussion of them

{\bf Appendix} includes application of the Characterisation Data Model to
the 3D spectroscopic datasets. Method for accessing those data in the
Virtual Observatory is proposed.

\section*{publication list:}

\begin{enumerate}

\item Chilingarian I. Object classification by SEDs. Moscow, MSU, Physics
Department, Division of Astropysics and Stellar Astronomy, 2003. Master
thesis. International conference for graduate, postgraduate students and
young scientists on fundamental sciences ''Lomonosov-2003''. Section:
Physics, proceedings of the conference, issued by Physics Department of MSU,
2003, pp.16-17.

\item Chilingarian I., Prugniel P., Sil'chenko O., Afanasiev V. Diffuse
elliptical galaxies, the first 3D spectroscopic observations. Proceedings
of JENAM-2004 (in press). Preprint: astro-ph/0412293

\item Prugniel P., Chilingarian I., Sil'chenko O., Afanasiev V. Internal
kinematics and stellar populations of dE galaxies: clues to their
formation/evolution. Proceedings of IAU Colloquium 198, edited by B.
Binggeli, H. Jerjen, 2005, p. 73; preprint: astro-ph/0510398

\item Chilingarian I., Prugniel P., Sil'chenko O., Afanasiev V. 3D
Spectroscopic studies of dE galaxies. Proceedings of IAU Colloquium 198,
edited by B. Binggeli, H. Jerjen, 2005, p. 105

\item Prugniel P., Chilingarian I., Popovic L. The history and dynamics of
the stellar population in the central kpc of active galaxies. Memorie della
Societa Astronomica Italiana Supplement, 2005, v.7, p.42

\item Chilingarian I., Bonnarel F., Louys M., McDowell J. Handling 3D data
in the Virtual Observatory. Proceedings of ADASS XIV, ASP Conference
Series, 2006, v. 351, p. 371

\item Koleva M., Bavouzet N., Chilingarian I., Prugniel, P. Validation of
stellar population and kinematical analysis of galaxies. Proceedings of ESO
Workshop ''Scientific Perspectives of 3D Spectroscopy'', in press, preprint:
astro-ph/0602362

\item Chilingarian I., Ferraz Lagana T., Cayatte V., Durret F., Adami C.,
Balkowski C., Chemin L., Prugniel P. Evolution of dE galaxies in Abell 496.
Kinematics and stellar populations of 46 galaxies. Proceedings of ''Mapping
the Galaxy and Nearby Galaxies'' (in press).

\item Chilingarian I., Prugniel P., Sil'chenko O., Afanasiev V. Kinematics
and stellar populations of the dwarf elliptical galaxy IC~3653. 2006, MNRAS,
submitted.

\item Data Model for Astronomical DataSet Characterisation, version 0.9,
edited by J. McDowell, F. Bonnarel, I. Chilingarian, M. Louys, A. Micol, and
A. Richards; IVOA Note from May 5, 2006 by IVOA Data Model Working Group.

\item Chilingarian I., Sil'chenko O., Afanasiev V., Prugniel Ph. Young Nuclei
in Dwarf Elliptical Galaxies. 2006, accepted for publication in 
''Astronomy Letters''. Preprint: astro-ph/0611866

\end{enumerate}

\chapter{Stellar population fitting technique}

Strong starburst events or periods of quiescent star formation during
lifetime of galaxies result in various generations of stars which we observe
presently. Thus, present stellar populations contain a fossil record of a
galaxy evolution in the past, and studies of them should help to bring
additional constraints to the scenarios of evolution of galaxies.

Numerous methods exist to study stellar populations. For the nearest
objects, which can be resolved into stars using deep ground-based
observations or HST imagery (resolved stellar populations), the most
efficient way is to build and analyse so called colour-magnitude diagrams
(CMD, see e.g. Da Costa \& Armandroff, 1990; Aparicio, 1994). Depending on
the depth of CMD, different features can be used to estimate age and
metallicity of the stellar populations: main sequence turn point(s),
position and width of the red giant branch, asymptotic giant branch stars, etc.
The complete star formation history over several Gyr can be reconstructed by
fitting models based on stellar evolutionary tracks into CMD. In addition,
this is a very precise method for estimating accurate distances to the
nearby galaxies (Makarov \& Makarova, 2004). Using CMD
analysis it was shown that faint local group dwarf spheroidal galaxies
exhibit great variety of star formation histories (Carraro et al. 2001), but
all of them contain significant amount of relatively old stars.

For more distant galaxies, where distribution of the stars on the H-R
diagram cannot be built directly (unresolved stellar populations), various
techniques have been developed to recover SFH either from broad-, middle-,
or narrow-band colours, or spectra, integrated along a line of sight.
Photometry-based methods are dealing with colour-magnitude relations for
particular sets of filters, or with larger sets of multicolour data
represented as spectral energy distributions (SED). Individual colours in
the optical band are demonstrated to be extremely degenerated with respect
to age and metallicity, e.g. old metal-poor stellar population as in
globular clusters will look nearly the same as intermediate-age metal-rich
ones (as in many dE galaxies). On the other hand, SED may cover significant
wavelength domains, from far-UV to mid-IR, providing opportunity to
disentangle roles of metallicity and age.

Spectral data may contain considerably larger amount of information, and
many attempts of its usage have been considered. In order to minimize the
effects of possible errors in flux calibration and to deal with data having
different spectral resolution, a concept of ''index''\ -- parametrized
representation of a line strength -- was proposed yet in early 80th
(Burstein et al. 1984). Methods exploiting spectral indices evolved quite
significantly during last 20 years, and at present time they remain the most
widely used.

All methods dealing with unresolved stellar populations are based on
comparison of observations against models: empirical or theoretical. There
are two main directions to construct these models: population and
evolutionary synthesis. Compared to resolved stellar populations, normally
it is not possible to reconstruct SFH in details, but only give some its
parameters, usually, luminosity-weighted age, metallicity, and element
abundance ratios (e.g. [Mg/Fe]).

In case of population synthesis, a model is a superposition of several
''populations'', for instance, spectrum of galaxy is modelled by a linear
combination of several stellar spectra. In practice, an inverse problem
needs to be solved: contribution of every subpopulation has to be restored.
This problem is unstable with respect to the observational errors, so
different astrophysical constraints are put on the contributions in order to
find a solution having physical sense. First applications of population
synthesis to analyse stellar population of galaxies were made by Wood (1966)
and Faber (1972).

Evolutionary synthesis is an alternative approach, based on our knowledge of
stellar evolution (Tinsley, 1968, 1972a,b). Spectrum (or colour) of a galaxy
is computed as the double integral:
\begin{equation}
L(\lambda) = \int_0^T \int_{M_{min}}^{M_{max}} L(\lambda, M, \tau) N(M, \tau) dM d\tau,
\end{equation}
where $L(\lambda)$ is a luminosity of a galaxy at $\lambda$ wavelength,
$L(\lambda, M, \tau)$ is a luminosity of a star having mass $M$ and age $\tau$
at the same wavelength, $N(M, \tau)$ is a number of such stars in a galaxy,
$T$ is an age of stellar population -- free parameter; $M_{min}$ and
$M_{max}$ are minimal and maximal stellar masses. From the theory of stellar
evolution we know various parameters of a star (e.g. $T_{eff}$ and $g$) for
a given mass at a given moment of time (evolutionary track). From a library
of observed stellar spectra, where atmosphere parameters are measured, or
set of theoretical spectra, $L(\lambda, M, \tau)$ is known. $N(M, \tau)$ can
be obtained assuming some initial mass function of stars, for instance,
Salpeter IMF (Salpeter, 1955), and a star formation rate as function of
time.

We are referring to the Full Doctor thesis of Olga Sil'chenko (1992) for a
historical review of different population and evolutionary synthesis methods
known at that time. Here we will emphasize only fundamental steps in the
evolutionary synthesis, made the current work possible. 

Study of Worthey et al. (1994) was one of the most successful attempts of
applying evolutionary synthesis to the set of spectral indices (Lick
indices). Grid of models for a wide set of ages and metallicities was
presented. It was shown, that exploiting different sets of indices gave a
possibility to disentangle age and metallicity effects (see next chapter for
an example of application). Continuation of this work (Worthey \& Ottaviani,
1997) defined several new indices and corrected definitions of some existing
ones. It became clear (Worthey et al. 1992) that giant early-type galaxies
usually exhibit super-solar [$\alpha$/Fe] abundance ratios. This stimulated
construction of new models for non-solar abundance ratios. The most cited
work of this kind is Thomas, Maraston \& Bender (2003). In the 4th chapter
we apply models published there to a large sample of early-type galaxies in
order to study enrichment mechanisms in dE's.

Another family of approaches was to synthesize whole spectral energy
distribution, not only specific details, based on available libraries of
stellar spectra (observational or theoretical): Fioc \& Rocca-Volmerange,
1997; Vazdekis, 1999; Leitherer et al., 1999; Eisenstein et al., 2003;
Bruzual \& Charlot, 2003. 

In this work we will be dealing with the evolutionary synthesis models,
providing whole spectral energy distribution at high spectral resolution
(R=10000), computed with the new PEAGSE.HR (Le Borgne et al. 2004) code.
Hereafter we present a method of stellar population parameters determination
based on fitting of the whole spectrum, not only specific spectral
features in order to optimize usage of the information, contained in
observations.

Several techniques exist for extracting internal kinematics from
absorption-line spectra. Historically, the first method was Fourier Quotient
(Sargent et al. 1977). A spectrum of galaxy is deconvolved with a spectrum of
template star in the Fourier space. This method takes into account the
instrumental broadening of the spectrograph, however it does not work very
well for low signal-to-noise spectra (Bottema, 1988), and it is quite
sensitive to the template mismatch (Bender, 1990).

The second method is a cross-correlation (Tonry \& Davis, 1979), where
cross-correlation function of two spectra: galaxy and template is built and
then analysed in the pixel space. It works quite well for low
signal-to-noise ratios and is less sensitive to the template mismatch than FQ
technique. At the same time, in order to get high contrast of the
correlation peak, it is necessary to remove continuum, which is not always
straightforward. Another practical problem is that only a certain region
around the peak has to be fitted, and changing this region might result in
biased estimates of velocity dispersion.

The third method, proposed by Bender (1990) is a combination of first two.
It is the Fourier Correlation Quotient, and its main idea is to
deconvolve the correlation peak of template-galaxy correlation function with
the peak of the autocorrelation function of the template.

In 1992-1994 there was a tendency toward development of methods for fitting
line-of-sight velocity distribution (LOSVD) directly in the pixel space (Rix
\& White, 1992; Kuijken \& Merrifield, 1993; van der Marel, 1994; Saha \&
Williams, 1994). The main reason was that in pixel space it became easy to
exclude gas emission lines or bad pixels from the fit, and take continuum
matching directly into account.

Van der Marel \& Franx (1993) introduced deviations of the galaxy's LOSVD
from Gaussian by using Gauss-Hermite polynomials of the 3rd and 4th order,
responsible for asymmetry of the profile ($h_3$), and its symmetric
deviations from Gaussian (narrower for positive or wider for negative $h_4$
respectively), and even higher order deviations ($h_5$, $h_6$).  From a
mathematical definition, $h_3$ is correlated with the radial velocity, and
$h_4$ is anticorrelated with velocity dispersion. This makes quite difficult
to expect unbiased values of the kinematical parameters for the case of low
signal-to-noise ratio, and undersampled LOVSD.

In order to improve the situation, penalization factor, depending on $h_3$
and $h_4$ was proposed to be applied to $\chi^2$ value during minimization
(Cappellari \& Emsellem, 2004), in order to fit $h_3$ and $h_4$ only if they
are statistically significant. This dramatically improves the quality of
fitting for low signal-to-noise ratios, that is demonstrated by the authors.
Penalized pixel fitting (ppxf) method now is the most advanced technique for
extracting internal kinematics and it is widely used in the community
(e.g. Emsellem et al., 2004).

\section{Description of the method}
Various methods have been developed to determine the star formation history
(SFH) directly from observed spectra (Ocvirk et al. 2003, Moultaka et al.
2004, de Bruyne et al. 2004, Ocvirk et al. 2006a, 2006b). The procedure that
we are proposing here, population pixel fitting, is derived from penalized
pixel fitting method developed by Cappellari \& Emsellem (2004) to determine
the LOSVD.

The observed spectrum is fitted in pixel space against the population model
convolved with a parametric LOSVD. The population model consists of one or
several star bursts, each of them parametrized by some of its characteristics,
typically age and metallicity for a single burst while the other
characteristics, like IMF, remain fixed. This method returns in a single
minimization the parameters of LOSVD and those of the stellar
population.

Ideally, we would like to reconstruct SFH, over all the life of the galaxy.
This means, disentangle internal kinematics and distribution in the HR
diagram from the integrated-light spectrum. This problem has been discussed
in several places (e. g. de Bruyne et al. 2004, Ocvirck et al. 2006a,b), it
is clearly extremely degenerated and solutions can be found only if a
simplified model is fitted.

In this work we discuss only the simplest case of SSP 
characterised by two parameters: age and 
metallicity. We do not discuss complex SFH, because signal-to-noise ratios
of our data are not sufficient for such elaborated studies.

The $\chi^2$ value (without penalization) is computed
as follows:
\begin{equation}
	\chi^2 = \sum_{N_{\lambda}}\frac{(F_{i}-P_{1p}(T_{i}(t,Z) \otimes
	\mathcal{L}(v,\sigma,h_3,h_4) + P_{2q}) )^2}{\Delta F_{i}^2},
\label{chi2eq}
\end{equation}
where $\mathcal{L}$ is LOSVD; $F_{i}$ and $\Delta F_{i}$ are observed flux and
its uncertainty; $T_{i}$ is the flux from a SSP spectrum, convolved according
to the line-spread function of the spectrograph (LSF, 
see next subsubsection); $P_{1p}$ and $P_{2q}$ are
multiplicative and additive Legendre polynomial of orders $p$ and $q$ for
correcting a continuum; $t$ is age, $Z$ is metallicity, $v$, $\sigma$,
$h_3$. and $h_4$ are radial velocity, velocity dispersion and
Gauss-Hermite coefficients respectively (Van der Marel \& Franx, 1993).
Normally we used no additive polynomial continuum, and 5-th (for MPFS) or
9-th (for Giraffe) order multiplicative one, and since dwarf galaxies
observed with MPFS had insufficient sampling of the LOSVD due to low
velocity dispersion, and Giraffe fibers were pointed to the centres of
galaxies, where one would not expect asymmetries of the LOSVD profiles, we
did not fit $h_3$ and $h_4$. There are two main reasons for including
multiplicative polynomial continuum terms in the minimization: (1) internal
extinction in the observed galaxy, (2) imperfections of the absolute flux
calibration. Additive terms may be included to eliminate effects of
improper subtraction of night sky emission or diffuse light in the
spectrograph.

The problem can be partially linearized: in
particular, fitting of additive polynomial continuum, and relative
contributions of sub-populations constituting
$T_{i}$ (in case it is not a SSP spectrum) is done linearly on each
evaluation of the non-linear functional. Thus we end up with the following
parameters to be found by the minimization procedure: $t$, $Z$, 6 or 10
coefficients for $P_{mult 5}$ ($P_{mult 9}$),
$v$, and $\sigma$.

The main technical part of our method is a non-linear minimization procedure for
$\chi^2$ difference between observed spectrum and template one, parametrized by
LOSVD and SFH parameters.
The parametric stellar population is made by interpolating a grid of
high-resolution ($R=10000$) synthetic SSP spectra,
computed with the PEGASE.HR,  with 25
steps in age (10~Myr to 20~Gyr, step increases from 5~Myr to 2~Gyr) 
and 10 steps in metallicity ([Fe/H] from -2.5 to
1.0). Minimization is done on the logarithm of age.
Because the minimization procedure requires that the derivatives
of the functions are continuous, we used a two-dimensional spline interpolation.
For processing the non-linear minimization we exploit the MPFIT 
package (by Craig B. Markwardt, NASA 
\footnote{http://cow.physics.wisc.edu/~craigm/idl/fitting.html}) implementing
constrained variant of the Levenberg-Marquardt minimization, thus we are able to fix
any of the LOSVD/SFH parameters.

\section{Line spread function of the spectrograph}
Before comparing a synthetic spectrum to an observation, it is required
to transform it as if it was observed with the same spectrograph
and setup, i. e. to degrade its resolution to the actual resolution 
of the observations. Actually the spectral resolution changes both 
with the position in the field of view and with the wavelength (thus it
is not a mere operation of convolving with the LSF). Taking into account
these effects is particularly critical when, as it is the case here,
the physical velocity dispersion is of the same order or smaller
than the instrumental velocity dispersion.

The procedure for properly taking into account the LSF goes in two steps.
First, determine the LSF as a function of the position in the field
and of the wavelength. Second, inject this LSF in the grid of SSP.

Therefore we made an exhaustive analysis of the LSF of our observations. For
MPFS spectrograph, where was a previous study of the change of the
resolution over the field of view (Moiseev 2001), which qualitatively
agrees with our results.

To measure the LSF change over the field of view of MPFS or across Giraffe
fibers we use the spectra
of standard stars (HD~135722 and HD~175743) and twilight sky (Solar spectrum).
We analyse these spectra using penalized pixel fitting procedure.
The  high-resolution spectra ($R=10000$)
for the corresponding stars (the Sun for the twilight spectra) taken from the 
ELODIE.3 library (Prugniel \& Soubiran 2001, 2004), were used as templates.
Since these spectra have exactly the same resolution as the PEGASE.HR
SSPs, the 'relative' LSF that we determined in this way can be directly
injected to the grid of SSP to make it consistent with observations.
We parametrize LSF using $v,
\sigma, h3$ and $h4$.

The whole wavelength range of the spectrograph is splitted into several
parts, overlapping by 10 per cent, and the LSF parameters are extracted in each
part independently in order to derive the wavelength dependence of the LSF.

Finally, to inject the LSF in the grid of SSPs, we applied the following
steps to every spatial element (IFU fiber of MPFS, MEDUSA fiber of 
FLAMES-Giraffe, or segment of the slit):
\begin{itemize}
 \item{Several convolved SSP grids were created using the
 LSF measured for all wavelength subsegments.}
 \item{The final grid was generated by linear interpolation at each wavelength
 point between the five grids of SSPs.}
\end{itemize}
It produces one grid of SSP for each spatial element of the spectrograph having
exactly the same LSF as the observations.

\section{Validation and error analysis}
\label{secvalid}

In this section we address questions concerning error analysis, stability of
the solutions, and possible biases for the MPFS and Giraffe data for
galaxies with relatively old stellar population (about 5~Gyr). Full
description of these aspects extended to any instrument and much wider range
of parameters will be described in details in the forthcoming paper. Here we
give only essential error analysis required for validation of the results
presented in this thesis and in the forthcoming papers based on MPFS and
Giraffe data for dwarf galaxies.

\subsection{Error analysis}
\label{subsecerr}
Error estimations for stellar population constraints 
for such a non-linear procedure as we are following is a
non-trivial task. A complete and detailed description of one of the
possible approaches to locate the alternate solutions using for a allied
inversion technique is given in Moultaka \& Pelat (2000).

We performed some Monte-Carlo simulations (about 10000 per spectrum for the
3-points binning for IC~3653, see next chapter) to demonstrate the consistency between
uncertainties on the parameters reported by the minimization procedure and
real error distributions. We have used best-fitting template spectra and
added Poisson noise to the fluxes, corresponding to signal-to-noise
ratios of the observed spectra.
These simulations have demonstrated that in case
of IC~3653 dataset, where there is neither significant template mismatch due
to element abundance ratios, nor strong mistake with subtraction of additive
terms (diffuse light and night sky), the uncertainties found from the
Monte-Carlo simulations using scattering of solutions in the
multidimensional parameter space coincide with values reported by the
minimization procedure being multiplied by $\chi^2$ values. Deviation of
$\chi^2$ from 1 might be caused either by poor quality of the fit
(template-mismatch) or by wrong estimations of absolute flux uncertainties
in the input data. 
We conclude that
in some cases our estimations of absolute flux uncertainties based on the
photon statistics are not ideal, that is not strange taking into account the
complexity of the data reduction. However, values of $\chi^2$ between 0.7
and 1.3 suggest that our error estimations are relatively precise.

To estimate the errors more accurately, locate possible alternate solutions and
search for degeneracies between kinematical and stellar population parameters
we perform direct scan of the $\chi^2$ space for age, metallicity, and velocity
dispersion. Here we present the procedure we followed for the analysis of
MPFS data for IC~3653. For the Giraffe data, the analysis was made
similarly. The procedure we followed includes the following steps:
\begin{itemize}
  \item We chose a grid of values for age, metallicity, and velocity dispersion,
  that was supposed to cover a reasonable region of the parameter space where we
  could expect to have solutions. In our case the grid was defined as: 2~Gyr $<$
  t $<$ 14~Gyr with a step of 200~Myr, -0.45$<$[Fe/H]$<$0.40 with a step of
  0.01~dex, 30~km~s$^{-1}$ $<$ $\sigma$ $<$ 100~km~s$^{-1}$ with a step of 0.5~km~s$^{-1}$ 
  \item At every node of t-Z grid we ran the pixel fitting procedure in order
  to determine multiplicative polynomial continuum, and to have the best
  fit for a given SSP
  \item Later $\chi^2$ was computed on a grid of values of $\sigma$ by fixing
  all other components of the solution that had been found in the previous step
\end{itemize}
This way of scanning $\chi^2$ space is stipulated by frequency effects of the
parameters (see Tab~\ref{freqef}): polynomial continuum and stellar populations
parameters have low-frequency effect on the template spectra, so we do need to
make a fit of the polynomial continuum if we want to avoid $N+1$ additional
dimensions of the $\chi^2$ space to scan for a case of $N$th order
multiplicative continuum. At the same time $\sigma$ is a high-frequency
parameter, thus we are safe when just computing $\chi^2$ values varying only
$\sigma$ and leaving other parameters fixed.

In other words using our procedure we compute a slice of the full $\chi^2$ by
the hypersurface defined as a set of minimal $\chi^2$ values for multiplicative
continuum terms and radial velocity values, and then reproject it onto ''t-Z''\ and
''t-Z-$\sigma$''\ hyperplanes. Result contains two arrays: 2D age-metallicity and
3D age-metallicity-velocity dispersion.

\begin{table}
\begin{tabular}{l c c c c c}
Parameter & $v$ & $\sigma$ & t & Z & $P_{1p}$ \\
\hline
Low-frequency & no & no & yes & yes & yes \\
High-frequency & yes & yes & yes & yes & no \\
\hline
\end{tabular}
\caption{Frequency effect of the parameters being minimized within the
pixel-fitting procedure (see Equation~\ref{chi2eq} for details).
\label{freqef}
}
\end{table}

Values of line-of-sight radial velocity obtained during the fitting procedure on
the t-Z grid are equal to each other within errors reported by the minimization
procedure, suggesting that scanning of the $\chi^2$ hyperspace on $v$ is not
necessary, though it is relatively easy to do by scanning the 2D ($v$, $\sigma$)
grid instead of 1D on ($\sigma$) on every point of the t-Z grid.

\begin{figure}
\hfil
 \includegraphics[width=17cm]{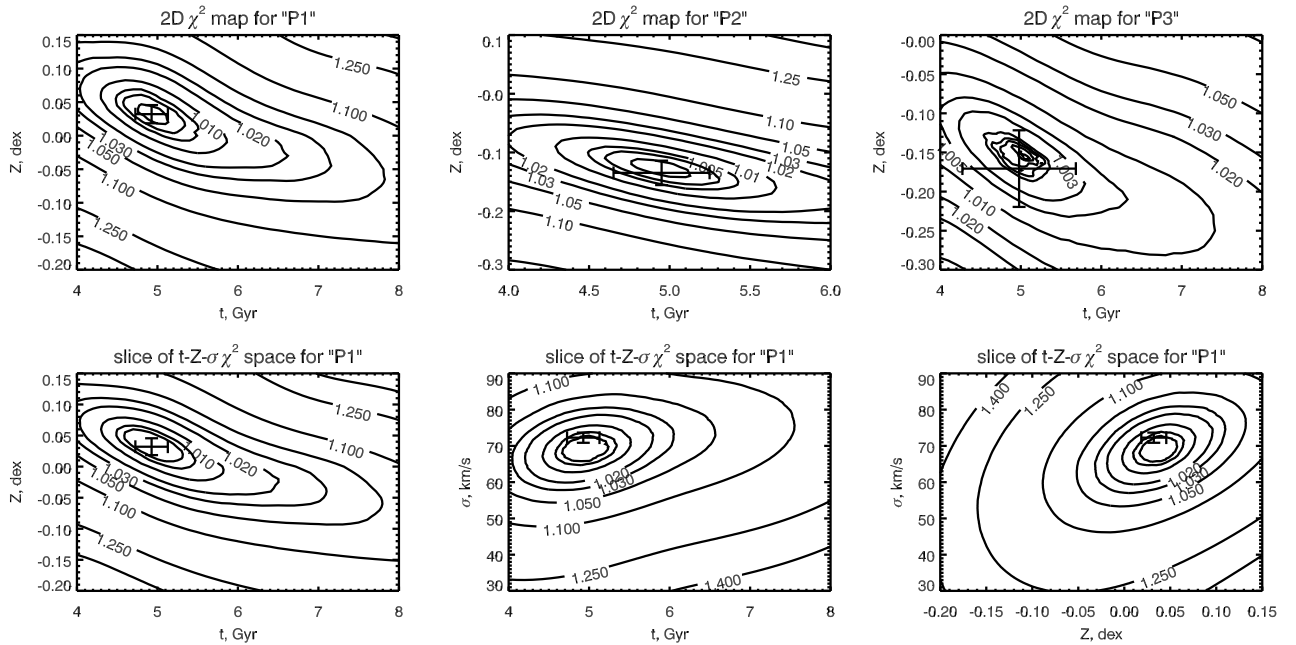}
 \caption{2-dimensional maps of $\chi^2$ distributions for the 3-points binning
 (upper row), and slices of the 3-dimensional $\chi^2$ distribution 
 (t-Z-$\sigma$) for the ''P1''\ bin.
\label{figchi2map}
}
\hfil
\end{figure}

In Fig~\ref{figchi2map} (upper line) we present the maps of $\chi^2$ for the
3-points binning of the MPFS data for IC~3653 (see next Chapter) on the
''t-Z''\ plane (all other parameters are fitted). One can see elongated
shapes of the minima, corresponding to well known age-metallicity
degeneracy. Three plots on the bottom of Fig~\ref{figchi2map} represent
slices of the 3D $\chi2$ space scan (t-Z-$\sigma$) for the ''P1''\ bin. One
can notice that the width of the minimum on ''t-Z''\ plane has decreased due
to a correlation between metallicity and velocity dispersion, that is
clearly seen on the ''Z-$\sigma$''\ slice. This degeneracy between velocity
dispersion and metallicity can be clearly explained: higher metallicity in
the template than in the observed spectrum increases depth of the absorption
lines, that can be compensated by stronger widening, i.e. using higher
velocity dispersion. This is a very important result. Thus, if one uses
cross-correlation or pixel fitting with the template having a metallicity
very different from observed spectrum, velocity dispersion measurements will
be biased. This might also produce artificial features of velocity
dispersion profiles/maps if the same template is used for regions of galaxy
having different metallicities, e.g. metallicity gradient.

To illustrate how $Z - \sigma$ degeneracy can affect velocity dispersion
maps, we are presenting 2D distributions of velocity dispersion for NGC~3412
obtained by the pixel fitting algorithm for two cases: (a) stellar
population parameters are fixed ($t=4.5$~Gyr, [Fe/H]=-0.05~dex) and only
kinematical parameters ($v, \sigma, h_3, h_4$) are fitted; (a)
SSP-equivalent stellar population parameters is fitted together with
kinematical ones. MPFS data for NGC~3412 were obtained in a frame of studies
of nearby lenticular galaxies, P.I.: Olga Sil'chenko, and kindly provided by
her.

NGC~3412 is a giant lenticular galaxy, exhibiting relatively high central
metallicity (up-to [Fe/H]=+0.2), and exactly solar [Mg/Fe] abundance ratio
(Sil'chenko, 2006), so fitting PEGASE.HR template spectra is not expected to
produce systematic errors due to template mismatch. The galaxy is known to
contain a counter-rotating core (Aguerri et al. 2003). NGC~3412 was one of
the first objects, where velocity dispersion dip in the central region was
found. Age distribution across the galaxy is almost flat with a mean value
of 4.5~Gyr. However, there is a sharp gradient in the metallicity in the
inner 3~arcsec, where it changes from -0.05 to +0.21.

\begin{figure}
\begin{tabular}{c c}
(a) & (b) \\
\includegraphics[width=7cm]{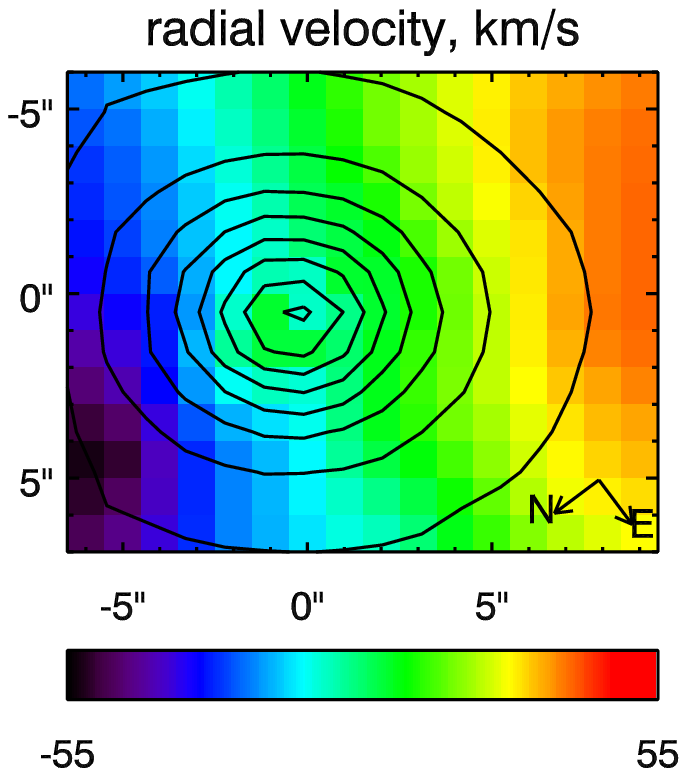} &
\includegraphics[width=7cm]{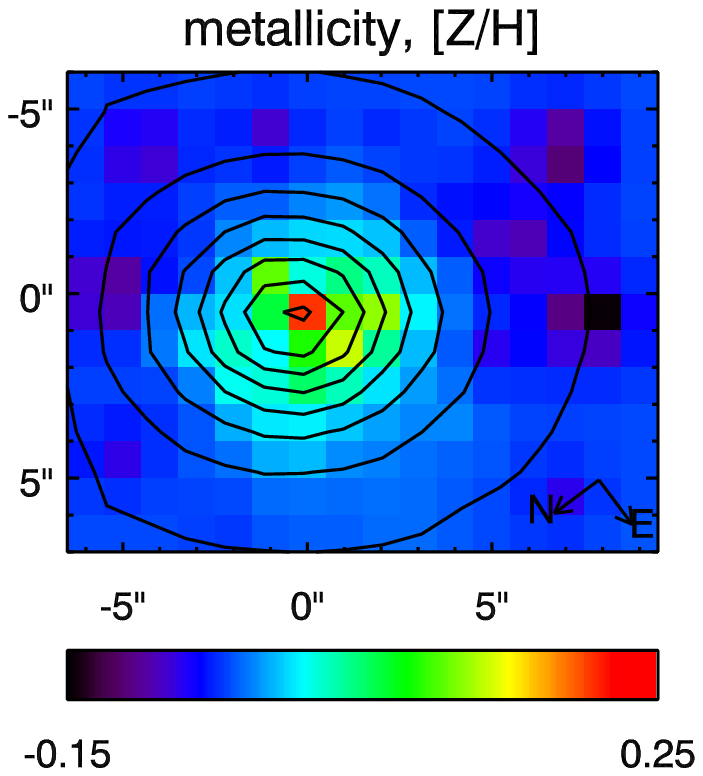} \\
(c) & (d) \\
\includegraphics[width=7cm]{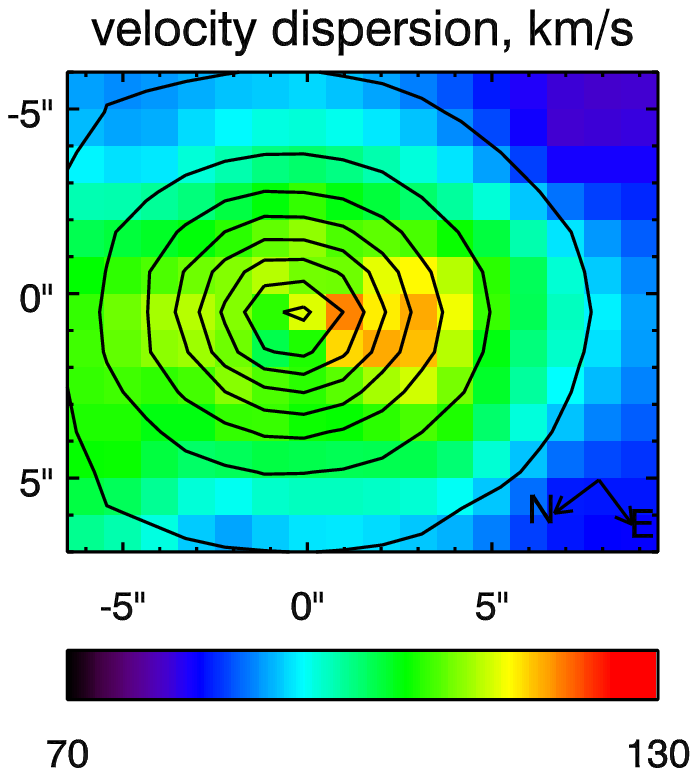} &
\includegraphics[width=7cm]{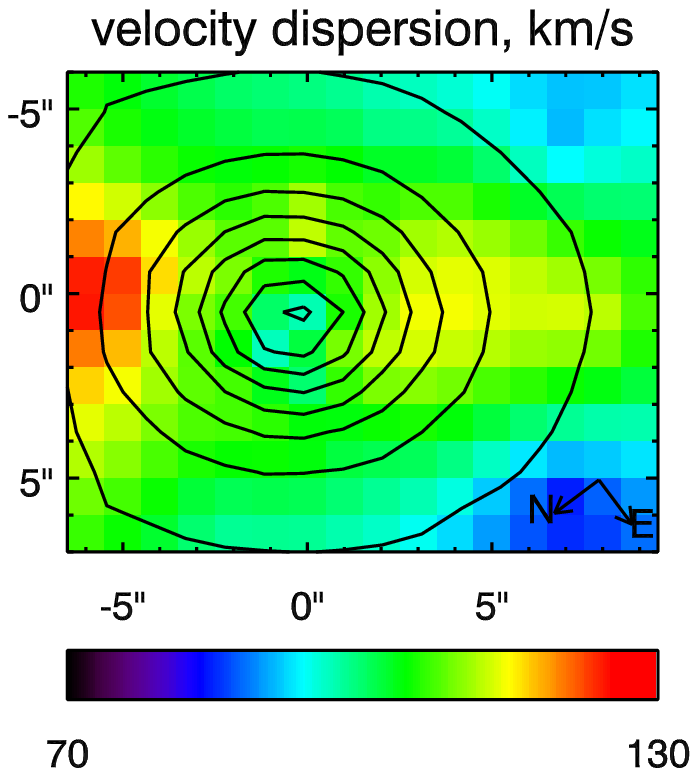} \\
\end{tabular}
\caption{Kinematics and metallicity of NGC~3412. 
Radial velocity field and SSP-equivalent metallicity map are shown in (a)
and (b) panels respectively; (c) and (d) represent stellar velocity
dispersion maps for a case of variable and fixed stellar population.
Counter-rotating core signature is clearly seen in the radial velocity
field. \label{fign3412}}
\end{figure}

In Fig.~\ref{fign3412} 2-dimensional radial velocity field
(Fig.~\ref{fign3412}a), metallicity distribution (Fig.~\ref{fign3412}b), and
two velocity dispersion fields for variable (Fig.~\ref{fign3412}c) and fixed
(Fig.~\ref{fign3412}d) stellar population parameters are shown. All fits are
made for the adaptively binned data using Voronoi tessellation for a target
S/N=40 (see next chapter for details). One can notice a valuable central dip
in the velocity dispersion distribution (down to 95~km~s$^{-1}$) obtained
when fitting fixed stellar population, whereas it remains above
110~km~s$^{-1}$ when stellar population parameters are fitted together with
kinematics.

We have checked three galaxies, included in the SAURON sample, exhibiting
velocity dispersion dips (Emsellem et al. 2004): NGC~2768, NGC~3384
(observations with MPFS at 6-m telescope), and NGC~4150 (GMOS-N at Gemini,
archival data). In all three cases velocity dispersion dips either
disappeared completely (NGC~3384 and NGC~4150), or became less significant in
case of NGC~2768 -- 70~km~s$^{-1}$ compared to 120~km~s$^{-1}$. In NGC~2768
central metallicity exceeds +0.5~dex. For such high metallicities, quality
of synthetic spectra if far from ideal due to lack of metal-rich stars in
the stellar libraries, used to construct them. Thus, we might expect further
increase of the velocity dispersion values in the central part of NGC~2768,
whenever quality of spectral synthesis gets improved.

\subsection{Stability of solutions}
\label{subsecstab}
Stability of solutions is a crucial point for every method dealing with
multiparametric non-linear minimization. We studied the stability with respect to
initial guess, wavelength range being used, and degree of the multiplicative
polynomial continuum.

\subsubsection{Initial guess}
\label{subsubsecstabini}
We have made several dozens of experiments with different initial guesses in
order to inspect the stability of convergence. We found no problems with
starting guess of age, metallicity and velocity dispersions in quite a wide
range of values. The only critical parameter is radial velocity -- the
initial guess needs to be within 2 values of velocity dispersion from the
solution that is around 100-150~km~s$^{-1}$ in case of MPFS data for IC~3653
(see next chapter).

For the 3-points binning solutions do not converge to exactly the same point
of the parameter space, but scattering of values (standard deviation) is
negligible: around 3~Myr for age, 0.0003~dex for metallicity, 0.02~km~s$^{-1}$ for
velocity dispersion, and 0.002~km~s$^{-1}$ for radial velocity.

\subsubsection{Wavelength range}
\label{subsubsecwlr}
We ran two series of experiments: one with $\lambda > 4700$\AA, and another one
with the full wavelength range, but regions of Balmer lines (H$\gamma$ and 
H$\beta$) masked. The reasons for the first experiment is: a region between
$4150 $\AA$ < \lambda < 4700$\AA\ contains a lot of strong absorption features
related to metals, thus one might expect to have metallicity estimations biased
in case of unknown problems with the algorithm and/or presence of additive
continuum (for instance, due to incorrect subtraction of night sky, or diffuse
light) varying with wavelength. There is a similar reason for the second
experiment because of age: Balmer lines are known to be good age estimators
(Worthey at al. 1994, Vazdekis\& Arimoto 1999).

\begin{table}
\begin{tabular}{l c c c}
 & P1 & P2 & P3 \\
\hline
$v$,~km~s$^{-1}$ &  601.8$\pm$   1.0 &  603.4$\pm$   1.4 &  603.8$\pm$   3.0 \\
 &  600.9$\pm$   1.0 &  603.1$\pm$   1.7 &  603.7$\pm$   3.4 \\
\hline
$\sigma$,~km~s$^{-1}$ &   70.9$\pm$   1.6 &   67.3$\pm$   2.2 &   52.1$\pm$   5.0 \\
 &   71.8$\pm$   1.6 &   65.3$\pm$   2.6 &   52.1$\pm$   5.7 \\
\hline
t, Gyr &  4.855$\pm$ 0.218 &  4.728$\pm$ 0.289 &  4.629$\pm$ 0.694 \\
 &  4.714$\pm$ 0.235 &  4.448$\pm$ 0.403 &  4.238$\pm$ 0.930 \\
\hline
Z, dex &   0.01$\pm$  0.02 &  -0.14$\pm$  0.02 &  -0.15$\pm$  0.05 \\
 &   0.03$\pm$  0.01 &  -0.13$\pm$  0.02 &  -0.15$\pm$  0.05 \\
\hline
\end{tabular}
\caption{Stability of the solutions for the 3-points
 binning with respect to the wavelength range. First lines for every parameter
 correspond to $\lambda > 4700$\AA, second ones to full range with Balmer lines
 excluded. See also Tab.~\ref{tabcomptz3b}
\label{stabwlr}
}
\end{table}

One may notice, that cutting the blue part of the spectrum does not affect
the results, but increases uncertainties of the determination of parameters.
The second set of experiments shows similar results. Errors of age are quite
large and they become comparable to the precision of Lick indices (see next
Chapter, Tab.~\ref{tabcomptz3b}). However,
values themselves coincide with the results of the fit for full wavelength
range within 1$\sigma$. It is quite a remarkable result, that demonstrates
that even without Balmer lines it is possible to give estimations of age of
the stellar population, because in case of pixel fitting the usage of the
information contained in the spectrum is by far much more optimal than in
case of Lick indices.

\subsubsection{Order of multiplicative polynomial continuum}
\label{subsubsecmult}
\begin{figure}
 \includegraphics[width=17.0cm]{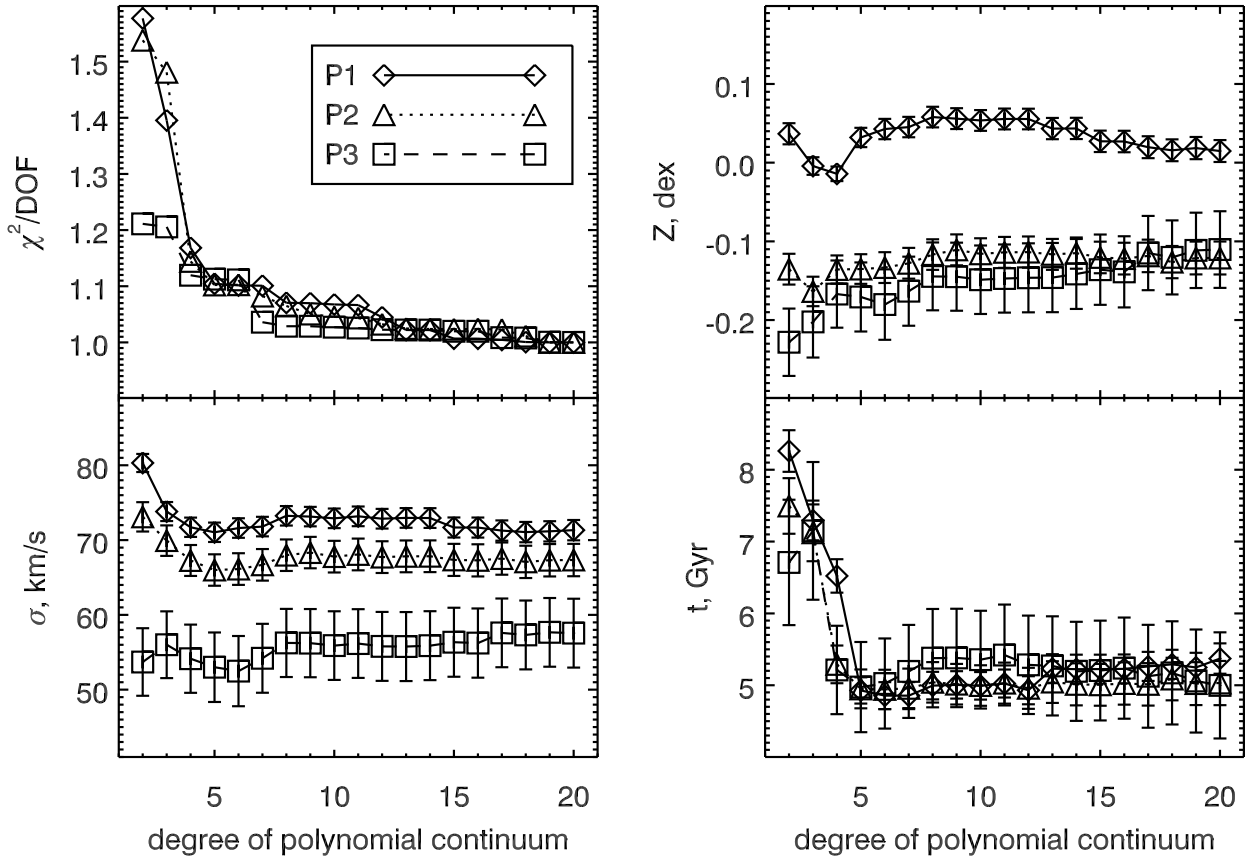}
\caption{Stability of the parameters with respect to the order of the
multiplicative polynomial continuum. Different plotting symbols correspond to
three bins of the 3-points binning.
\label{figstabmult}
}
\end{figure}

We also explored the stability of the method with respect to the order of
the multiplicative polynomial continuum. The results (for MPFS data) are
shown in Fig~\ref{figstabmult}. One can see that for n$>$5 there is neither
significant changes of the estimations of kinematical and stellar population
parameters, nor of $\chi2$ value. Time of computation is growing with $n$
quite rapidly, because it is minimized non-linearly. Thus we chose $n=5$ for
all our data analysis of MPFS datasets. Using similar technique we found
$n=9$ as the optimal value for Giraffe data analysed in this work.

\subsection{Possible biases}
\label{subsecbias}
There are several possible sources of systematic errors on the parameters: (1)
additive systematics of the flux calibration due to under- or oversubtraction
of the night sky, (2) imperfections of the models, one of the most important
among those is non-solar abundance ratios of chemical elements.

\subsubsection{Additive terms}
Accurate subtraction of the night sky emission is quite a challenging step of
the data reduction for low-surface brightness objects. Basically, night sky
emission consists of continuum emission, that might include scattered solar light
as well, and several bright emission lines. Under- or oversubtraction of night
sky brings additive component resulting in changing the depths of absorption
spectral features (equivalent widths). This will affect results of the pixel
fitting procedure, and conclusions based on measurements of Lick indices as
well.

\begin{figure}
 \includegraphics[width=17.0cm]{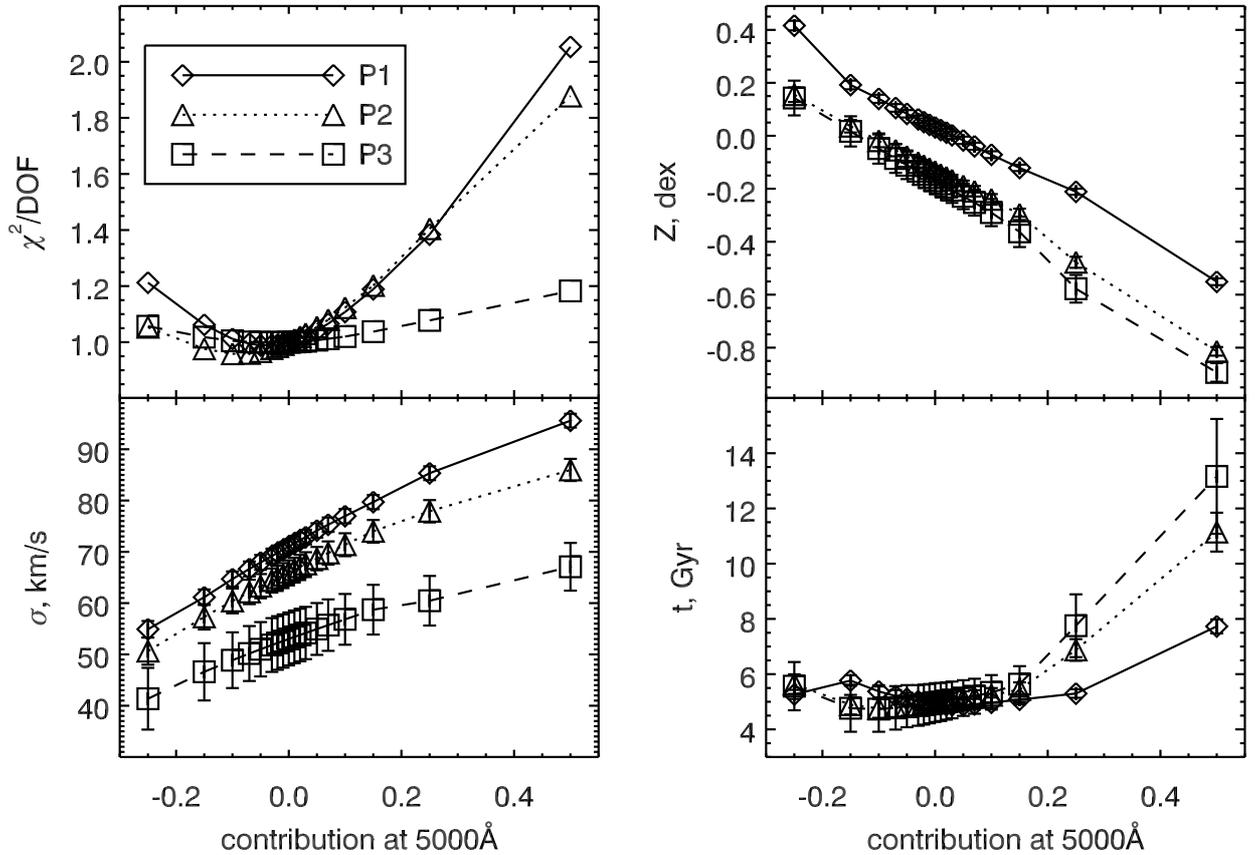}
\caption{Effects of additive terms on the results. Abscissa represents the
contribution of the constant level at $\lambda$=5000\AA. Different plotting
symbols correspond to three bins of the 3-points binning.
\label{figaddcont}
}
\end{figure}

We have conducted two series of experiments: (1) adding a constant term or
(2) heavily smoothed spectrum itself (smoothing window of 300 pixels) to
emulate the diffuse light in the spectrograph. In every series the fraction
of the additive term was between -20 and +50 per cent to model over- and
undersubtraction. Additive polynomial terms were not included in the fit.
Results appear to be virtually the same. The results for
the constant term as a fraction of flux at 5000\AA\ (for MPFS data) are
shown in Fig~\ref{figaddcont}. One may notice that
$\chi^2$ reaches minimum on slightly negative (over-subtraction) values of the
additive term. It is easily understandable taking into account that we did not
change flux uncertainties during our experiments. The remarkable result is
stability of age estimations on a wide range of additive components (-25 to
15 per cent). This is quite an important advantage of the pixel-fitting technique over
Lick indices, because additive terms will always bias age estimations based on
Lick indices, since all index measurements, including H$\beta$, will be biased
in the same way for obvious reasons. Metallicity and velocity dispersion exhibit
expected behaviour: growth of $\sigma$ and fall of Z. Indeed within a range of
contribution between -5 and 5 per cent changes are quite small ($\sim$8 per
cent for $\sigma$, and $\sim$0.1~dex for Z) though significant.

In order to test the consequences of bad sky subtraction we made two
additional experiments: we tried to fit the data (IC~3653 dataset), where
the sky spectrum was represented by a low-order polynomial continuum and
with no sky subtraction at all. We excluded four regions of the spectrum
containing bright emission lines: HgI $\lambda=$4358\AA, 5461\AA, [NI]
$\lambda=$ 5199\AA, and [OI] $\lambda=$ 5577\AA. The experiments were made
for the 3-points binning of the data, demonstrating the effects for high,
intermediate, and low surface brightness (see Tab~\ref{tabpar3bin}).
Basically we found no significant difference for the ''P1''\ and ''P2''\ bins
between the parameters for the correct sky subtraction and subtraction of
the low-order polynomial model of the sky (see Tab~\ref{tabnsky}). ''P3''\ bin
gives younger age and higher metallicity, but the estimations are in
agreement with the normal sky subtraction within 2$\sigma$ However, as
expected, when sky is not subtracted at all we find valuable bias on
$\sigma$, age, and metallicity, and velocity dispersion estimations for the
''P2''\ bin, and even stronger effect for ''P3''. Due to additive continuum
metallicities are found to be lower, ages older, and velocity dispersions
higher than expected. These experiments demonstrate that for the surface
brightness down to $\mu_B=20$~mag~arcsec$^{-2}$ features of the night sky
spectrum do not affect the results of the pixel fitting procedure, and very
rough sky subtraction is sufficient to obtain the realistic estimations of
kinematical and stellar population parameters.

\begin{table}
\begin{tabular}{l c c c}
 & P1 & P2 & P3 \\
\hline
$v$,~km~s$^{-1}$ &  604.3$\pm$   1.0 &  606.0$\pm$   1.5 &  609.4$\pm$   3.2 \\
 &  604.4$\pm$   1.0 &  605.9$\pm$   1.6 &  610.3$\pm$   3.7 \\
\hline
$\sigma$,~km~s$^{-1}$ &   71.5$\pm$   1.5 &   64.9$\pm$   2.4 &   54.4$\pm$   5.2 \\
 &   80.5$\pm$   1.5 &   88.3$\pm$   2.2 &  105.5$\pm$   4.6 \\
\hline
t, Gyr &  4.868$\pm$ 0.210 &  4.547$\pm$ 0.310 &  3.956$\pm$ 0.731 \\
 &  4.972$\pm$ 0.185 &  7.203$\pm$ 0.390 & 12.982$\pm$ 1.641 \\
\hline
Z, dex &   0.04$\pm$  0.01 &  -0.10$\pm$  0.02 &  -0.06$\pm$  0.04 \\
 &  -0.08$\pm$  0.01 &  -0.50$\pm$  0.02 &  -0.94$\pm$  0.03 \\
\hline
\end{tabular}
\caption{Determination of the kinematical and stellar population parameters
 for a case of polynomial night sky model (first line for every parameter)
 and no sky subtraction (second line for every parameter).
\label{tabnsky}
}
\end{table}

\subsubsection{Non-solar [$\alpha$/Fe] ratios}
In order to assess reliability and precision of the stellar population
parameters found by the pixel fitting procedure under different
circumstances, we have conducted a number of tests using Monte-Carlo
simulations and real published datasets. Taking into account quite a high
fraction of massive objects exhibiting supersolar values of [Mg/Fe] in the
sample of early-type galaxies in Abell~496 cluster, the most principle
questions for the validation of results are: does our technique produces
biased estimations of SSP-equivalent ages and metallicities in case of
non-solar [Mg/Fe] abundance ratio? do they depend on the presence of
H$\beta$? If there are biases, is it still possible to apply some empirical
corrections?

Up to now there was no attempts to model spectral energy distribution of
synthetic stellar populations for non-solar $\alpha$-element abundance
ratios. Therefore we have to use published spectral data, where age and
metallicity can be estimated using both: Lick indices and pixel fitting. 
Since there is a tight correlation between [Mg/Fe] abundance ratio and
central velocity dispersion (and luminosity as well) of the galaxies:
significantly positive [Mg/Fe] ratios are observed in galaxies with
$\sigma>120$~km~s$^{-1}$, spectral resolution of the dataset does not need
to be very high. We decided to use spectral data from Nearby Field Galaxy
Survey (Jansen et al. 2001) obtained with the FAST spectrograph at the F. L.
Whipple Observatory's 1.5 m Tillinghast telescope. Data have very wide
spectral range: 3600\AA\ to 7500\AA, spectral resolution is about 6\AA in
the middle of the range ($\sigma_{inst}=155$~km~s$^-1$ at 5200\AA). We have
also used twilight spectra available through NOAO FAST archive observed with
the same setup of the spectrograph and at the same periods, as the NFGS data
were obtained. Our goal was to determine LSF of FAST and its variations
along the wavelength range. Nearly all 200 galaxies observed in a frame of
NFGS presented by two spectra: nuclear, representing inner part of a galaxy,
and total spectrum integrated along the slit.

We have selected only spectra of early type galaxies (E, S0, Sa) with no
visible emission lines. 

To measure Lick indices we have degraded spectral resolution of the NFGS
by convolving original spectra with the Gaussian countour having width
$\sigma_{degr} = \sqrt{\sigma_{Lick}^2 - \sigma_{FAST}^2 - \sigma_{g}^2}$,
where $\sigma_{g}$ is velocity dispersion of the galaxy (in a given
spectrum), $\sigma_{FAST}$ is a width of FAST LSF depending on a given Lick
index, and $\sigma_{Lick}$ is the resolution needed to measure Lick
indices, also depending on a given index. If the value under the square root
turned to be negative, no degradation was done. Instead, $\sigma$-correction
according to Kuntschner (2004) was applied to the measurements of Lick
indices using $\sigma_{corr} = \sqrt{\sigma_{FAST}^2 + \sigma_{g}^2 -
\sigma_{Lick}^2}$.

After having measured Lick indices, selection was restricted to objects
having values of H$\beta$ index between 1.4 and 2.6\AA, and $<$MgFe$>$ index
between 2.0 and 4.2\AA: in this range models by Thomas et al. (2003) in a
range of $Z=$-0.5~...~+0.5~dex, and $t=$2~...~15~Gyr form unequivocally
reversible grids for any [Mg/Fe]. Resulting sample contains 49 NFGS spectra:
25 nuclear and 24 integrated ones.

Every spectrum was fitted three times in different wavelength ranges: (1)
between 4300 and 5600\AA\ (full range of MPFS), (2) between 4800 and 5600\AA
(Giraffe spectral range for Abell~496 galaxies including H$\beta$), (3)
between 4880 and 5600\AA (the same, but excluding H$\beta$). Three tests
were conducted in order to assess stability of age estimations for objects
with non-solar [Mg/Fe] abundance ratios with respect to the wavelength range
being used.

Values of ages and metallicities obtained by pixel fitting have been
compared to the results obtained by inverting a grid of Lick indices:
H$\beta$ and $<$MgFe$>$. For 38 of 49 spectra (77 percent) age estimations
coincide within 1$\sigma$ confidence level. We found no correlation between
[Mg/Fe] and $t_{fit} - t_{Lick}$. This is a strong argument for using pixel
fitting technique with PEGASE.HR models to estimate ages of stellar
populations even for non-solar [Mg/Fe] abundance ratios.

\chapter{dE galaxies in the Virgo cluster}
In this chapter we present 3D spectroscopic observations of Virgo cluster
dE galaxies in order to bring further observational constraints to the
evolutionary scenarios of dE's formation. Velocity fields and spatial
distribution of the stellar population are most needed to check if
counterparts of the observed kinematical sub-structures can be detected.

\section{IC~3653}
In this section we are presenting the first 3D observations of a dE. IC~3653
is a bright dE galaxy belonging to the Virgo cluster (Binggeli et al.
1985). In Tab~\ref{tabic3653params} we summarize its main characteristics.
IC~3653 was chosen because it
is amongst the most luminous dE in Virgo and has a
relatively high surface brightness. It is located 2.7~deg from the center
of the cluster, i. e. 0.8~Mpc in projected distance. Its radial velocity 588$\pm4$~km~s$^{-1}$
(this work) confirms its membership to the Virgo cluster, the velocity difference from the
mean velocity of Virgo (1054~km~s$^{-1}$, HyperLeda, Paturel et al. 2003 
\footnote{http://leda.univ-lyon1.fr/}) is nearly -470~km~s$^{-1}$. 
IC~3653 is located some 100~kpc in the projected distance from NGC~4621, a
giant elliptical galaxy having a similar radial velocity value
(410~km~s$^{-1}$, HyperLeda) With other low luminosity Virgo cluster
members, in particular IC~809, IC~3652 for which the radial velocities have been
measured, they may belong of a physical substructure of Virgo, crossing the
cluster at 500 km~s$^{-1}$.

Velocity and velocity dispersion
profiles from by Simien \& Prugniel (2002) show some rotation. ACS/HST
archival images from the Virgo cluster ACS survey (C\^ot\'e et al.2004) 
are also available and will be discussed here.

\begin{table}
  \begin{tabular}{l l}
\hline
Name & IC3653, VCC1871 \\
Position & J124115.74+112314.0 \\
B & 14.55\\
Distance modulus & 31.15\\
A(B) & 0.13\\
M(B)$_{corr}$ & -16.78\\
Spatial scale & 82 pc~arcsec$^{-1}$ \\
Effective radius, $R_e$ & 6.7 arcsec $\equiv$ 550 pc \\
$\mu_B$, mag~arcsec$^{-2}$ & 20.77 \\
Ellipticity, $\epsilon$ & 0.12\\
S\'ersic exponent, $n$ & 1.2\\
Heliocentric cz, km~s$^{-1}$ & 588 $\pm$ 4 \\
$\sigma_{cent}$, km~s$^{-1}$ & 80 $\pm$ 3 \\
$V_{max}$, km~s$^{-1}$  & 18 $\pm$ 2\\
$V_{max}/\sigma$ & 0.27$\pm$0.08\\
$t$, Gyr (lum. weighted) & 5.2$\pm$0.2\\
$[Z/\mbox{H}]$, dex (lum. weighted) & -0.06$\pm$0.02\\
\hline
  \end{tabular}
  \caption{General characteristics of IC~3653. S\'ersic exponent, kinematical
  and stellar population parameters are obtained in this work, other properties
  are taken from HyperLEDA and Goldmine databases, and from Ferrarese et al. 2006.
  Uncertainties given for age and metallicity correspond to the measurements
  on co-added spectra.
  \label{tabic3653params}
}
\end{table}

\subsection{Spectroscopic observations and data reduction}
\label{secobs3653}
The spectral data we analyse were obtained with the MPFS integral-field
spectrograph. 

The Multi-Pupil Fiber Spectrograph (MPFS), operated on the 6-m telescope
Bolshoi Teleskop Al'tazimutal'nij (BTA) of the Special Astrophysical Observatory
of the Russian Academy of Sciences, is a fibre-lens spectrograph with a
microlens raster containing $16 \times 16$ square spatial elements together
with 17 additional fibres transmitting the sky background light, taken
four arcminutes away from the object. The size of each element is 1''$
\times$1''. We used the grating 1200 gr mm$^{-1}$ providing the reciprocal
dispersion of 0.75~\AA~pixel$^{-1}$ with a EEV CCD42-40 detector.

Observations of IC~3653 were made on 2004 May 24 under good
atmosphere conditions (seeing 1.4''). The total integration time was 2
hours. The
spectral resolution, as determined by analysing twilight spectra, 
varied from $R=1300$ to $R=2200$ over the field of view
and the selected spectral range (4100\AA -- 5650\AA).
The resolution is lower in the centre of the field and it slightly increases
toward top and bottom; there is also a smooth increase of the resolution in the
red end of the wavelength range (Moiseev, 2001).

The following calibration frames were taken during the observations
of IC~3653 with MPFS (as for any absorption line spectra):
\begin{enumerate}
\item BIAS, DARK.
\item ''Etalon'': 17 night-sky fibres illuminated by the incandescent bulb.
This frames are used to determine positions of spectra on the frame.
\item ''Neon''\ (arc lines): by exposing the spectral lamp filled with
Ar-Ne-He to perform a wavelength calibration.
\item The internal flat field lamp.
\item A spectrophotometric standard ($Feige~56$ for our observations), used
to turn the spectra into absolute flux units.
\item A standard for Lick indices and radial velocity ($HD~137522$ and 
$HD~175743$), used also to measure instrumental response: asymmetry and width of
the line-spread function.
\item ''SunSky'': twilight sky spectra for additional corrections of the 
systematic errors of the dispersion relation and transparency
differences over the fibres.
\end{enumerate}

\subsubsection{Data reduction}
\label{subsecdatared}
The data reduction for integral-field spectroscopy (3D spectroscopy) is a
quite elaborated procedure. We use the original IDL software package created
and maintained by V. Afanasiev. We introduced some modifications in the
package: error frames are created using photon statistics and then processed
through all the stages to have realistic error estimates for the fluxes in
the resulting spectrum. Besides we included an option to get spectra
logarithmically rebinned in a wavelength, because it was necessary for the
extraction of kinematics, and it allowed to avoid resampling the spectra
twice.

The primary reduction process (up-to obtaining flux-calibrated data cube)
consists of:
\begin{enumerate}
\item Bias subtraction, cosmic ray cleaning.
   Cosmic ray cleaning implies the presence of several frames. Then they are
   normalized and combined into the cube
   (x,y,Num). The cube is then analysed in each pixel through ''Num''\ frames.
   All counts exceeding some level (5-$\sigma$) are replaced with the robust
   mean through the column. Then the cleaned cube is summed.
\item Creation of the traces of spectra in the ''etalon''\ image.
   Accuracy of the traces is usually about 0.02 or 0.03 pixels.
\item Flat field reduction and diffuse light subtraction.
   Flat field is applied to the CCD frames before extracting
   the spectra. The scattered light model is also constructed and
   subtracted from the frames during this step. It is made using parts
   of the frames not covered by spectra and then interpolated with
   low-order polynomials.
\item Creation of the traces for every fibre.
   On this step the traces are determined for each fibre in the
   microlens block (presently 256 fibres) using the night sky fibres
   traces created on the 2-nd step and interpolation between them
   using the tabulated fibre positions.
\item Spectra extraction.
   Using the fibre traces determined in the previous steps, spectra
   are extracted from science and calibration frames using fixed-width
   Gaussian (usually with FWHM=5~px for the present configuration of the
   spectrograph). The night sky spectra are also extracted from the science
   frames.
\item Creation of dispersion relations.
   Spectral lines in the arc lines frame are identified and dispersion relations
   are computed independently for every fibre.
\item Wavelength rebinning.
   All the spectra of night sky, object and standard stars are rebinned
   independently into logarithm of wavelength.
   The sampling on the CCD varies between 0.65 and 0.85~\AA\ and we rebinned
   to a step of 40~km s$^{-1}$, i. e. 0.55 to 0.75~\AA, corresponding to the
   mean oversampling factor 1.2.
\item Sky subtraction.
   Median vector of the night sky is computed using spectra of 17 night sky 
   fibres. Then it is subtracted from each fibres after applying correction
   computed from dome flat and twilight sky flat. This is necessary because
   apertures of night sky fibers are twice the size of object fibres.
\item Determination of the spectral sensitivity.
   Using the spectrophotometric standard star, the ratio between counts and
   absolute flux is calculated and then approximated with a high-order
   polynomial function over the whole wavelength range.
\item Flux calibration of the data cube.
   The spectral sensitivity curve is used to perform the flux calibration.
   Finally, the values in the data cube correspond to the $F_{\lambda}
   [erg \cdot cm^{-2} \cdot s^{-1} \cdot $\AA$^{-1}]$.
\end{enumerate}

\subsubsection{Spatial adaptive binning}
Dwarf elliptical galaxies exhibit a decrease of surface brightness
in peripheral parts, making difficult to obtain precise measurements of
kinematics and Lick indices in every spatial element. In our data
$\mu_{B}$ changes from 18.5~mag~arcsec$^{-2}$ in the centre down to 
21.2~mag~arcsec$^{-2}$ in the outer parts of the field of view.
At the same time, the
signal-to-noise ratio at the central part is high enough ($\sim 30$) for detailed
analysis, so smoothing the whole field spatially with the same window might destroy
some important details there. To avoid these problems, the Voronoi adaptive
binning procedure (Cappellari \& Copin, 2003) can be applied. This technique was
especially developed to work with the data coming from panoramic spectroscopy and its
main idea is usage of variable bin size to achieve equal signal-to-noise
ratio in every bin.

The result of Voronoi 2D binning procedure is a set of 1D spectra, for those
all further steps of the analysis might be done independently. For the
kinematical analysis we will use a target signal-to-noise ratio of 15, and
for stellar population analysis we will use 30.

Besides we will be using a tessellation of the dataset containing only three
bins (3-points binning hereafter): central condensation (3 by 3 arcsec region
around the centre of the galaxy), elongated disky substructure (14 by 7 arcsec)
oriented according to kinematics (see subsection 4, Fig~\ref{spec4}, illustrating
locations of bins and demonstrating spectra integrated in them) with the central
region excluded, and the rest of the galaxy. Such a physically-stipulated
tessellation allows to gain high signal-to-noise ratios in the bins in order to
have high quality estimations of the stellar population parameters in the
regions where populations are expected to differ. In Tab~\ref{tabpar3bin} we
present the parameters of the resulting bins.

\begin{table}
\begin{tabular}{l c c c c}
Bin & N$_{spax}$ & m(AB) & $\mu$(AB) & S/N \\
\hline
P1 &    9 &  16.3& 18.7& 69\\
P2 &   77 &  15.2& 19.9& 49\\
P3 &  122 &  15.7& 20.9& 21\\
\hline
\end{tabular}
\caption{Parameters of the ''3-points''\ binning: number of spatial
elements, mean AB magnitude, mean AB surface 
brightness (mag~arcsec$^{-2}$), and mean signal-to-noise ratio at 5000\AA.
\label{tabpar3bin}
}
\end{table}

\subsection{SSP age and metallicity derived from Lick indices}
\label{seclick}
A classical and effective method of studying stellar population properties 
exploits diagrams for different pairs of Lick indices (Worthey et al., 1994). 
A grid of values, corresponding
to different ages and metallicities of single stellar population models 
(instantaneous burst, SSP), is plotted
together with the values computed from the observations. 
A proper choice of the pairs of indices, sensitive to mostly age or metallicity
like  H$\beta$ and Mg$b$, allows to determine SSP-equivalent age and 
metallicity.

\begin{figure}
\hfil
\begin{tabular}{c c}
 (a) & (b) \\
 \includegraphics[width=8cm,height=8cm]{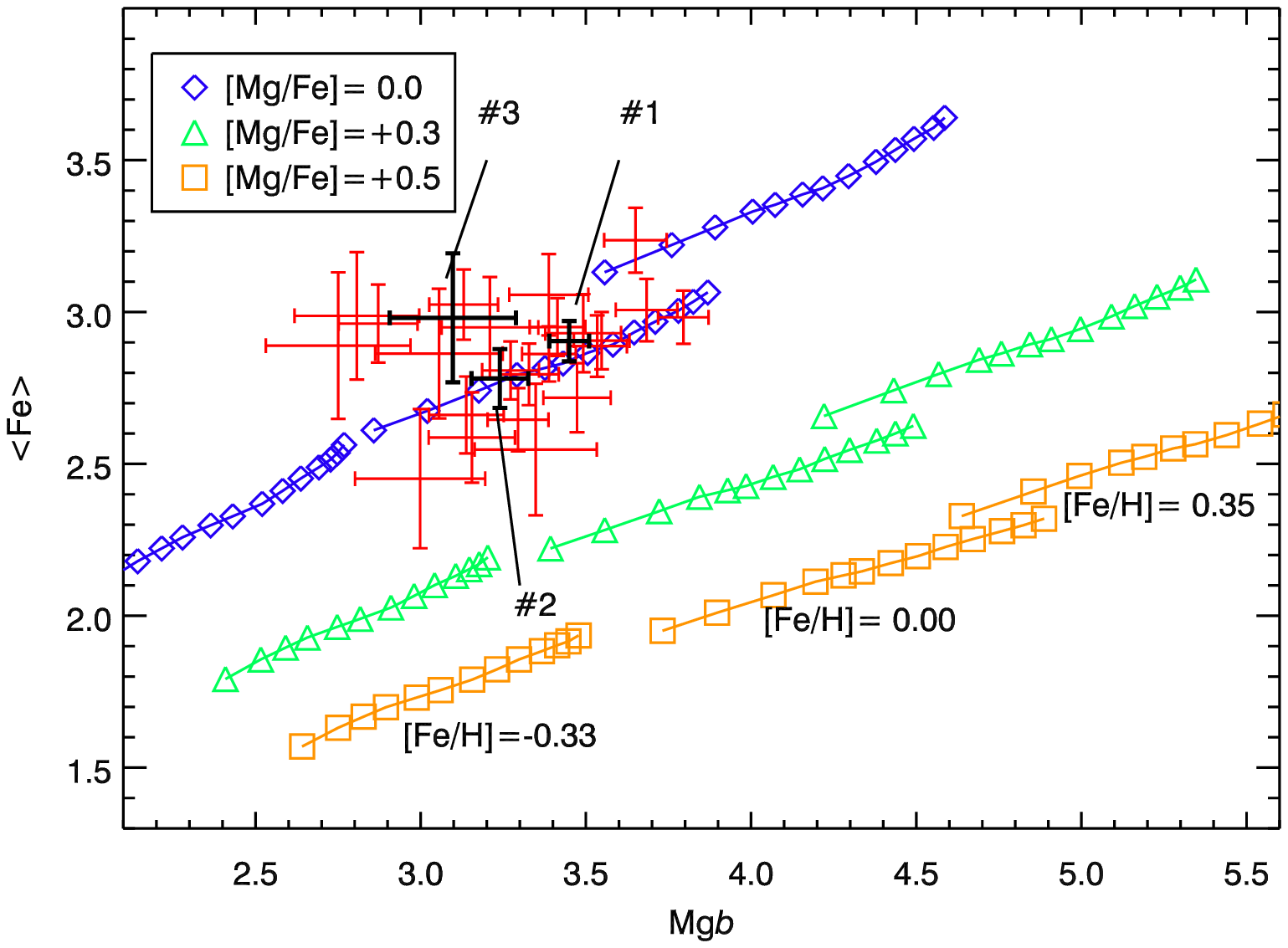} &
 \includegraphics[width=8cm,height=8cm]{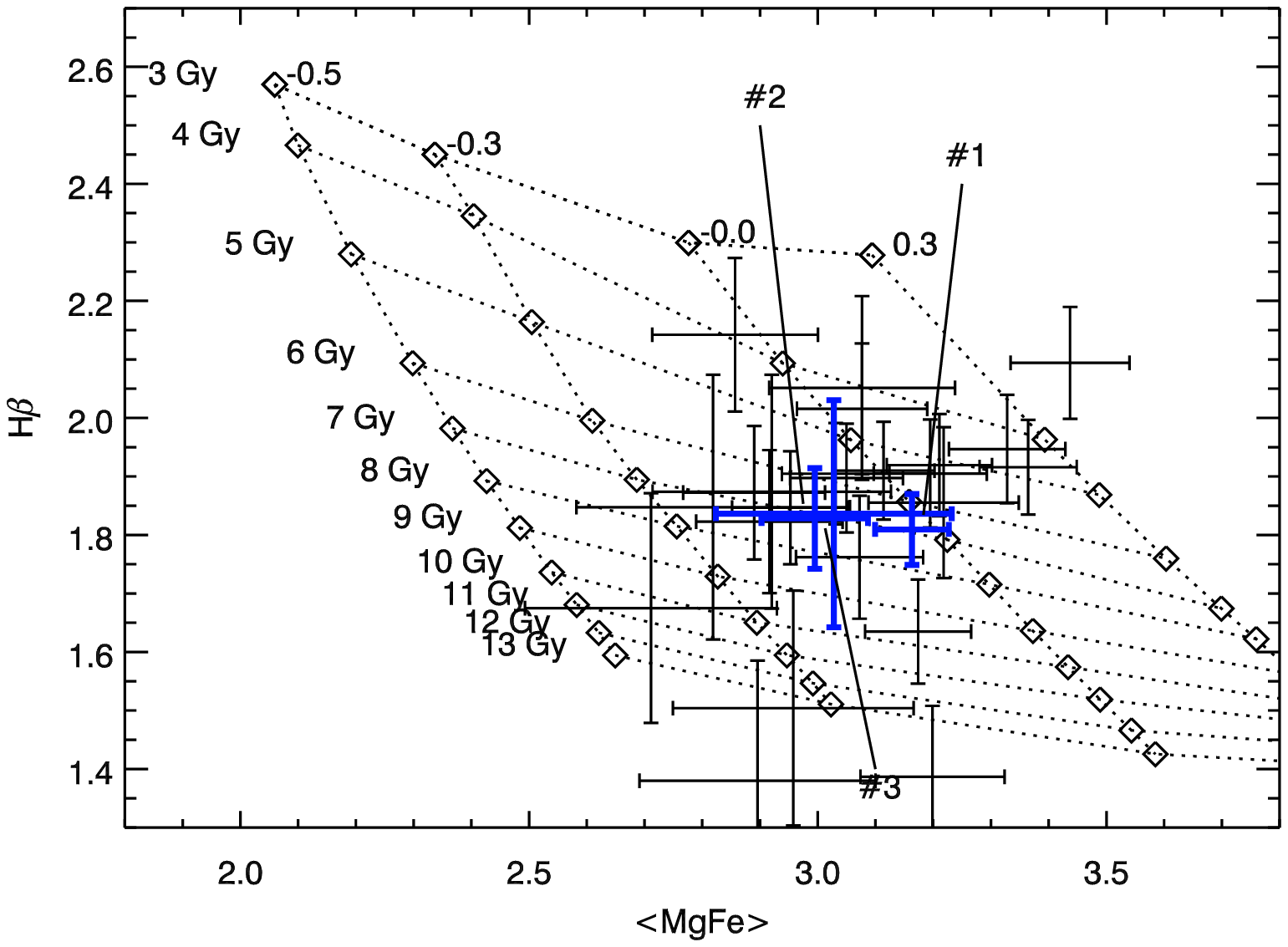} \\
 (c) & (d) \\
 \includegraphics[width=8cm,height=8cm]{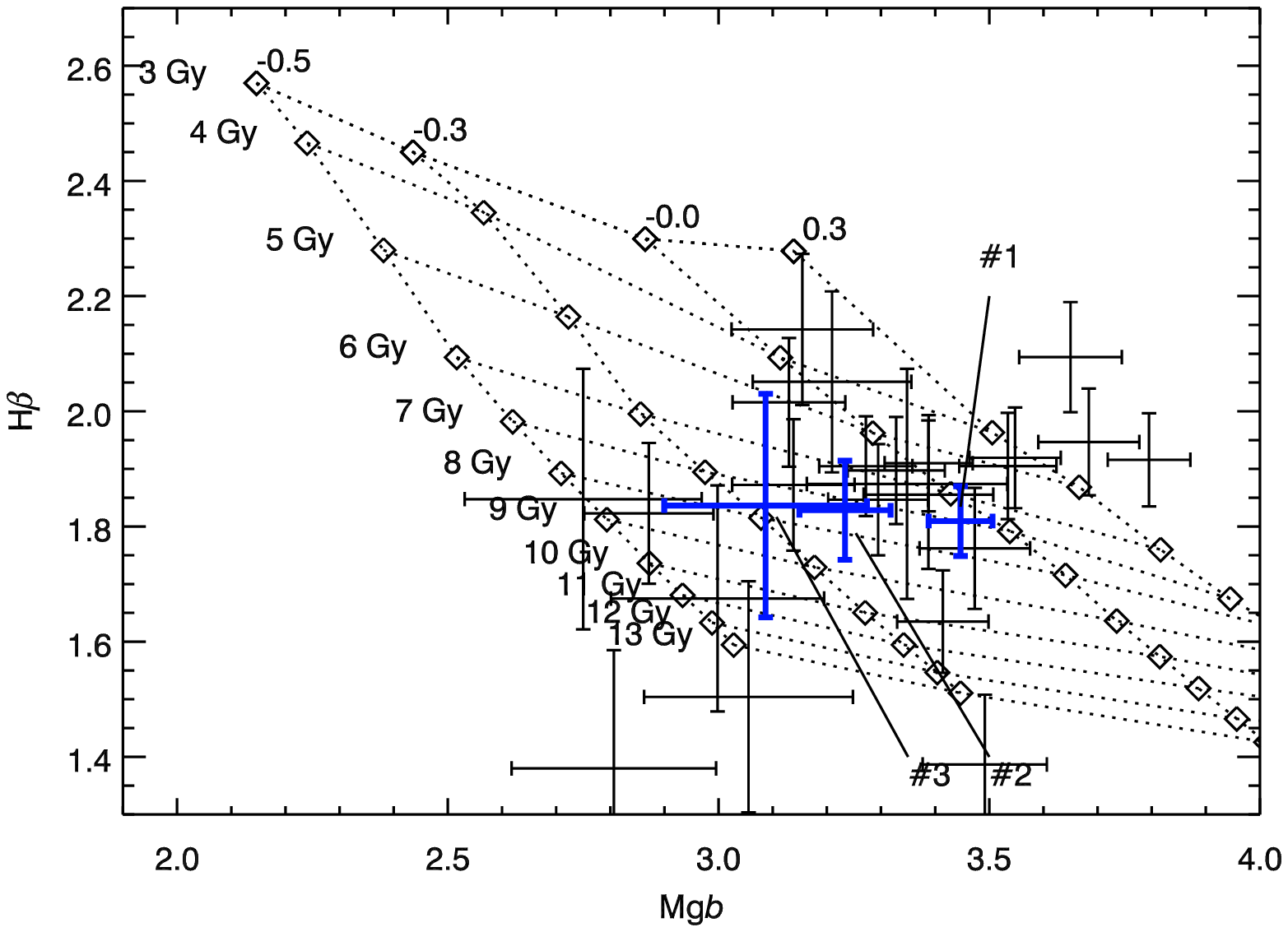} &
 \includegraphics[width=8cm,height=8cm]{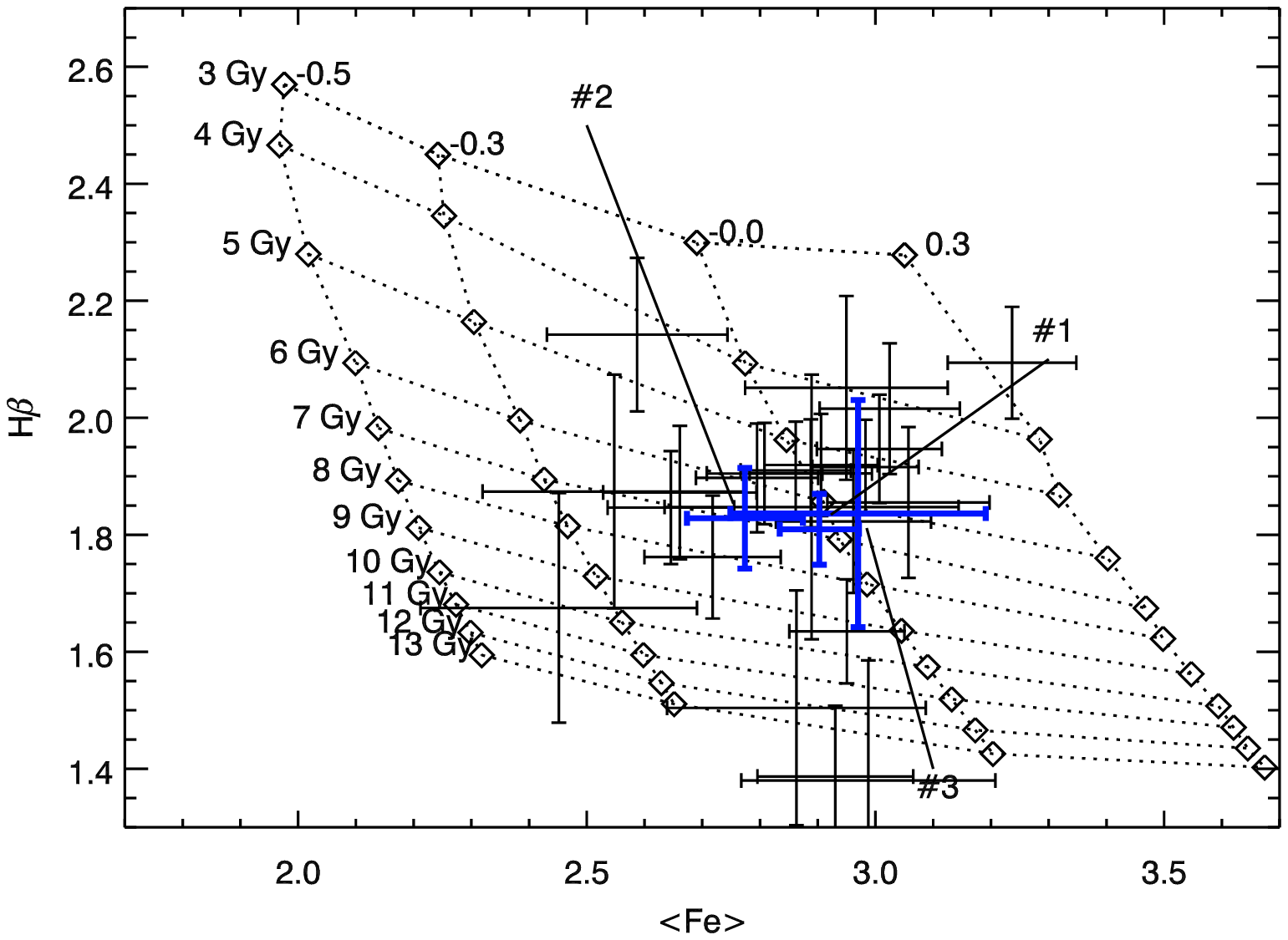} \\
\end{tabular}

\caption{The Mg$b$ - $<$Fe$>'$ (a), H${\beta}$ - [MgFe] (b), H${\beta}$ -
Mg$b$ (c)
and H${\beta}$ - $<$Fe$>'$ (d) diagrams. On (a) the models from Thomas et al. (2003)
are plotted. On (b), (c), and (d) the displayed grid is constructed of the
values of the Lick indices for PEGASE.HR synthetic spectra (SSP) for different
ages and metallicities. Bold crosses with pointers represent measurements for 3
regions of the galaxy (see text), thin crosses are for individual bins for
Voronoi tessellation with target S/N=30.
\label{lickdiag}
} \hfil
\end{figure}

We use a grid of models computed with the evolutionary synthesis code:
PEGASE.HR (Le Borgne et al., 2004). These models are based on the 
empirical stellar library ELODIE.3 (Prugniel \& Soubiran 2001, 2004)
and are therefore bound to the [Mg/Fe] abundance pattern of the solar 
neighborhood (see Chen et al. (2003) and references in it).
To show that this limitation is not critical for our (low-mass) galaxies,
Fig~\ref{lickdiag}a presents the Mg$b$ versus $<$Fe$>$ diagram with the models
by Thomas et al. (2003) for different [Mg/Fe] ratios overplotted. These data
allow to conclude that IC~3653 has solar [Mg/Fe] abundance ratio with a
precision of about 0.05~dex.

We tried to use different metallicity tracers among the ''standard''\ set of Lick
indices to see possible effects of abundance ratios: Mg$b$, combined iron index
$<$Fe$>'=0.72$Fe$_{5270} + 0.28$Fe$_{5335}$, and  ''abundance-insensitive''\  
[MgFe]$ = \sqrt {\mbox{Mg} <\mbox{Fe}>'}$ (Thomas et al. 2003). The statistical
errors on the measurements of Lick indices were computed according to 
Cardiel et al. (1998).

On Fig~\ref{lickdiag}c the popular pair, H${\beta}$-Mg$b$ is
presented. The grid is constructed from the values of the corresponding indices
measured on PEGASE.HR SSPs.

One of important disadvantages of Lick indices is their high sensitivity to
missed/wrong values in the data, for example due to imperfections of the
detector, or uncleared cosmic ray hits. A simple interpolation of
the missed values (e.g. linear or spline) cannot be used, because if some
important detail in the spectrum, e.g. absorption line, is affected,
the final measurement of the index will be biased. Pseudo-continuum
and index regions are defined as mean fluxes without possibility of
weighting individual pixels (see equations 1, 2, and 3 in Worthey et al. 1994)
Due to a defect of the detector, our data  have
a 3pixel wide bad region (hot pixels) in the middle of the blue continum
of Mg$b$. So, strictly speaking, we could not measure Mg$b$ at all,
neither H${\beta}$ on a significant part of the field of view.

As a workaround, we replaced all the missing or flaged values in the
data cube by the corresponding values of the best-fitting model determined
as explained in the next subsection.

In Tab~\ref{tablick3b} we present measurements of selected Lick indices
for the 3-points binning. We see almost no population difference among three bins
within the precision we reach. Age is around 6~Gyr, metallicity is about solar
for ''P1'', and slightly subsolar for ''P2''\ and ''P3''.

\begin{table}
\begin{tabular}{l l l l}
Name & bin 1 & bin 2 & bin 3 \\
\hline
Ca4227& 1.062 $\pm$  0.080 & 0.874 $\pm$  0.190 & 0.689 $\pm$  0.700 \\
 & 1.096              & 1.020              & 1.092              \\
G4300& 5.106 $\pm$  0.135 & 5.042 $\pm$  0.313 & 6.923 $\pm$  1.040 \\
 & 4.995              & 4.820              & 4.979              \\
Fe4383& 5.794 $\pm$  0.176 & 4.940 $\pm$  0.389 & 6.251 $\pm$  1.243 \\
 & 4.861              & 4.458              & 4.696              \\
Ca4455& 1.187 $\pm$  0.089 & 1.096 $\pm$  0.186 & 0.750 $\pm$  0.561 \\
 & 1.336              & 1.235              & 1.313              \\
Fe4531& 2.858 $\pm$  0.125 & 2.326 $\pm$  0.260 & 1.595 $\pm$  0.777 \\
 & 3.499              & 3.367              & 3.457              \\
Fe4668& 6.070 $\pm$  0.181 & 5.776 $\pm$  0.361 & 5.685 $\pm$  1.025 \\
 & 5.114              & 4.602              & 4.808              \\
H$\beta$& 1.841 $\pm$  0.067 & 1.823 $\pm$  0.122 & 1.800 $\pm$  0.304 \\
 & 1.908              & 1.966              & 1.878              \\
Fe5015& 5.121 $\pm$  0.138 & 4.767 $\pm$  0.246 & 4.912 $\pm$  0.584 \\
(Ti) & 5.491              & 5.211              & 5.292              \\
Mg$b$& 3.575 $\pm$  0.065 & 3.550 $\pm$  0.116 & 3.771 $\pm$  0.278 \\
 & 3.361              & 3.203              & 3.334              \\
Fe5270& 3.016 $\pm$  0.072 & 2.969 $\pm$  0.132 & 3.362 $\pm$  0.309 \\
 & 3.059              & 2.886              & 2.963              \\
Fe5335& 2.708 $\pm$  0.084 & 2.551 $\pm$  0.155 & 2.526 $\pm$  0.367 \\
 & 2.675              & 2.524              & 2.598              \\
Fe5406& 1.687 $\pm$  0.065 & 1.667 $\pm$  0.121 & 1.542 $\pm$  0.286 \\
 & 1.840              & 1.720              & 1.779              \\
$<$Fe$>'$& 2.930 $\pm$  0.075 & 2.852 $\pm$  0.138 & 3.128 $\pm$  0.325 \\
 & 2.952              & 2.785              & 2.861              \\
$[$MgFe$]$ & 3.236 $\pm$  0.070 & 3.182 $\pm$  0.127 & 3.435 $\pm$  0.301 \\
 & 3.150              & 2.987              & 3.089              \\
\hline
\end{tabular}
\caption{Measurements of the selected Lick indices for the 3-points binning. All
values are in \AA. Two lines for each index correspond to the measurements made
on the real spectra and on the best-fitting optimal templates (see text).
''Ti''\ symbol indicates that Fe5015 index is strongly contaminated by titanium
absorptions (see e. g. Sil'chenko \& Shapovalova, 1989).
\label{tablick3b}
}
\end{table}

\begin{figure}
\hfil
\begin{tabular}{c c c c}
 (a) & (b) & (c) & (d) \\
 \includegraphics[width=4cm]{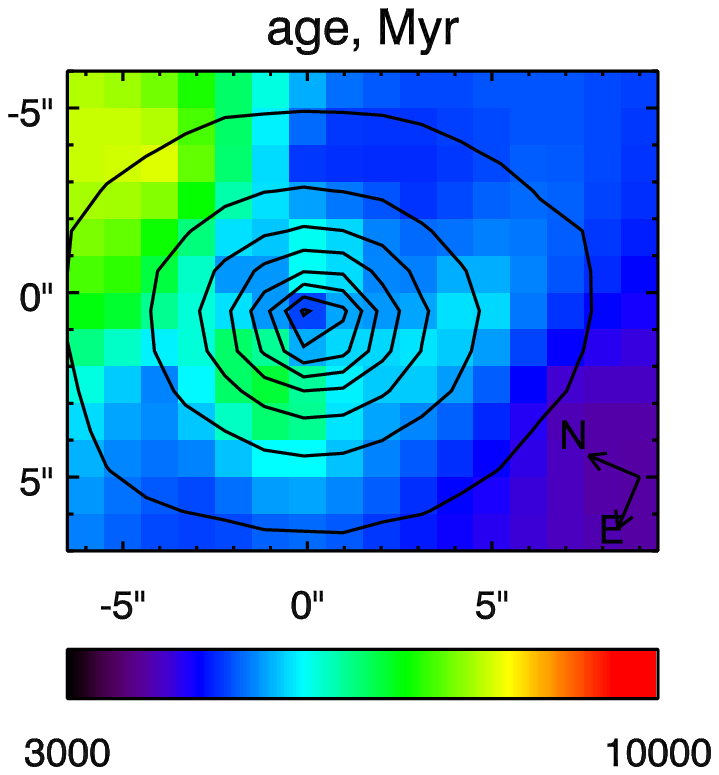} &
 \includegraphics[width=4cm]{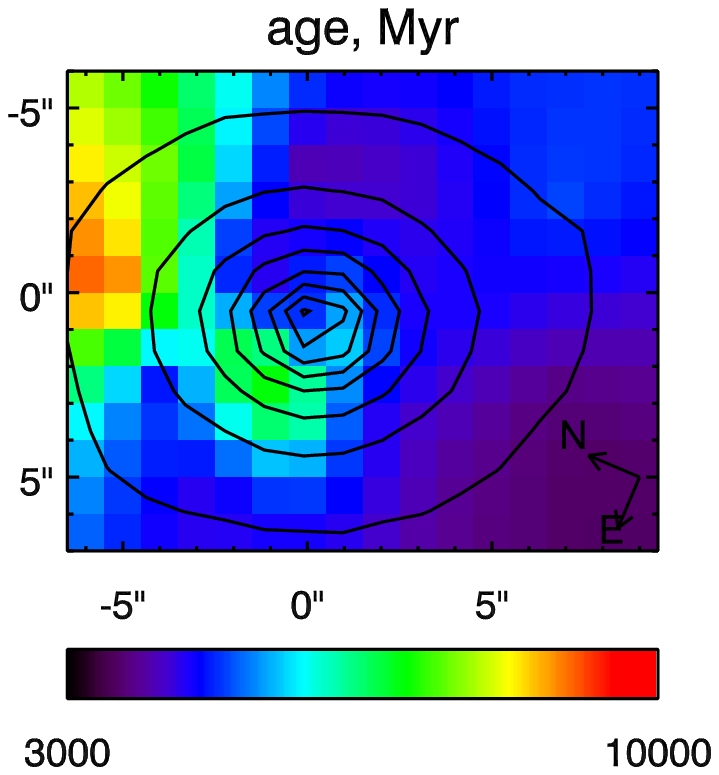} &
 \includegraphics[width=4cm]{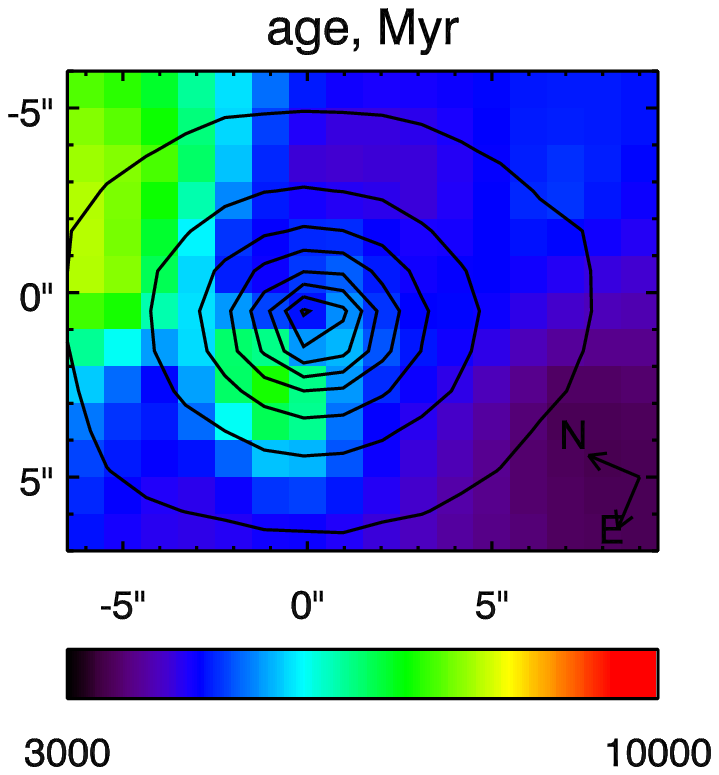} &
 \includegraphics[width=4cm]{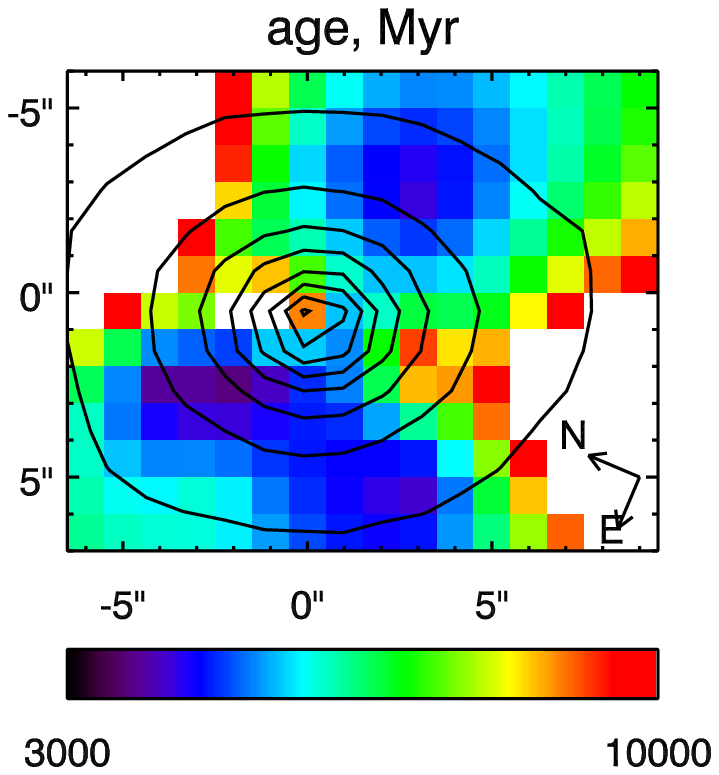} \\
 \includegraphics[width=4cm]{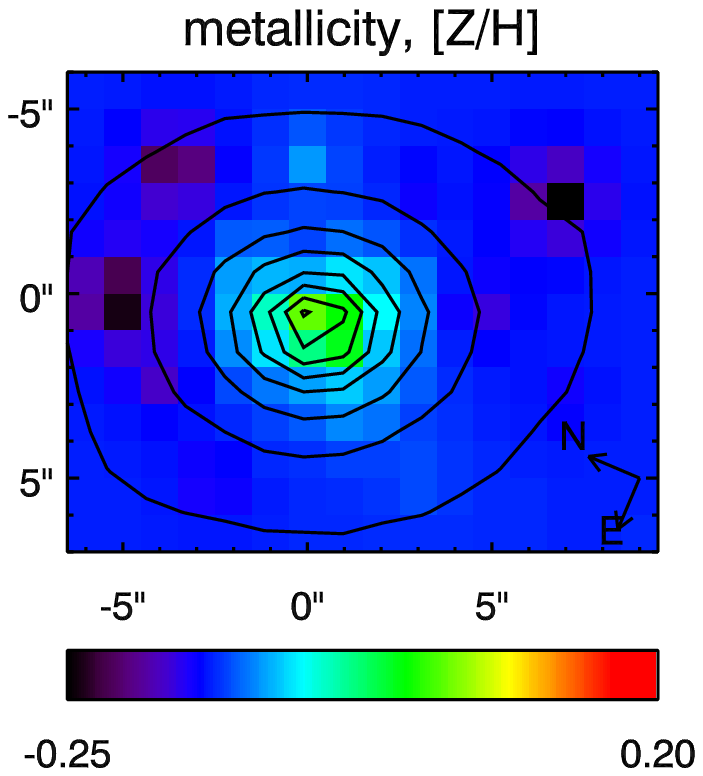} &
 \includegraphics[width=4cm]{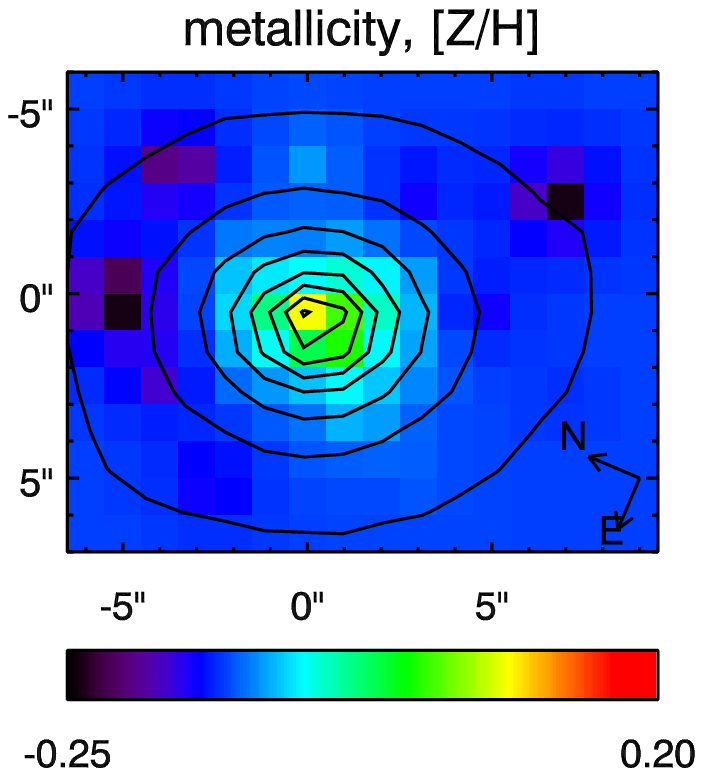} &
 \includegraphics[width=4cm]{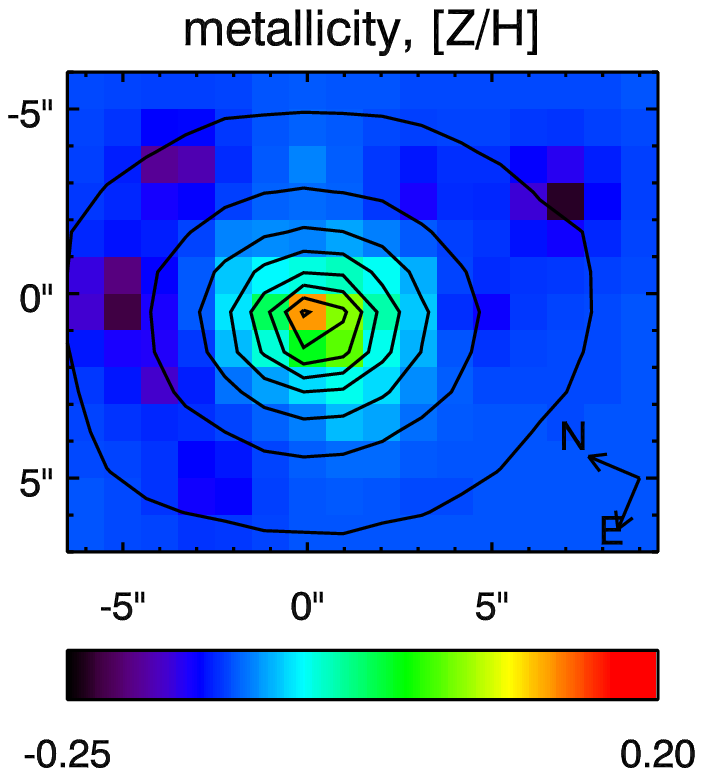} &
 \includegraphics[width=4cm]{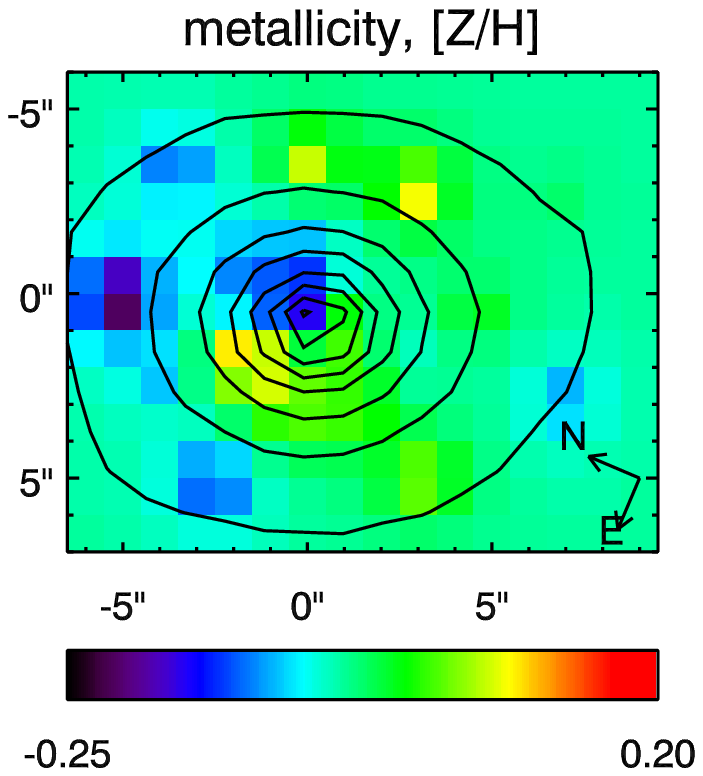} \\
\end{tabular}

\caption{Maps for age and metallicity obtained by inverting bi-index grids for
indices measured on optimal template spectra: (a) Mg$b$ - H${\beta}$, (b) 
[MgFe] - H${\beta}$, (c) $<$Fe$>$ - H${\beta}$; and on real data: 
(d) $<$Fe$>$ - H${\beta}$.
\label{lickfld}
}
\hfil
\end{figure}

There is no good age tracers in the MPFS spectral range, beside H${\beta}$ and
H${\gamma}$. Good intermediate resolution age tracer,
H${\gamma}$+Mg+Fe$_{125}$ (Vazdekis \& Arimoto 1999) cannot be used
in our case, because the required signal-to-noise ratio of about 100 at around 
$\lambda = 4340$\AA\ cannot be achieved even after co-adding all the spectra in
the data cube due to efficiency degradation of the spectrograph in the blue end
of the spectral range. 

Measurements of H${\beta}$ are quite scattered (see Fig~\ref{lickdiag}),
resulting in age estimations in a wide range of values: from 4~Gyr to 13~Gyr. 

One may notice also quite strong scattering of the points on the Mg$b$ -
H${\beta}$ diagram caused by the problems in the data (not all the spikes/dark
pixels were marked as bad data and interpolated with the model).

One of the reasons (beside unmasked bad values in the data) might be presence
of weak nebular emission lines in the spectrum of a galaxy: H${\beta}$ index
might be affected by emission in H${\beta}$, Mg$b$ -- by [NI]
($\lambda=5199$\AA) laying on the red continuum definition region. Though we do
not see significant emission line residuals when we subtract the best fitting 
model, we can not exclude completely this effect.

To improve the situation, we decided to measure Lick indices on the optimal
templates fitted to the data. This approach may produce biased results in case
of inconsistent abundance rations in the models and in the real stellar
population, when Lick indices measurements will be bound to the model abundances.
However, IC~3653 exhibits solar Mg/Fe abundance ratio (see Fig~\ref{lickdiag}a),
thus we do not expect biases.

We made inversions of the bi-index grids for three combinations of indices:
Mg$b$ - H${\beta}$, [MgFe] - H${\beta}$, $<$Fe$>'$ - H${\beta}$. The results
are presented in Fig~\ref{lickfld}. We used target signal-to-noise ratio of 30
at 5300\AA\ for the tessellation procedure. The maps shown represent
interpolated values of the parameters between intensity-weighted 
centres of the bins.

Metallicity distribution shows slight gradient from -0.15~dex at peripheral
parts to +0.10 in the very centre (average error-bar on metallicity measurements
using [MgFe] - H${\beta}$ is 0.15~dex).

Age map contains no significant details, the median value for age using [MgFe]
- H${\beta}$ pair is $6 \pm 2.5$~Gyr. [MgFe] and $<$Fe$>'$ indices are not very
age sensitive, thus age estimations depend mostly on values of H${\beta}$, and
they are almost equal for all three inversions.

\subsection{Stellar populations and internal kinematics using pixel fitting}
\label{secssp}

\subsubsection{Results: kinematics, age and metallicity maps}
\label{subsecresssp}
We applied Voronoi adaptive binning procedure to our data, 
setting the target signal-to-noise ratio to 15. Resulting
tessellation includes 76 bins with sizes from 1 to 12 pupils. To get the better
presentation one may interpolate the computed values of each parameter over the
whole field of view using the intensity-weighted centroids of the bins as the
nodes.

We measured the systemic radial velocity $588\pm5$km~s$^{-1}$. The
uncertainty includes possible systematic effects not exceeding 4~km~s$^{-1}$.

\begin{figure}
\hfil
\begin{tabular}{c c c}
 \includegraphics[width=5.5cm]{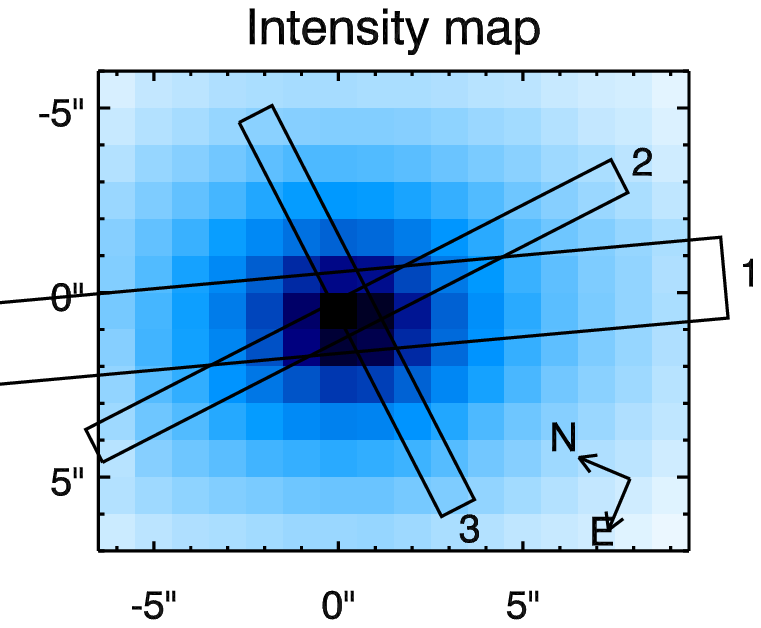} &
 \includegraphics[width=5.5cm]{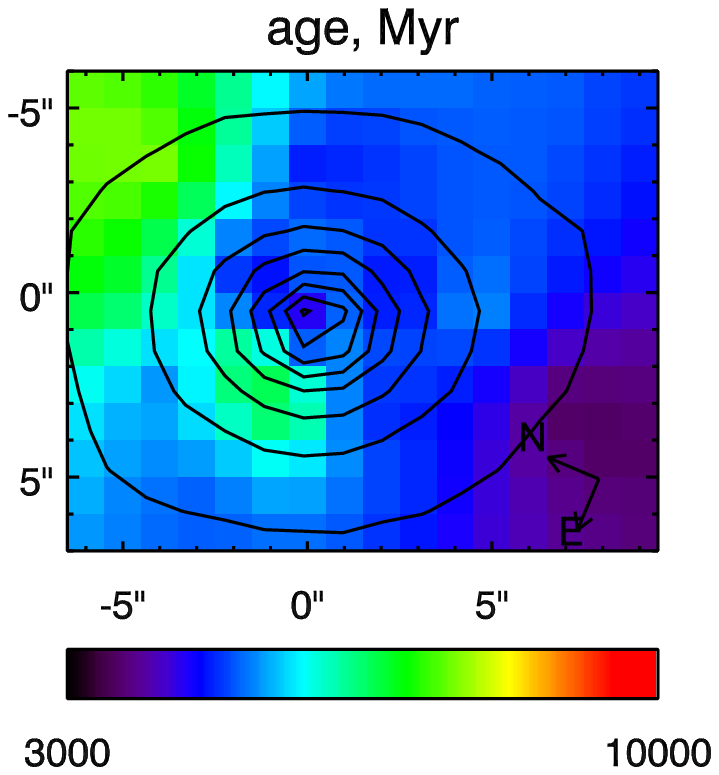} &
 \includegraphics[width=5.5cm]{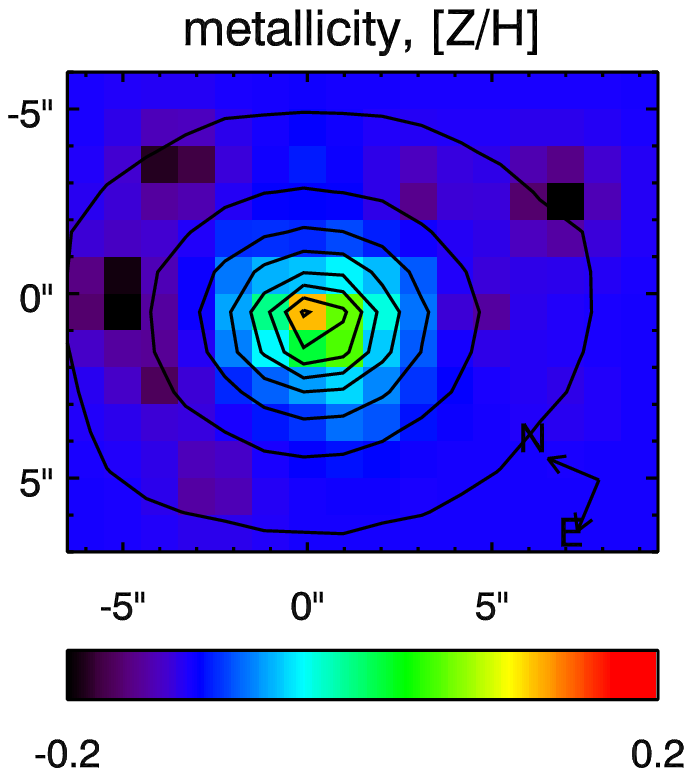} \\
 (a) & (b) & (c) \\
 \includegraphics[width=5.5cm]{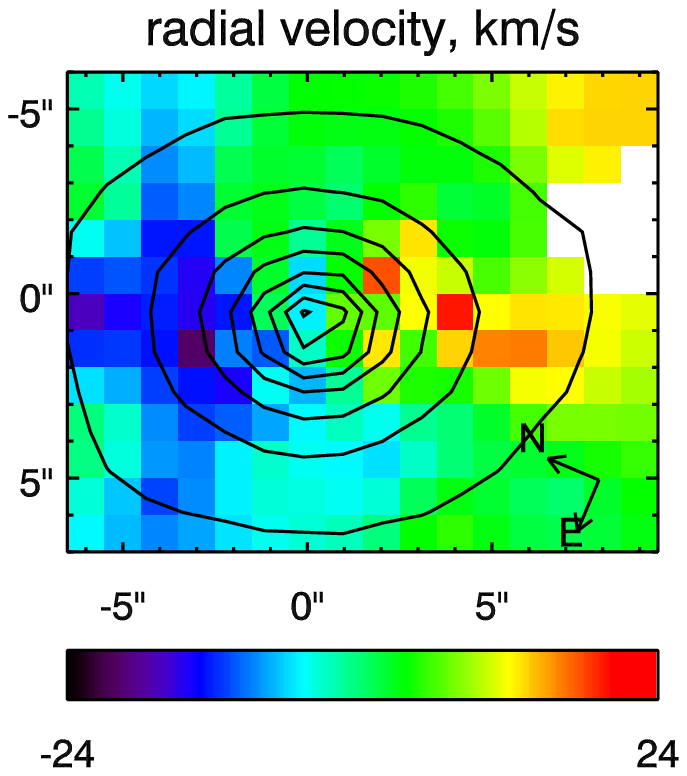} &
 \includegraphics[width=5.5cm]{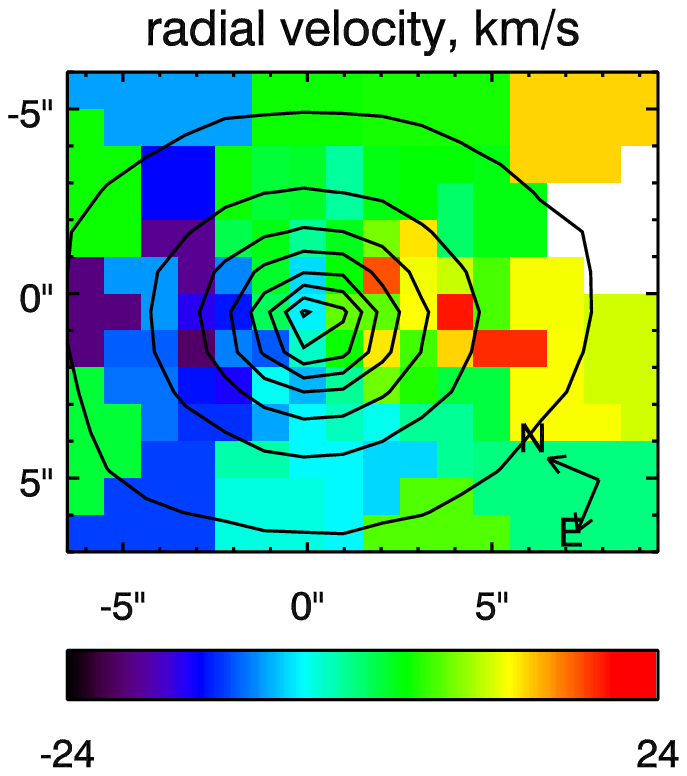} &
 \includegraphics[width=5.5cm]{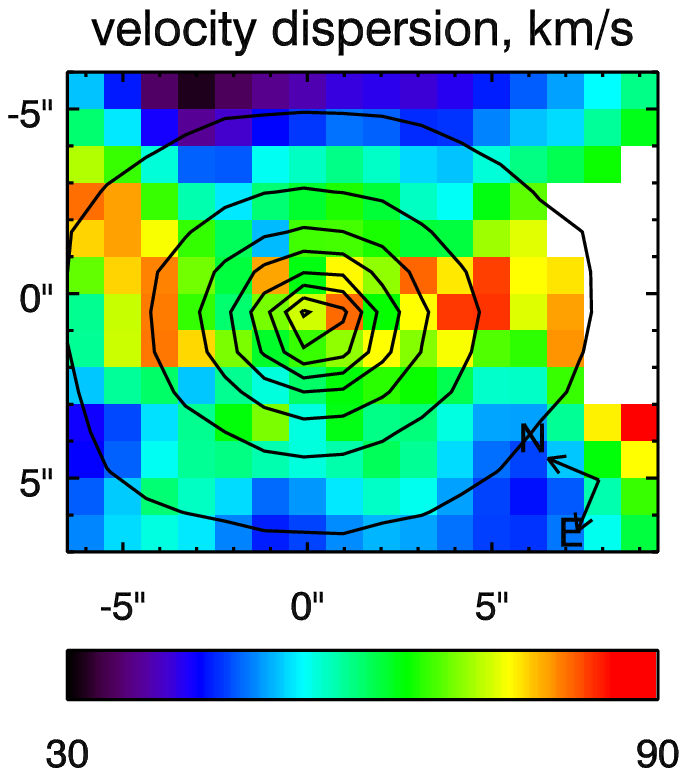} \\
 (d) & (e) & (f) \\
\end{tabular}
 \caption{The generalized view of the kinematical and stellar population data.
(a) intensity map with the positions the kinematical profiles are presented
for: (1) position of slit in Simien \& Prugniel (2002), (2) and (3) positions
of major and minor axis of embedded stellar disc; 
(b) map of the luminosity-weighted age distribution (in Myr);
(c) map of the luminosity-weighted metallicity 
distribution ($[Z/\mbox{H}]$, dex);
(d) radial velocity field, interpolated between the nodes of the Voronoi
tessellation; (e) radial velocity field: values exactly correspond to the
Voronoi tessellae; (f) velocity dispersion field, interpolated.
\label{fields}
}
\hfil
\end{figure}

The fields of radial velocity and velocity dispersions are presented in
Fig~\ref{fields}(d,e,f). The galaxy shows significant rotation and highly inclined disc-like
structure. The uncertainties of the velocity measurements were estimated using
Monte Carlo simulations and confirmed by a direct scan of the 3-dimensional
$\chi^2 (t, Z, \sigma)$ space: fitting only multiplicative polynomial continuum
on a grid of values of age, metallicity and velocity dispersion (see
appendix for details).
They depend on the signal-to-noise ratio and change from 2.5~km~s$^{-1}$ for the
$S/N=30$ to 8~km~s$^{-1}$ for the $S/N=10$.

The velocity dispersion distribution shows a gradient from 45-50~km~s$^{-1}$
near the maxima of rotation to 75~km~s$^{-1}$ in the core. Note a sharp peak of
the velocity dispersion of 88~km~s$^{-1}$ slightly shifted to the south-west of the
photometric core. The uncertainties of the velocity dispersion measurements are
3.8~km~s$^{-1}$ for the $S/N=30$, 5.5~km~s$^{-1}$ for the $S/N=20$, and 11~km~s$^{-1}$ for the
$S/N=10$.

\begin{figure}
 \includegraphics[width=17cm,height=9cm]{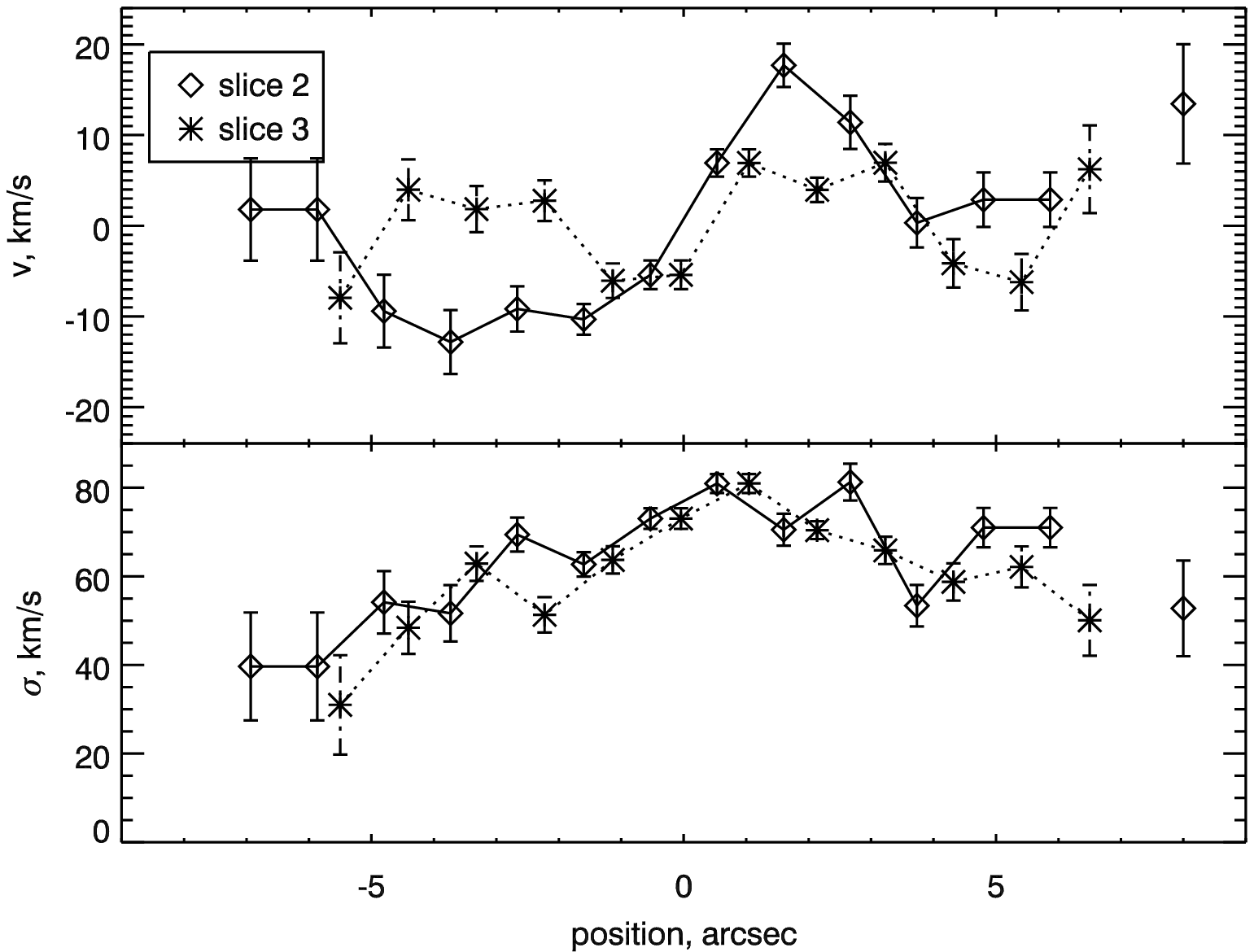}
 \includegraphics[width=17cm,height=9cm]{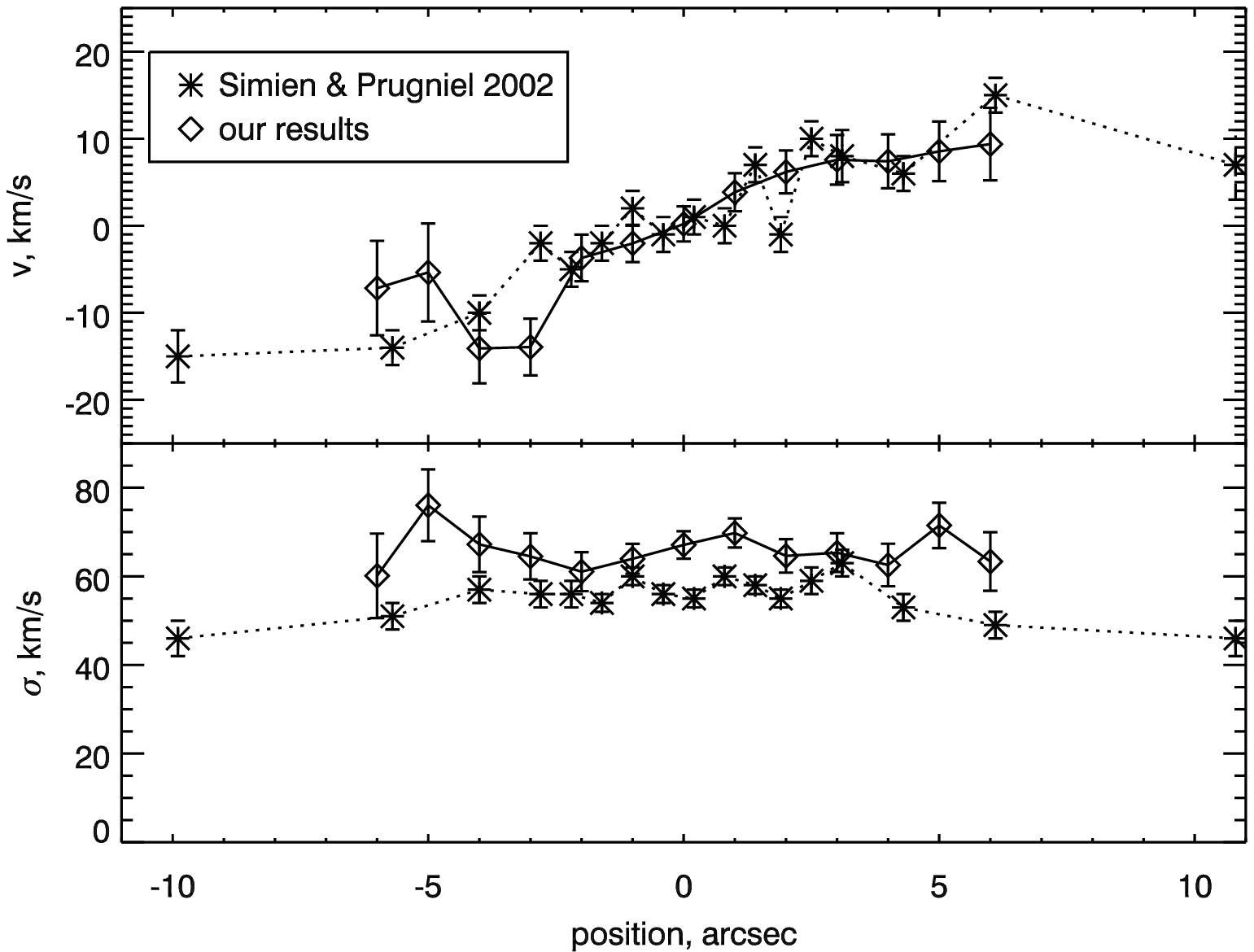}
 \caption{Top pair of plots:
The radial velocity profile for the slice ''2''\ (major axis of the embedded
disc) and velocity dispersion profiles for the slices ''2''\ and ''3'' (major
and minor axes of the embedded disc).  Bottom pair of plots: comparison of
the kinematical profiles (slice ''1'', major axis of main galactic body) to
Simien \& Prugniel (2002): radial velocity and velocity dispersion.
\label{prof}
}
\end{figure}

\begin{figure}
\hfil
\begin{tabular}{c c}
 \includegraphics[width=4cm]{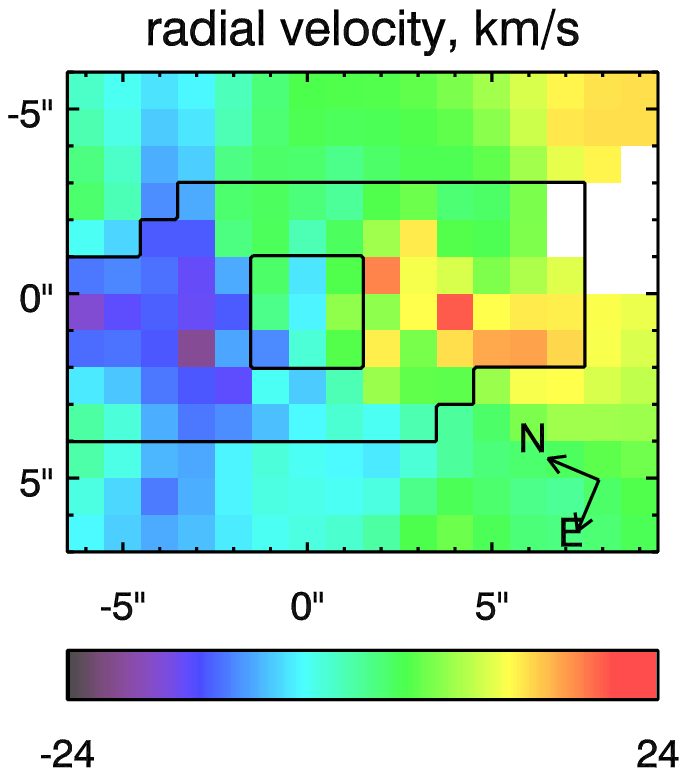} &
 \includegraphics[width=13cm]{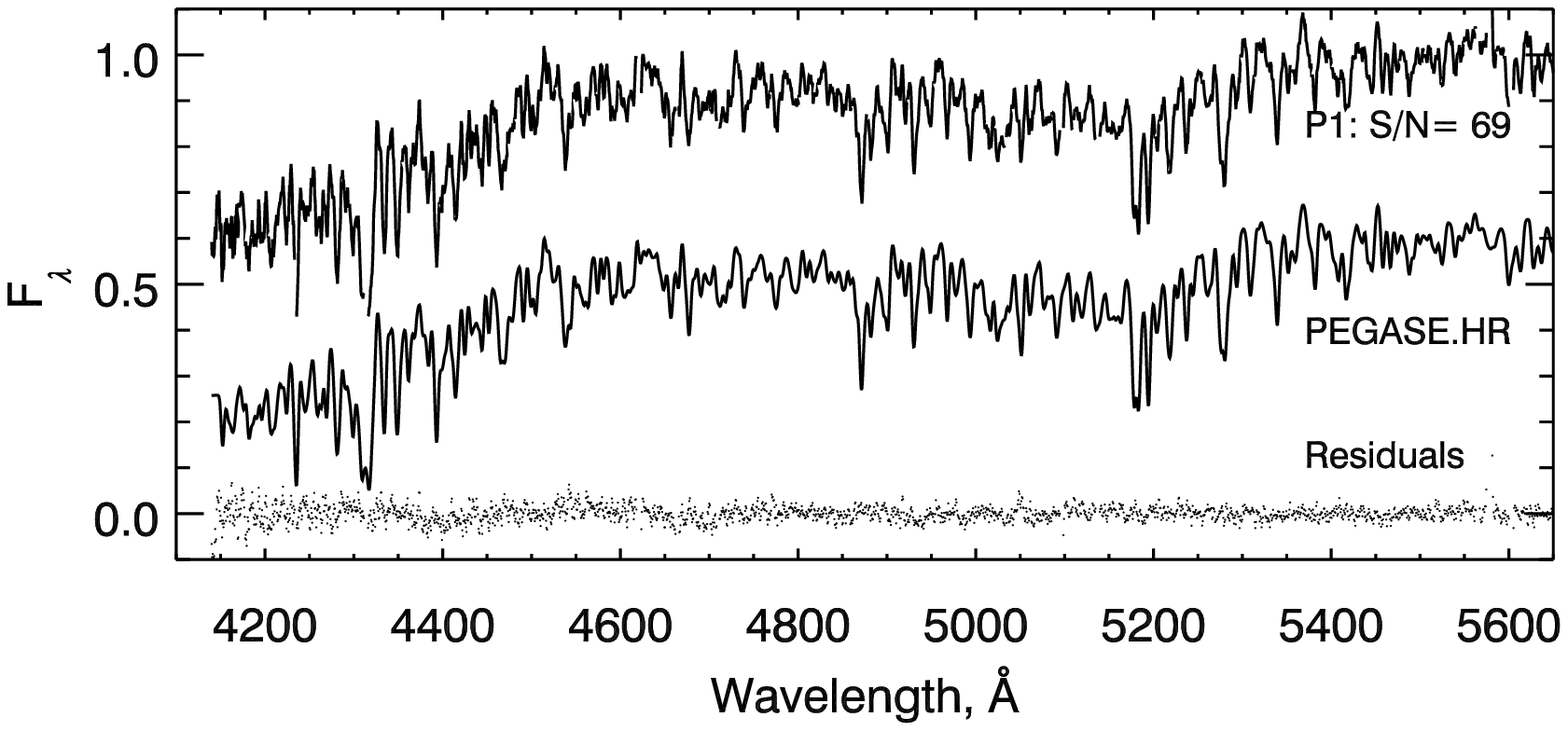} \\
 (a)&(b)\\
\end{tabular}
\begin{tabular}{c}
 (c)\\
 \includegraphics[width=18cm]{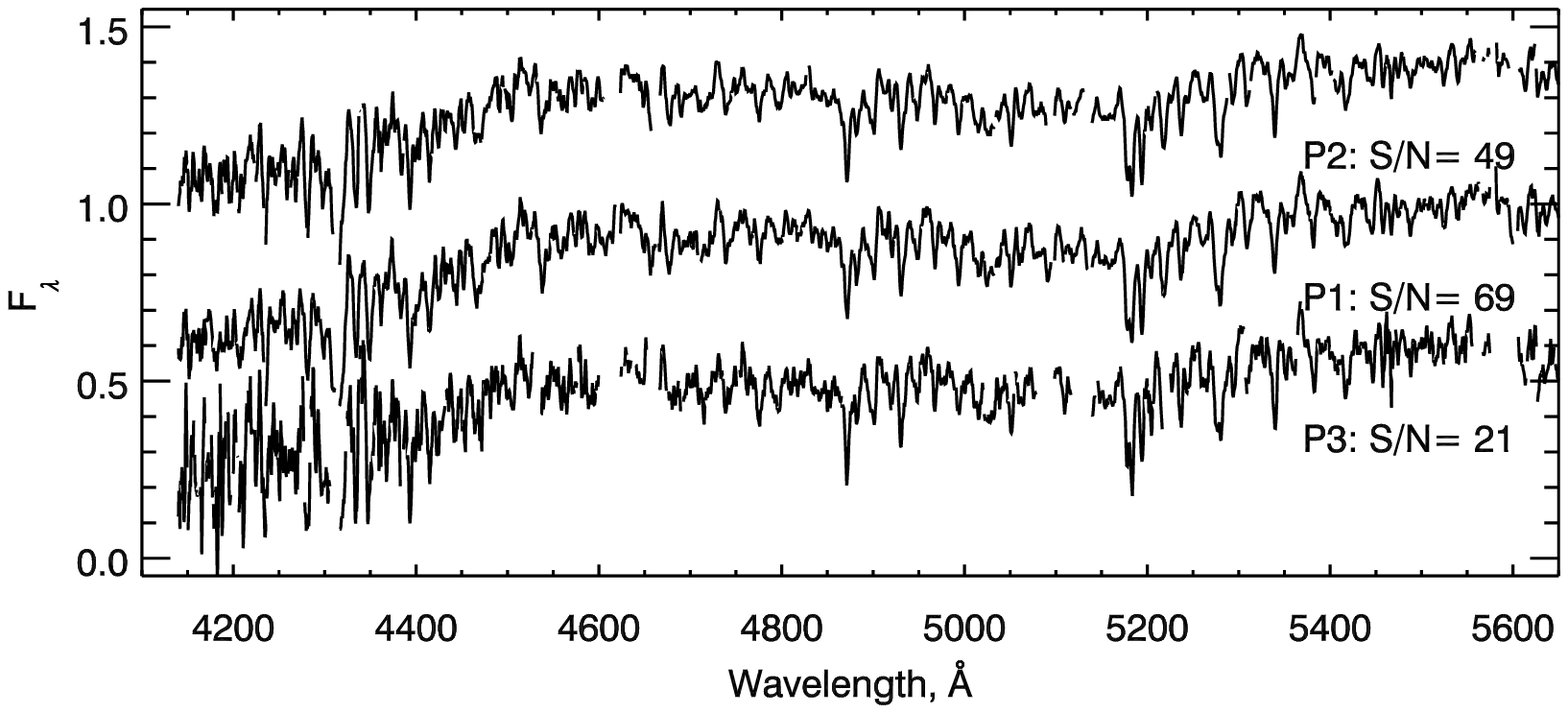}\\
\end{tabular}
 \caption{(a)The radial velocity field with 3 bins overplotted (3-points binning,
 see text); (b) fit of the ''P1''\ bin, representing the spectrum, ''PEGASE.HR''\
 marking the template spectrum (shifted by -0.3 on the Flux axis from its real
 position, and ''Residuals''\ showing the difference between the fit and observed
 spectrum; (c) ''P1'', ''P2'', and ''P3'': co-added spectra for the 3-points binning.
\label{spec4}
}
\hfil
\end{figure}

The rotation velocity and velocity dispersion profiles are shown in
Fig~\ref{prof} (top pair).

The previous studies of IC~3653 were made using long-slit
spectroscopy (Simien \& Prugniel, 2002). No measurable rotation was found.
This can be explained by poor atmosphere conditions during the observations (6
arcsec seeing). After the proper degrade of the spatial resolution of the MPFS
data one could see a very good agreement with (Simien \& Prugniel, 2002)
both for radial velocity and velocity dispersion profiles (Fig~\ref{prof}, 
bottom pair).

\subsection{Photometry and morphology from ACS images}
\label{secphot}

We have used ACS images from the HST archive, proposal 9401,
''The ACS Virgo Cluster Survey''\ by Patrick C\^ot\'e.
In C\^ot\'e et al. 2004 the first analysis is given, but IC~3653 is not
included. We have converted ACS counts into corresponding ST magnitudes
according to the ACS Data Handbook, available on-line on the web-site of STScI.

We have fitted two-dimensional S\'ersic profile using the GALFIT package
(Peng et al. 2002). We can see significant positive residuals, representing the
nucleus in the very centre (around 1.5 arcsec in size, central surface
brightness ST$_{475} \sim 16.25$~mag~arcsec$^2$,
slightly asymmetric and offcentered with
respect to the centre of the S\'ersic profile having n=1.88, $R_e$=6.9~arcsec,
and $\epsilon$=0.11 (S\'ersic index, effective radius, and ellipticity
respectively; our values coincide with ones from Ferrarese et al. 2006).
There are faint large-scale residuals as well, that can be explained by
superposition of several components (at least two).

\begin{figure}
\includegraphics[width=17cm]{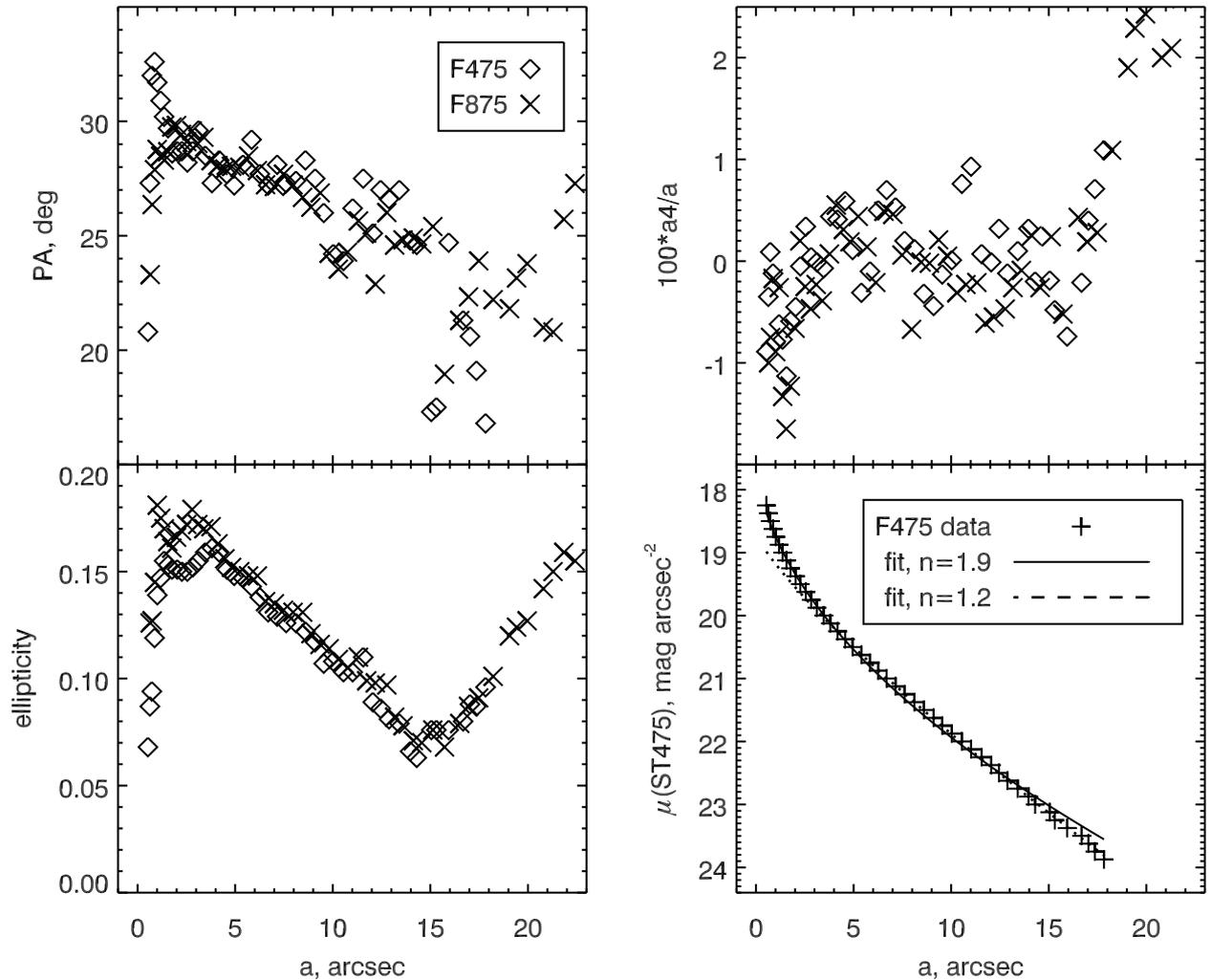}
\caption{Photometrical characteristics of IC~3653: position angle (upper
left), ellipticity of the isophotes (lower left), disky/boxy parameter a4 (upper
right) in two colours, and photometric profile in F475. 
\label{figphotpar}
}
\end{figure}

Then we have modeled images by elliptical isophotes with free center and
orientation. We see some isophote twist and change of the ellipticity in the
inner region of the galaxy. Main parameters of the model fitted are
presented in Fig~\ref{figphotpar}. The subtraction of the model from the
original image does not reveal any internal feature.

\begin{figure}
\includegraphics[width=17cm]{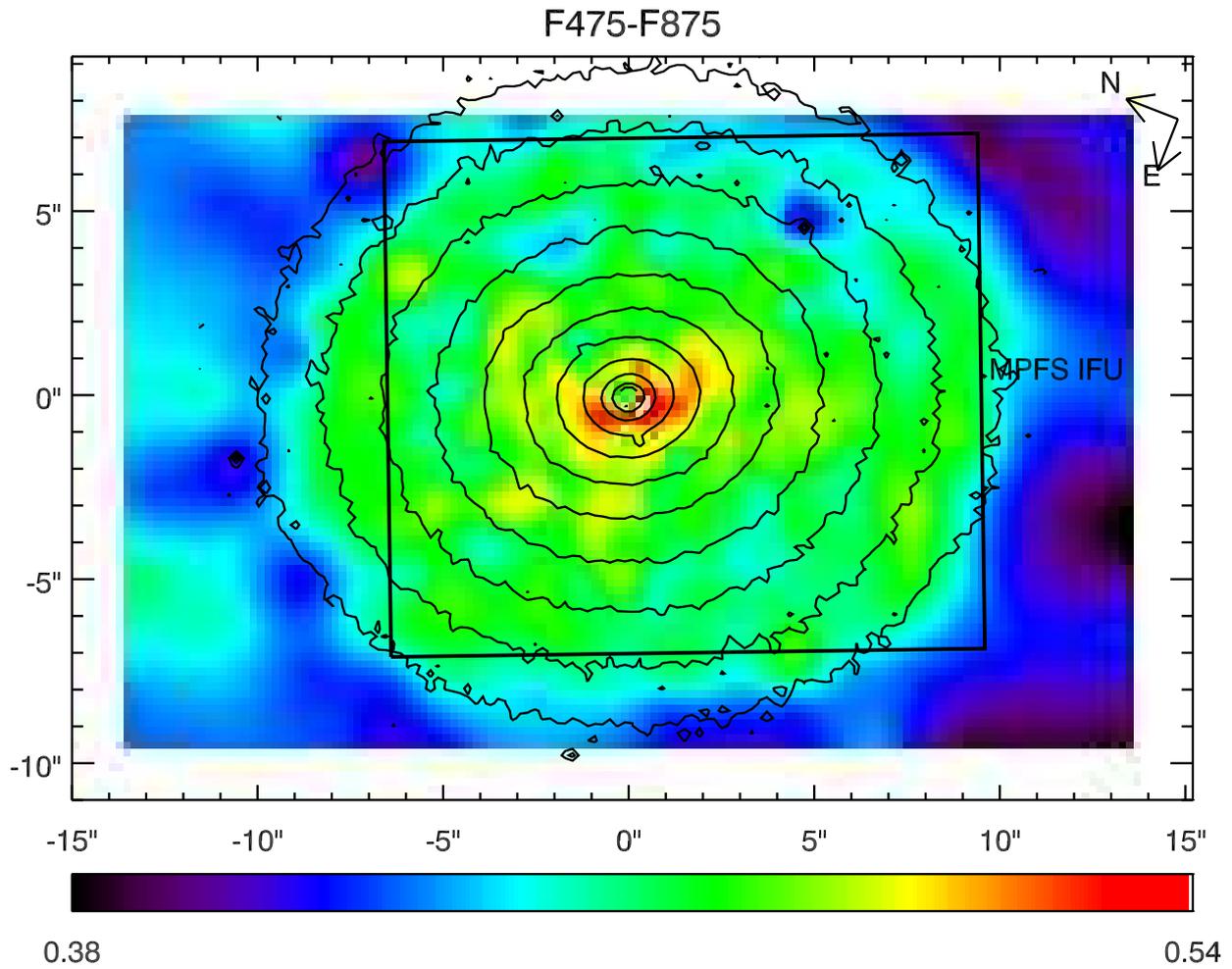}
\caption{F475-F875 colour map (HST ACS data). Isophotes of the F875 image are
overimposed. Position and size of MPFS IFU are shown.
\label{figcolmaphst}
}
\end{figure}

However, the F475-F875 colour map reveals the elongated structure (a/b $\sim$
3.5) having size (major axis) of about 7 arcsec, and orientation coinciding
with the kinematical disc-like feature. In Fig~\ref{figcolmaphst} the F475-F875
colour map is shown. It was obtained using Voronoi 2D binning technique applied
to the F875 image in order to reach the signal-to-noise ratio of 80 per bin.
Redder colour of the structure might be caused by slightly higher metallicity of
the sub-population contained in it. However we do not see the elongated
structure, but only metallicity gradient in maps obtained with MPFS, because of
the bin size used to create stellar population maps, that is larger than for
kinematics due to higher target S/N ratio used in the adaptive tessellation.

After that we re-fitted the surface brightness distribution with a S\'ersic law
with central region, corresponding to the disc-like structure, excluded. This
fit leaves no significant residuals, and S\'ersic exponent decreases to
1.22 -- close to the exponential distribution. This value is given in 
Tab~\ref{tabic3653params}.

On the lower right plot in Fig~\ref{figphotpar} the light profile in
F475 is shown with crosses. Solid line represents the best-fitting S\'ersic
profile for the whole galaxy excluding only very centre (inner 1 arcsec) with
n=1.9, and dashed line gives the best-fitting (n=1.2) for the peripheral parts of
the galaxy (beyond the disc found in the colour map).

\subsection{Discussion}
\label{secdisc3653}
Both line-of-sight velocity field extracted from the MPFS data cube and
colour map obtained from the HST imagery provide undoubted coinciding
arguments for a presence of a faint internal co-rotating stellar disc
embedded within a rotating spheroid. This is the main observational result
from our study of IC~3653, which may be regarded as an edge-on counterpart
of IC~3328, the dE with embedded spiral structure found by Jerjen et al.
(2000). Authors stated there that spiral structure had low amplitude,
about several per cent. However, estimating total mass of the small embedded
stellar disc is much more complex and model-dependent task. If the similar
stellar disc observed edge-on contains one to several tens of per cent of
the stellar mass of the spheroid, it will be easily detected kinematically. At
the same time, photometric appearance will strongly depend on the difference
of stellar populations in disc and main spheroid.

In this subsection we will first compare the results of two methods for estimating
stellar population parameters: Lick indices and pixel-fitting, then the
characteristics of IC~3653 to other dE galaxies. Finally, we will review the
different possible origins of the particular properties of this galaxy.

\subsubsection{Comparison of two techniques for estimating stellar population
characteristics}
In Tab~\ref{tabcomptz3b} we present comparison between SSP-equivalent age
and metallicity obtained with the pixel fitting and inversion
of bi-index grids for H$\beta$-Mg$b$, H$\beta$-$<$Fe$>'$, H$\beta$-[MgFe]
using index measurements on observed spectra and best-fitting templates.

One can notice particularly good agreement between the approaches. Ages
derived from Lick indices appear to be slightly older, but the difference is
not significant. The best agreement for both ages and metallicities is
reached between pixel fitting and measurements of H$\beta$ and the combined
[MgFe] index (Thomas et al. 2003). The internal precision of the parameters
derived from pixel fitting is better those from Lick indices by a factor
three to four, depending on the indices used.  This can be explained by more
optimal usage of the information, contained in the spectra, by the pixel
fitting procedure. Though it is difficult to assess the reliability of these
small error bars, the relative variations of age and metallicity can be
trusted, even when the signal-to-noise ratio is as low as 10 per pixel (with
MPFS spectral resolution and wavelength coverage).

\begin{table}
\begin{tabular}{l c c c}
 & ''P1'' & ''P2'' & ''P3'' \\
\hline
\hline
t$_{fit}$, Gyr &  4.93$\pm$ 0.20 &  4.95$\pm$ 0.30 &  4.97$\pm$ 0.70 \\
\hline
t$_{H\beta-Mgb}$ &  7.04$\pm$ 1.56 &  6.97$\pm$ 1.47 & 11.25$\pm$ 6.02 \\
t$_{H\beta-<Fe>'}$ &  7.02$\pm$ 1.33 &  7.28$\pm$ 2.20 &  6.08$\pm$ 3.94 \\
t$_{H\beta-\mbox{[MgFe]}}$ &  7.11$\pm$ 1.65 &  6.88$\pm$ 1.89 &  6.70$\pm$ 5.11 \\
\hline
t$_{H\beta-Mgb\mbox{mod}}$ &  5.27$\pm$ 1.56 &  5.13$\pm$ 1.47 &  4.85$\pm$ 6.02 \\
t$_{H\beta-<Fe>'\mbox{mod}}$ &  5.15$\pm$ 1.33 &  4.30$\pm$ 2.20 &  4.23$\pm$ 3.94 \\
t$_{H\beta-\mbox{[MgFe]mod}}$ &  5.22$\pm$ 1.65 &  4.12$\pm$ 1.89 &  4.15$\pm$ 5.11 \\
\hline
\hline
Z$_{fit}$, dex &  0.03$\pm$ 0.01 & -0.14$\pm$ 0.02 & -0.17$\pm$ 0.05 \\
\hline
Z$_{H\beta-Mgb}$ & -0.05$\pm$ 0.09 & -0.18$\pm$ 0.08 & -0.34$\pm$ 0.24 \\
Z$_{H\beta-<Fe>'}$ & -0.02$\pm$ 0.04 & -0.10$\pm$ 0.05 &  0.03$\pm$ 0.13 \\
Z$_{H\beta-\mbox{[MgFe]}}$ & -0.04$\pm$ 0.05 & -0.14$\pm$ 0.06 & -0.11$\pm$ 0.15 \\
\hline
Z$_{H\beta-Mgb\mbox{mod}}$ & -0.02$\pm$ 0.09 & -0.14$\pm$ 0.08 & -0.16$\pm$ 0.24 \\
Z$_{H\beta-<Fe>'\mbox{mod}}$ &  0.02$\pm$ 0.04 & -0.14$\pm$ 0.05 & -0.17$\pm$ 0.13 \\
Z$_{H\beta-\mbox{[MgFe]mod}}$ &  0.00$\pm$ 0.05 & -0.13$\pm$ 0.06 & -0.16$\pm$ 0.15 \\
\hline
\hline
\end{tabular}
\caption{Comparison of age and metallicity measurements for the 3-points binning
obtained with pixel fitting and based on different pairs of Lick indices grids: 
measured on real spectra and on best-fitting templates.
\label{tabcomptz3b}
}
\end{table}

\subsubsection{Properties and nature of IC~3653}

We computed the position of IC~3653 on the fundamental plane (FP, Djorgovski \&
Davis 1987). ''Vertical''\ deviation from FP ($d=-8.666 + 0.314 \mu_e + 1.14 \log
\sigma_0 - \log R_e$, Guzman et al. 1993) is $d=0.2$. Such a deviation places
IC~3653 into the centre of the cloud, representing dE galaxies in Fig~2~(left) in
De Rijcke et al. 2005, and exactly on the theoretical predictions by Chiosi \&
Carraro (2002) and Yoshii \& Arimoto (1987), overplotted on the same figure.

The mean age of the stellar population of IC~3653, $t=5$~Gyr, coincide with the
mean age of dE galaxies in Virgo ($t_{mean}=5$~Gyr, Geha et al. (2003);
$t_{mean}=5...7$~Gyr, Van Zee et al. 2004b). However metallicity of the main body,
$Z=-0.1$ is slightly higher ($Z_{mean}=-0.3$, Geha et al. (2003); $Z_{mean}=-0.4$,
Van Zee et al. 2004b) that does not look strange keeping in mind that IC~3653 is
more luminous than most of the galaxies in the samples of Geha et al. (2003), and
Van Zee et al. (2004b).

We see that fundamental properties of IC~3653 do not differ from typical dE
galaxies, though the effective radius is one of the smallest within the samples
of Virgo dE's presented in Simien \& Prugniel (2002), Geha et al. (2003), and
van Zee (2004a).

We derived the B-band mass-to-light ratio of IC~3653 following the method by
Richstone \& Tremaine (1986) as $M/L_B = 8.0\pm1.5 (M/L_B)_{odot}$. Based on the
model by Worthey (1994) the luminosity weighted age and metallicity we
found can be translated into the stellar mass-to-light ratio
$(M/L_B)_{*}=3.5\pm0.4$ assuming Salpeter initial mass function. This value
is more than twice lower than the dynamical estimate, meaning that either the
simple dynamical model overestimates mass by a factor of two, or IC~3653 has a
dark matter halo. 

Recent theoretical studies based on N-body modelling of the evolution of disc
galaxies within a $\Lambda$ CDM cluster by Mastropietro et al. (2005)
suggest that discs are never completely destroyed in cluster environment, even
when the morphological transformation is quite significant. Our discovery of the
faint stellar disc in IC~3653 comforts these results. Thus, one of the possible
origins of IC~3653 is morphological transformation in the dense cluster
environment from late-type disc galaxy (dIrr progenitor). Gas was removed by
means of ram pressure stripping and star formation was stopped. This process
must have finished at least 5~Gyr ago, otherwise we would have seen younger
population in the galaxy. However, duration of the star formation period must
have been longer than 1~Gyr, otherwise deficiency or iron (i.e. Mg/Fe
overabundance) would have been observed. Within this scenario, metallicity
excess in the disc can be explained by a slightly longer duration of the star
formation episode compared to the spheroid. But we cannot see the difference
in the star formation histories (even luminosity weighted age), because of
insufficient resolution on the stellar population ages.

Another way to acquire a disc having higher metallicity than the rest of
the galaxy is to experience a minor dissipative merger event (De Rijcke et
al. 2004).  
This is
rather improbable for a dwarf galaxy, but cannot be completely excluded. In
particular, kinematically decoupled cores, recently discovered in dwarf
and low-luminosity
galaxies (De Rijcke et al. 2004, Geha et al. 2005, Prugniel et al. 2005,
Thomas et al. 2006) can be explained by minor merger events.

The most popular scenario which is usually considered to explain
formation of embedded stellar discs in giant early-type galaxies is star
formation in situ after infall of cold gas onto existing rotating spheroid,
e.g. from a gas-rich companion (cross-fueling). Just this scenario has been
used by Geha et al. (2005) to explain counter-rotating core in NGC~770, a
dwarf S0 which is more luminous than IC~3653 ($M_B=-18.2$~mag) and located
close to a massive spiral companion NGC~772 ($M_B=-21.6$~mag) in a group. In
NGC~127 ($M_B=-18.0$~mag), another galaxy, being a nearby satellite of a giant
gas-rich NGC~128, we observe the process of cross-fueling at present
(see Chapter 3). Group environment, where relative
velocities of galaxies are rather low, favours of the interaction processes
on large timescales, such as smooth gas accretion.

Our data on IC~3653 cannot provide a decisive choice between those
alternatives. However, from a general point of view, dynamically hot
environment of the Virgo cluster with high relative velocities of
member galaxies does not
conduce to the slow accretion of cold gas. IC~3653 is not a member of a
subgroup including large galaxies, which can foster gaseous disc, so we
believe that in this particular case the scenario of slow accretion is not
applicable.

At present, the sample of objects where disc-like sub-structures were
searched either from images or integral field spectroscopy is still too
small to draw statistical conclusions. But it is quite probable that 
progenitors of the dE's were disc galaxies (pre-dIrr or small spiral galaxies)
and that they evolved due to feedback of the star formation and
environmental effects. Present dIrr also experienced feedback but kept
their gas, so it is unlikely that the feedback alone can remove the
gas. Therefore environmental effects are probably the driver of the
evolution of dE's, and the discovery of stellar discs in dE's is consistent
with this hypothesis.

\section{Young nuclei in Virgo dE's}
Young nuclei are a frequent phenomenon in the giant early-type galaxies
(Sil'chenko 1997, Vlasyuk \& Sil'chenko 2000) observed both in clusters
and groups (Sil'chenko 2006). However, there were no detections of young
circumnuclear structures in dwarf elliptical/lenticular galaxies, probably
because of difficulties observing them due to low surface brightness.
To reach high signal-to-noise ratios, sufficient for the stellar population
analysis using classical approach by measuring Lick indices
(Worthey et al. 1994), integration time must be order of several hours with 
large telescopes. In addition, integral field spectroscopy is an essential 
technique for reliable detections of such structures. However no attempts
have been made so far to observe even small samples of dE galaxies with IFU
spectrographs, but only individual objects (Geha et al. 2005).

We have started a project of observing a sample of dE galaxies in cluster
and groups using the Multi-Pupil Fiber Spectrograph on the Russian 6-m telescope.

\subsection{Observations and data reduction}
\label{secobs3dEV}
The spectral data we analyse were obtained with the MPFS integral field
spectrograph. We used the same setup of an instrument, as for observations
of IC~3653. Parameters of the observations are summarized in the
Table~\ref{tabobsdEV}.

\begin{table}
\begin{tabular}{lccc}
Object & Dates & Total exp.time & Seeing \\
\hline
IC~783 & 21,23 May 2004 & 3.5h & 2'' \\
IC~3468 & 20 March 2004 & 2.5h & 1.5'' \\
IC~3509 & 10,12 May 2005 & 3.5h & 1.7'' \\
\hline
\end{tabular}
\caption{Parameters of observations \label{tabobsdEV}}
\end{table}

Data have been reduced and analysed using techniques, described in a section 
devoted to IC~3653. Target signal-to-noise ratios used in the Voronoi 2D
binning procedure were between 15 and 20.

''2-bins''\ tesselation was also applied to all three galaxies. It contains
two bins: central young embedded structure, and the rest of the galaxy.

For two galaxies: IC~3468 and IC~3509 we have used results of analysis of
ACS images from the HST archive, proposal 9401, ''The ACS Virgo Cluster Survey''\ 
(P.I.: P. C\^ot\'e), presented in Ferrarese et al. (2006). For IC~783 we have
used light and colour profiles available through the GOLDMine database
(Gavazzi et al. 2003).

\subsection{Stellar population and internal kinematics}

\begin{table}
\begin{tabular}{lccc}
Object & $t$ Gyr & $Z$ dex & $(M/L)_{B*}$ \\
\hline
IC~783 (core) & $3.3\pm0.4$ & $-0.35\pm0.04$ & $2.1\pm0.2$ \\
IC~783 (out.) & $12.8\pm4.0$ & $-0.79\pm0.12$ & $5.2 \pm 1.5$ \\
\hline
IC~3468 (disc) & $5.3\pm0.4$ & $-0.40\pm0.05$ & $2.8\pm0.4$ \\
IC~3468 (out.) & $8.6\pm0.9$ & $-0.60\pm0.05$ & $4.0\pm0.6$ \\
\hline
IC~3509 (core) & $4.1\pm0.4$ & $-0.05\pm0.05$ & $3.1\pm0.4$ \\
IC~3509 (out.) & $7.8\pm0.8$ & $-0.40\pm0.10$ & $4.3\pm0.5$ \\
\hline
\end{tabular}
\caption{Luminosity-weighted parameters of the stellar populations of
three dE galaxies: age, metallicity [Fe/H], and mass-to-light ratios
of the stellar population according to Worthey (1994)}
\end{table}
     
\subsubsection{IC~783}
This galaxy was found to have a remarkable spiral structure (Barazza et al.
2002). IC~783 exhibits rotation ($v_{rot} \sim 20$~km~s$^{-1}$, see also
Simien \& Prugniel, 2002). Velocity dispersion field is flat
($\langle \sigma \rangle =35\pm 10$~km~s$^{-1}$)
and does not show any significant features.

We discover a young nucleus in this galaxy, having luminosity-weighted
stellar population parameters: $t=3.3\pm0.4$~Gyr,
$Z=-0.35\pm0.04$~dex, compared to the main body
(containing the spiral structure) $t=12.8\pm4.0$~Gyr, $Z=-0.79\pm0.12$~dex. 
Young nucleus in IC~783 remains spatially
unresolved. The B-band mass-to-light ratio for the stellar populations
according to Worthey (1994) are: $(M/L)_{B*} = 2.1 \pm 0.2$ for the core,
and $(M/L)_{B*} = 5.2 \pm 1.5$ for the rest.

\begin{figure}
\hfil
\begin{tabular}{c c}
 (a) & (b) \\
 \includegraphics[width=7cm]{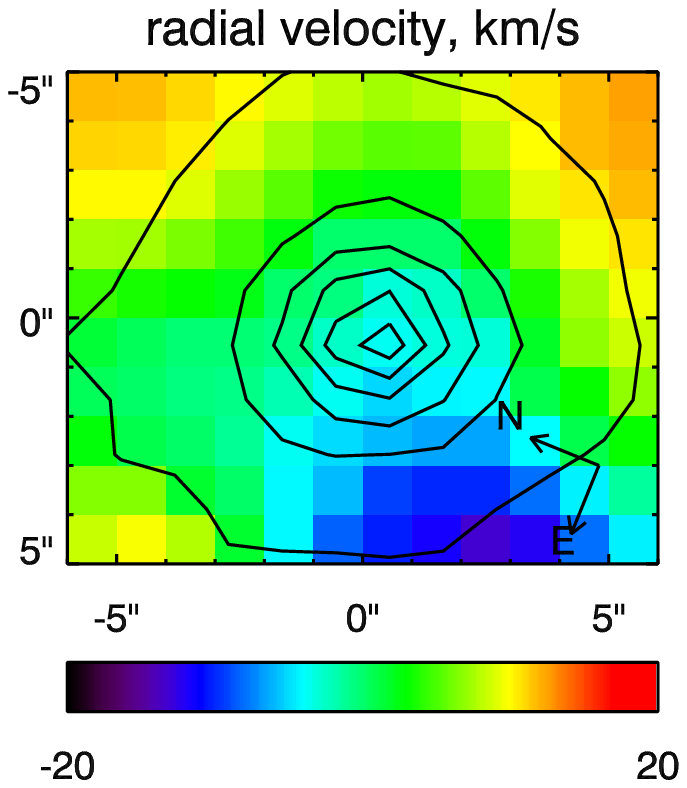} &
 \includegraphics[width=7cm]{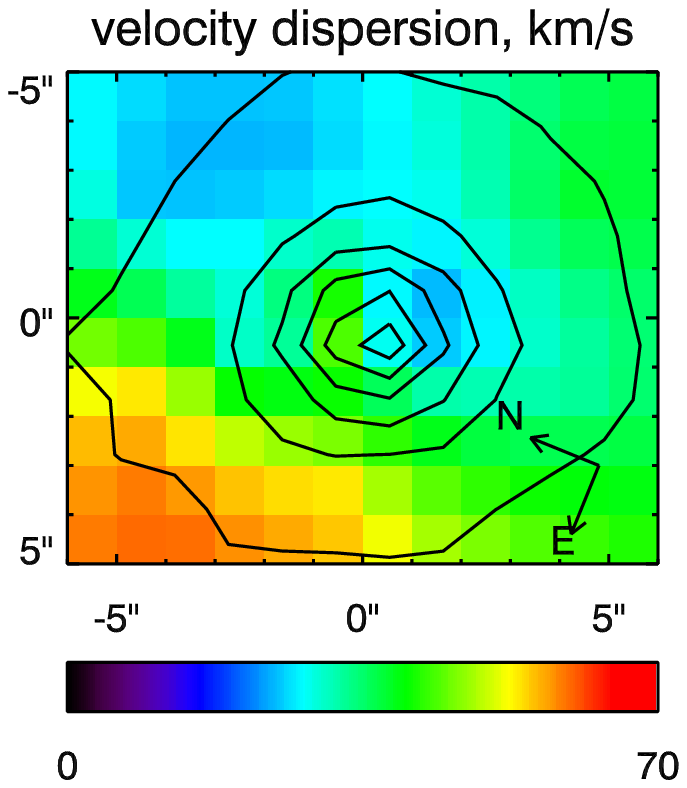} \\
 (c) & (d) \\
 \includegraphics[width=7cm]{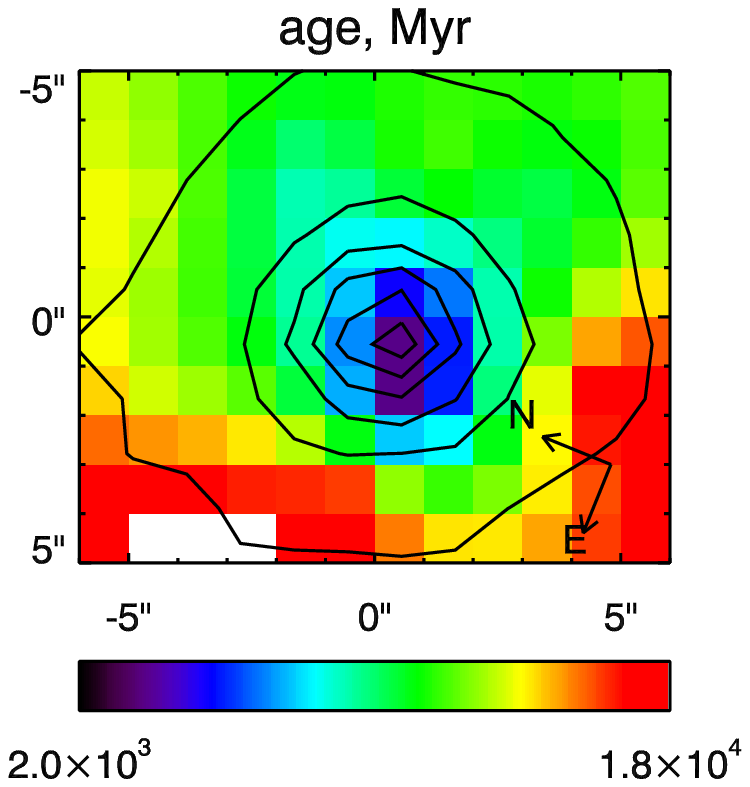} &
 \includegraphics[width=7cm]{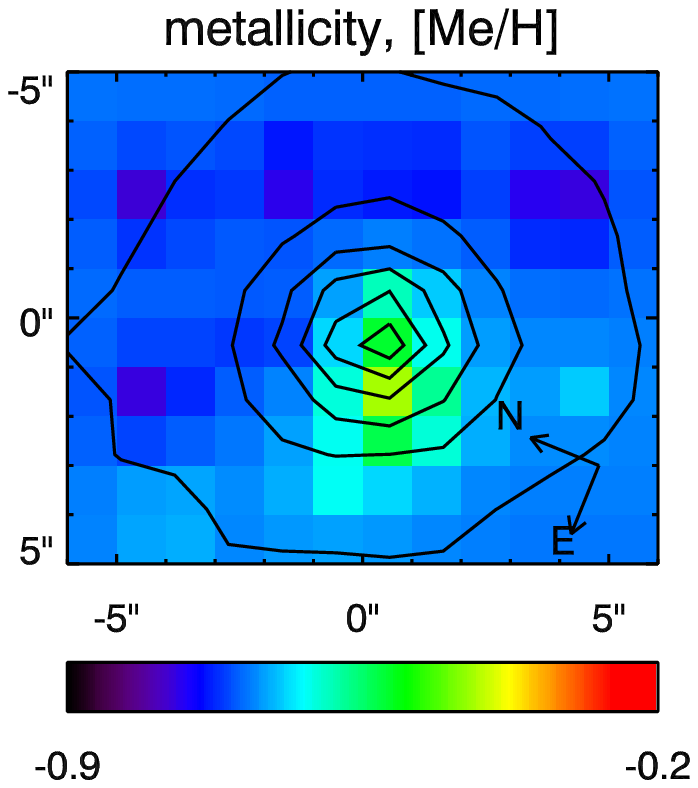} \\
\end{tabular}
\caption{Kinematics and stellar population of IC~783. Maps of internal kinematics
and stellar population parameters are built for a Voronoi tessellation with a
target S/N ratio or 15. (a) line-of-sight stellar velocity,
(b) stellar velocity dispersion, (c)
luminosity-weighted age, (d) luminosity-weighted metallicity
\label{figic783}}
\end{figure}

\begin{figure}
\hfil
\includegraphics[width=17cm,height=8cm]{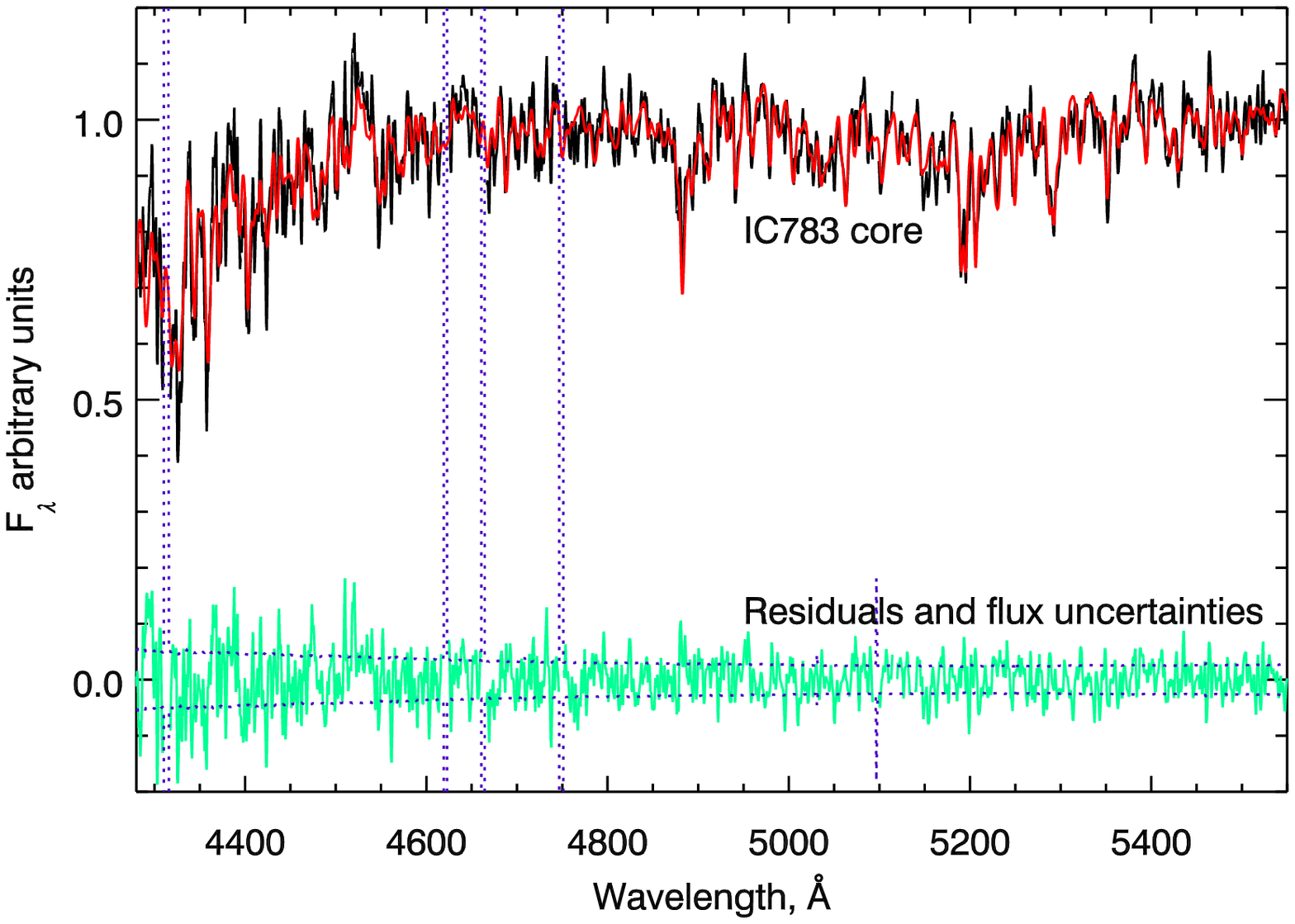}\\
\includegraphics[width=17cm,height=8cm]{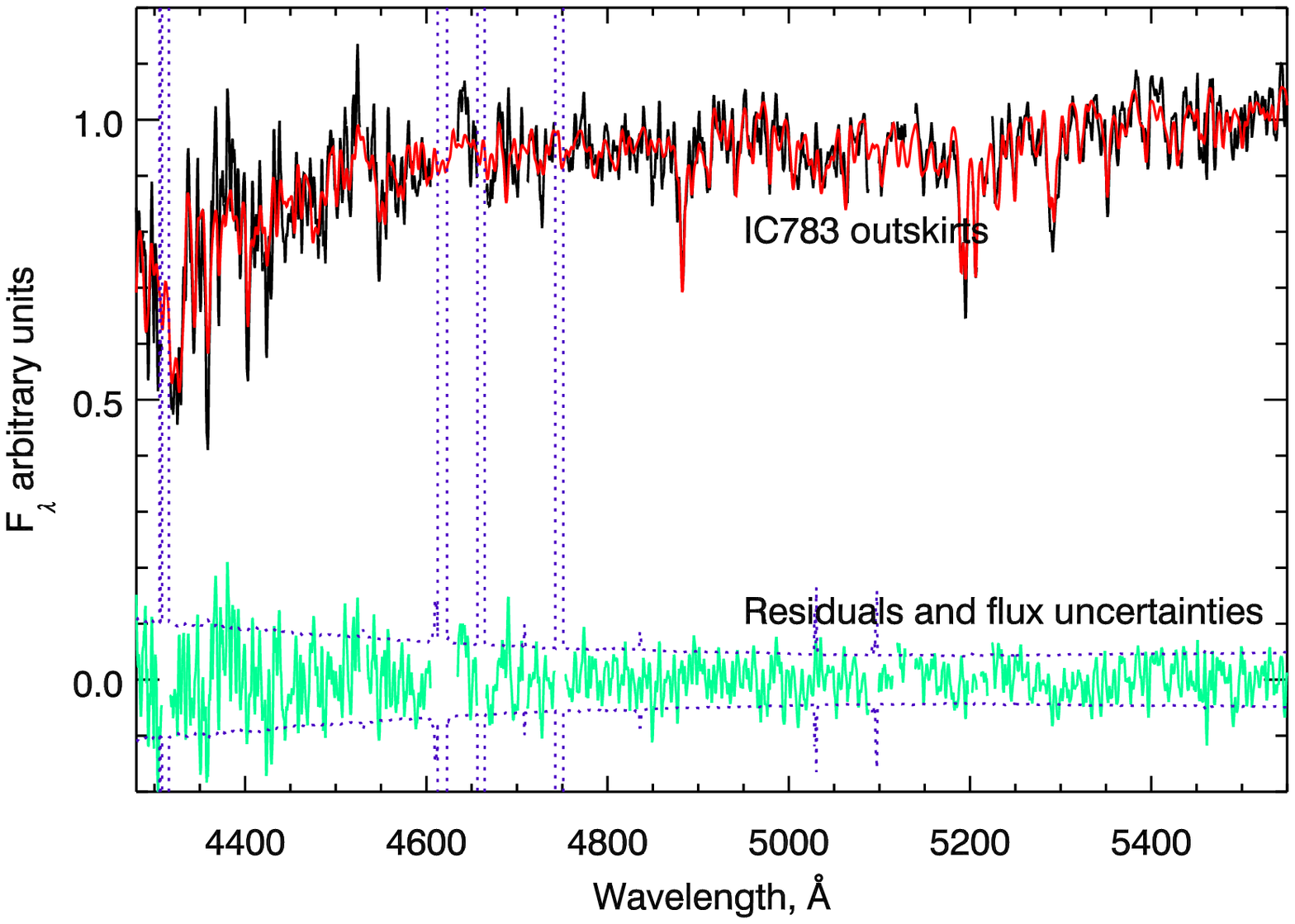}
\caption{
Spectra of central and peripheral parts of the galaxy, with the best-fitting
PEGASE.HR single stellar populations: observed and synthetic spectra
are shown as black and red solid lines, $\pm 1 \sigma$ flux uncertainties
are shown as dashed violet lines, and residuals of the fit as green
solid lines respectively.
\label{figspecic783}}
\end{figure}

\subsubsection{IC~3468}
An embedded structure is known to present in IC~3468 (Barazza et al. 2002).
The question on the nature of this substructure was left open, because
no rotation was detected in long-slit spectroscopy by Simien \& Prugniel (2002).
However, we found a complex kinematics -- rotation along two non-perpendicular
directions (NW-SE, and NE-SW separated by $\sim$ 60 degrees), none
of them coinciding with the position of the slit from Simien \& Prugniel
(2002). Unsharp masking of HST ACS imagery reveals elongated
structure in the central region of the galaxy. By varying the smoothing radius
for unsharp masking, different parts of this structure can be revealed.

In the map of luminosity-weighted age, an elongated substructure having
$t=5.3\pm0.4$~Gy, coinciding with one of the rotating components (NW-SE) is
clearly seen. It is roughly 3.3~Gy younger than the rest of the galaxy
($t=8.6\pm0.9$~Gy). One may also notice a ''blue stripe''\ in the velocity
dispersion distribution (lower by $\sim$ 10~km~s$^{-1}$ compared to mean
values) having the same locus. Based on these results we conclude that
NW-SE rotation corresponds to a moderately inclined stellar disc
($i \sim 60^{\circ}$). Rotational velocity is $17\pm4$~km~s$^{-1}$ at 7~arcsec
from the centre, but we cannot be sure to reach the maxima of rotation --
wider-field observations are needed.

Surprisingly, this disc almost does not affect the metallicity distribution.
Luminosity-weighted metallicity exhibits relatively smooth map
($Z=-0.60\pm0.05$~dex) with a slight gradient towards the centre
(up to $Z=-0.40\pm0.05$~dex).

The B-band mass-to-light ratios of the stellar populations are:
$(M/L)_{B*} = 2.8 \pm 0.4$ for the disc and $(M/L)_{B*} = 4.0 \pm 0.6$ for
outskirts. Our estimates of the luminosity-weighted age and metallicity in the
centre of IC~3468 are consistent with the g'-z' colour of the compact nucleus
(averaged within the host galaxy over the aperture corresponding to the seeing
conditions) provided in Ferrarese et al. (2006)

\begin{figure}
\hfil
\begin{tabular}{c c}
 (a) & (b) \\
 \includegraphics[width=7cm]{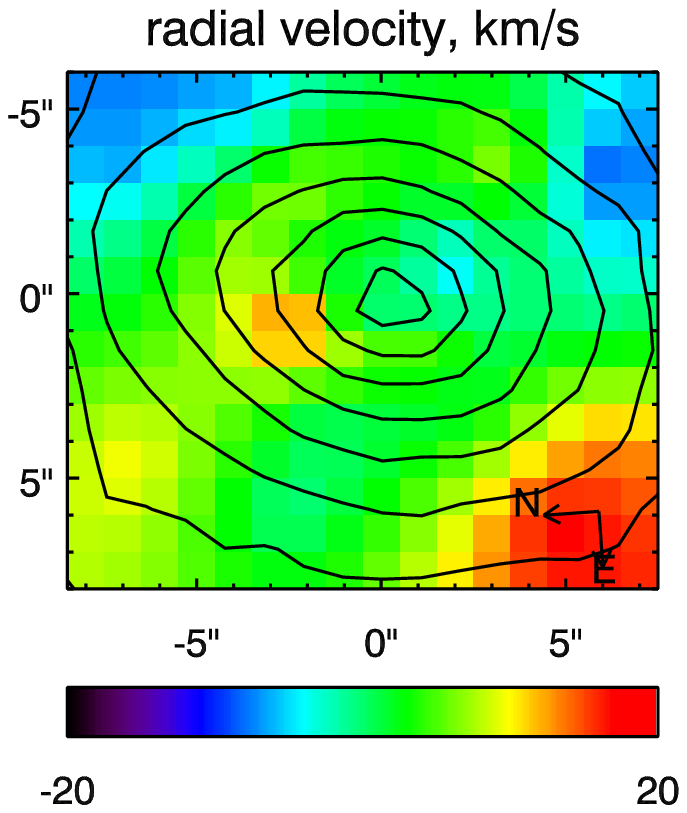} &
 \includegraphics[width=7cm]{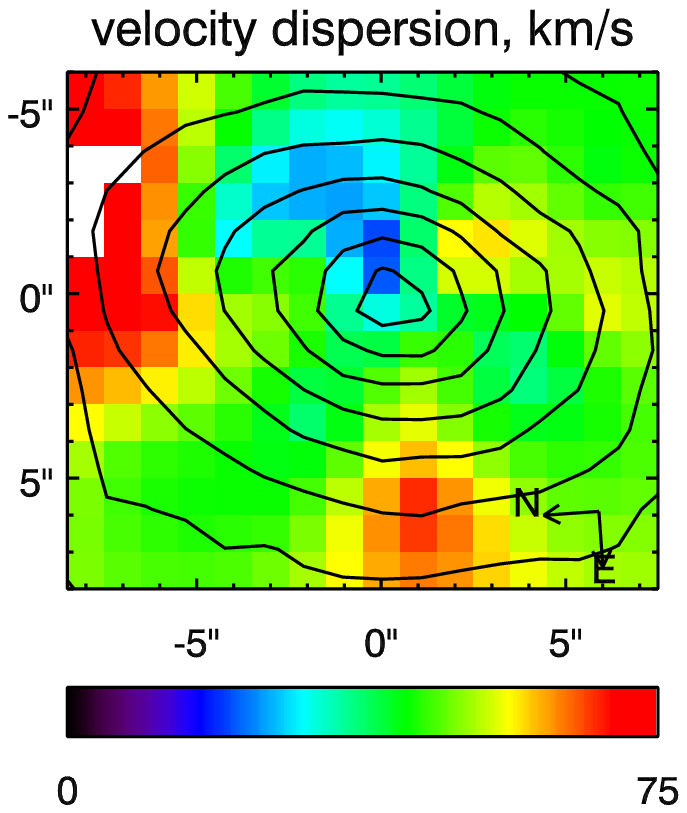} \\
 (c) & (d) \\
 \includegraphics[width=7cm]{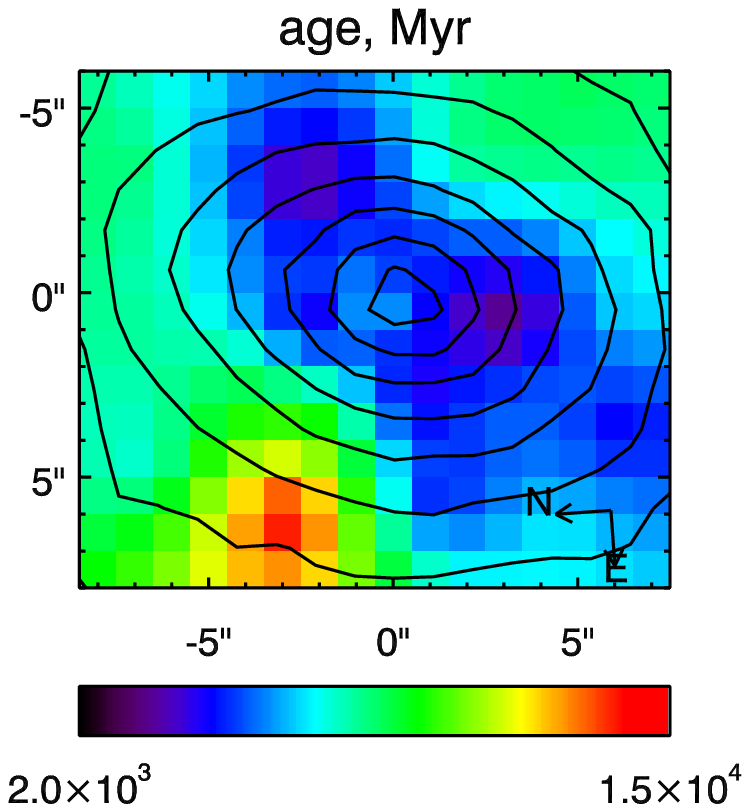} &
 \includegraphics[width=7cm]{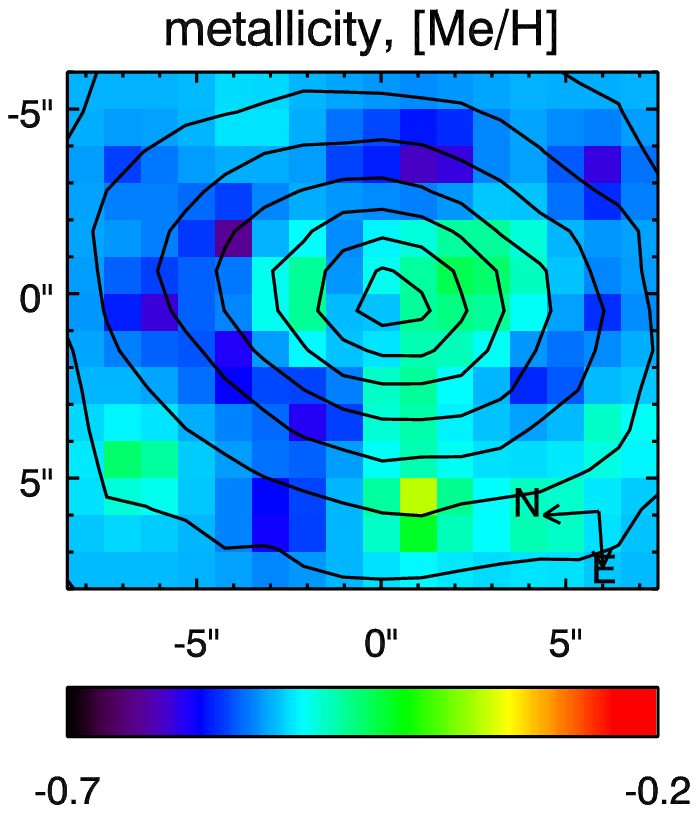} \\
\end{tabular}
\caption{Kinematics and stellar population of IC~3468. Maps of internal kinematics
and stellar population parameters are built for a Voronoi tessellation with a
target S/N ratio or 15. (a) line-of-sight stellar velocity, (b) stellar velocity
dispersion, (c) luminosity-weighted age, (d) luminosity-weighted metallicity.
\label{figic3468}}
\end{figure}

\subsubsection{IC~3509}
When we were selecting the targets for observations, IC~3509 was chosen as a
''prototypical''\ dE galaxy, classified as a galaxy without nucleus in
Binggeli et al. (1985). We did not expect to find unusual kinematics
and/or stellar
populations in this object. However, we detected a kinematically decoupled
central region, rotating ($v_{rot} \sim 10$~km~s$^{-1}$) in the perpendicular direction
to the major axis, where significant rotation is also seen ($v_{rot} \sim 20$~km~s$^{-1}$).
This structure is associated with a dip in the velocity dispersion distribution
(50~km~s$^{-1}$ compared to 75~km~s$^{-1}$) and a metallicity gradient of about 0.2~dex per
4~arcsec. Stellar population of the galaxy is relatively old and metal-poor
($t=7.8\pm0.8$~Gyr, $Z=-0.40\pm0.10$~dex, $(M/L)_{B*} = 4.3 \pm 0.5$)
In the very centre of the galaxy we see a spatially unresolved 
young ($t=4.1\pm0.4$~Gyr) metallic ($Z=-0.05\pm0.05$~dex) nucleus
($(M/L)_{B*} = 3.1 \pm 0.4$).

We applied unsharp masking technique to the HST imagery available from the
Virgo ACS Survey (Cot\^e et al. 2004) with different smoothing radii.
No fine structures have been revealed.

Kinematical appearance quite similar to IC~3509 was observed earlier in giant
early-type galaxies, for example, in NGC~5982 (Statler 1991). An explanation
was proposed, which did not require presence of dynamically distinct structures --
projection of orbits in the triaxial potential. Based on quite regular (except
the very centre) maps of stellar population parameters of IC~3509 we
conclude that the galaxy outside the core region can be represented by a
single-component triaxial ellipsoid. 

As in the case of IC~3468, for IC~3509 nucleus, the g'-z' colour reported by
Ferrarese et al. (2006) is consistent with our stellar population parameter
estimations.

\begin{figure}
\hfil
\begin{tabular}{c c}
 (a) & (b) \\
 \includegraphics[width=7cm]{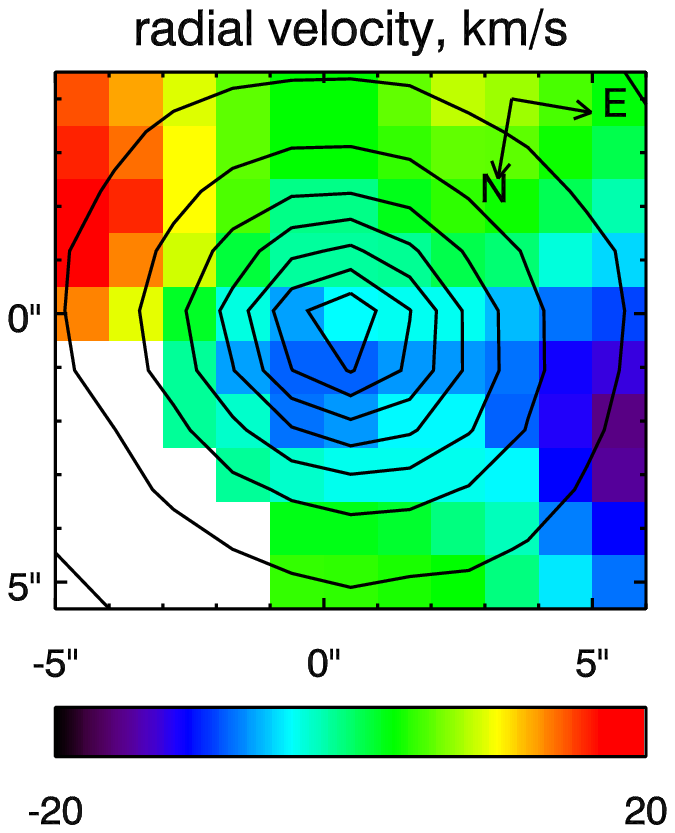} &
 \includegraphics[width=7cm]{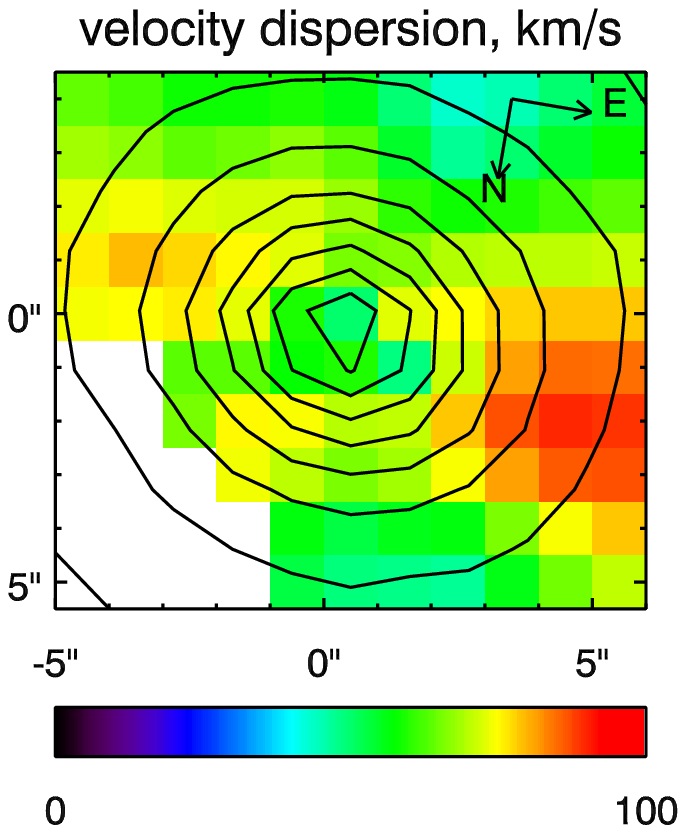} \\
 (c) & (d) \\
 \includegraphics[width=7cm]{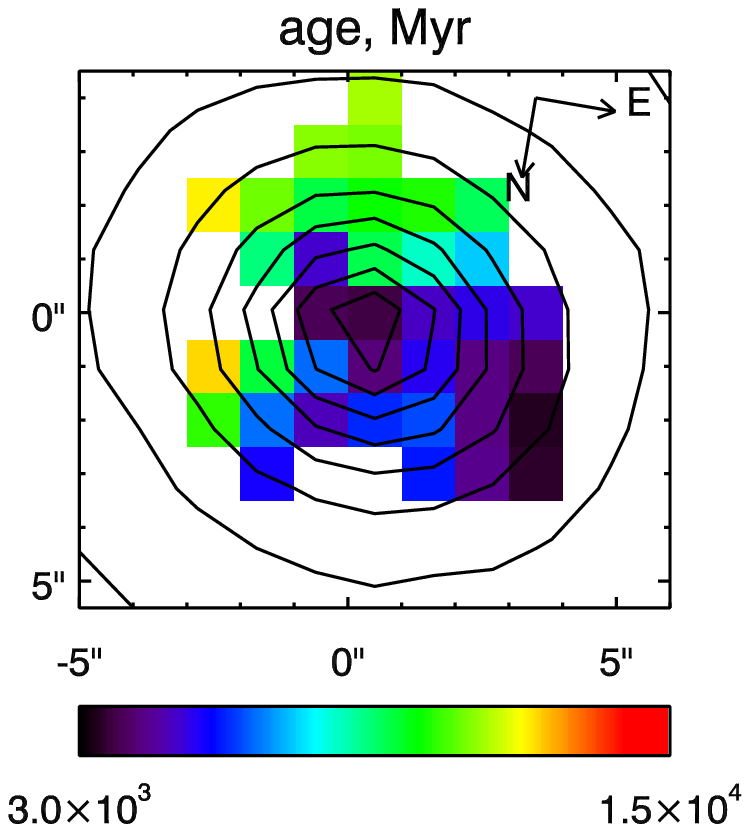} &
 \includegraphics[width=7cm]{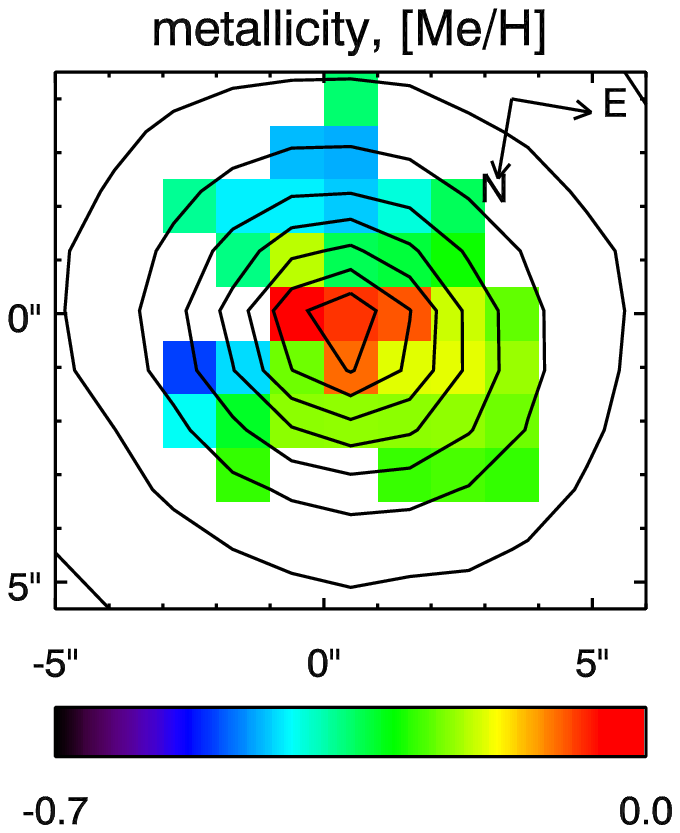} \\
\end{tabular}
\caption{Kinematics and stellar population of IC~3509. Maps of internal kinematics
and stellar population parameters are built for a Voronoi tessellation with a
target S/N ratio or 10. (a) line-of-sight stellar velocity,
(b) stellar velocity dispersion, (c)
luminosity-weighted age, (d) luminosity-weighted metallicity. 
\label{figic3509}}
\end{figure}

\subsection{Discussion}
\label{secdisc3dEV}
Since first discoveries of chemically (Sil'chenko et al. 1992) and evolutionary 
(Sil'chenko 1997, Vlasyuk \& Sil'chenko 2000) decoupled cores in giant
early-type galaxies, no attempts of modelling their formation and evolution were
made. Usual explanation of this phenomenon is a dissipative merger event. Whereas
merger is an established scenario of formation for giant early-type galaxies, 
normally it is considered as improbable for dwarfs because of their little sizes.


However, presence of embedded disc in IC~3468 is an important argument for a
merger scenario. Kinematically decoupled structures, associated with young
metal-rich stellar population is consistent with a hypothesis of dissipative
merger event which took place several Gyr ago. Dissipative merger is expected
to trigger a starburst that consumes available gas, leading to a KDC that is
younger than the host galaxy.

Another possibility is effects of ram pressure stripping. Efficiency of ram
pressure stripping depends on the density of the region being stripped: higher
density results in lower efficiency (Gunn \& Gott, 1972; Abadi et al. 1999).
Thus, it might happen that in the dense nucleus of a dwarf galaxy gas will
not be removed. A similar phenomenon of gaseous disc truncation is observed
in giant spiral galaxies in the Virgo cluster (Cayatte et al. 1994,
Kenney \& Koopmann 1999) and modelled by Abadi et al. (1999).

A possible scenario to acquire a structure observed in IC~783 (full absence
of gas and young nucleus without obvious evidences of kinematical decoupling) is
multiple crossings of the cluster centre. IC~783 is located on the projected
distance of 1.1~Mpc from the centre of the Virgo cluster. Thus its orbital period
is at least 4.5~Gyr (assuming mass of the cluster of $10^{14}$~M$_{\odot}$).
Gas of IC~783 can be depleted in the disc during the first passage, but
preserved in the inner dense nucleus, because intercluster medium density
and/or velocity of the
galaxy might be not sufficient to remove gas completely. 

To remove gas and stop star formation in the core we may assume that during
the second crossing several Gyr later the intracluster orbit of the galaxy
might be transformed into more elongated one, say due to casual encounter
with a massive galaxy, so $v_{cross}$ would increase (and $\rho$ as well,
because the galaxy would pass
closer to the centre of the cluster), resulting in ram pressure
$P = \rho v^2$ reaching sufficient value to strip the nuclear region of
the galaxy completely.

Alternative possibility of varying efficiency of ram pressure stripping
for IC~783 might be explained by its belonging to Messier~100
group. IC~783 is located at some 90~kpc of projected distance from M~100,
and radial velocity difference of $\sim$270~km~s$^{-1}$ is an argument for
their interaction. The M~100 group is believed to have crossed
the Virgo cluster centre recently (Binggeli et al. 1987). Assuming that it is passing
near its apocentre now the orbital period of M~100 group in the cluster
turns to be about 5--8~Gyr, thus the recent passage of the cluster centre
took place some 3--4~Gyr ago, and the previous one around 8--12~Gyr ago.
On the other hand, the orbital period of IC~783 with respect to
M~100 should be around 1~Gyr. Thus if the orbital velocity of IC~783 was
counter-directed to the orbital motion of M~100 in the cluster during the
first passage of the central region, and co-directed during the second
one, the ram pressure value $P = \rho v^2$ might differ by a factor
of 3.5 (assuming the maximum velocity of M~100 with respect to
the Virgo intracluster medium to be $\sim 1000$~km~s$^{-1}$ and the orbital
velocity of IC~783 with respect to M~100 to be $\sim 300$~km~s$^{-1}$).
The coincidence of the estimated dates of the cluster centre crossing
by the M~100 group with the ages of two subpopulations in IC~783
is a strong argument for this scenario.

Ram pressure stripping during repetitive crossings of the cluster centre may be
considered as a possible explanation of young metal-rich cores in the
low-luminosity early type galaxies. Depending on the orbital parameters for a
particular galaxy, one would expect large scatter of ages/metallicities of these
substructures with respect to their host galaxies.

Good agreement between our estimations of the stellar population parameters and
g'-z' colours in the nuclei of IC~3468 and IC~3509 may be considered as an
evidence for the presence of young metal-rich stellar populations in all
compact blue nuclei of dE galaxies. However, this hypothesis may only be
proved by the forthcoming observations.

\chapter{dE galaxies in groups}
In this chapter we present 3D spectroscopic observations of two quite unusual
dwarf galaxies in groups to demonstrate how different environmental
conditions may affect their evolution.

Parameters of observations presented in this chapter are summarised in
Table~\ref{tabobsgroup}.

\begin{table}
\centering
\begin{tabular}{llrrcr}
\hline
Name & Date & seeing & $t_{exp}$ & S/N$_{cent}$ & $\sigma_{cent}$ \\
\hline
NGC~770 & 2004/Oct/07 & 2'' & 2h & 50 & 100 \\
NGC~126 & 2006/Sep/07 & 1.3'' & 1h 20m & 25 & 100 \\
NGC~127 & 2005/Oct/01 & 1.5'' & 2h & 20 & 90 \\
NGC~130 & 2005/Oct/02 & 2'' & 1h 30m & 25 & 120 \\
\hline
\end{tabular}
\caption{Parameters of observations of low luminosity elliptical and
lenticular galaxies in groups. \label{tabobsgroup}}
\end{table}

\section{NGC~770 (NGC~772 group)}
NGC~770 is a low-luminosity elliptical ($M_B$ = 18.2) companion to the large
spiral galaxy NGC~772, and the brightest satellite galaxy identified in this
system by Zaritsky et al. (1997). The parent spiral galaxy NGC~772 ($M_B =
21.6$) is listed in the Atlas of Peculiar Galaxies (Arp~78; Arp, 1966) and
known to contain a prominent asymmetric spiral arm. Radial velocity of NGC~772
according to Zaritsky et al. (1997) is 2468~km~s$^{-1}$ results in a
distance of 33~Mpc assuming $H_0=75$~km~s$^{-1}$~Mpc$^{-1}$.

Luminosity of NGC~770 (M$_B=-18.2$) and its velocity dispersion
(110-120~km~s$^{-1}$) place it between giants and dwarfs. Spectroscopy of
NGC~770 does not reveal presence of emission lines, though neutral hydrogen
is detected (Geha et al. 2005). Very good signal-to-noise ratio in the MPFS
data allows precise measurements of all the parameters ($\Delta v \sim 1$
km~s$^{-1}$). Good sampling of LOSVD due to high velocity dispersion allows to
measure $h3$ and $h4$ Gauss-Hermite coefficients.

We see impressive kinematically decoupled core in this object. This
kinematical decoupling has been recently reported by Geha et al. (2005).
There are following evidences for considering it to be a counter-rotating
young metal-rich highly inclined stellar disc aligned almost with the major
axis of the galaxy:
\begin{itemize}
\item unsharp masking reveals presence of embedded structure (Geha et al.
2005);
\item velocity dispersion map shows larger values on the ''switches'' of the
velocity, there is a stripe with lower values orthogonal to the structures
in age/metallicity distributions;
\item map of $h3$ coefficient shows regions with positive and negative values
before and after ''switch'' of the velocity;
\item age and metallicity distributions clearly show the decoupled structure
aligned almost along the major axis having younger age ($t \sim 4.5$ Gyr vs
7 Gyr) and higher metallicity ($[Fe/H] \sim -0.05$ vs $-0.2$) than the
surrounding spheroid.
\end{itemize}

We consider two possibilities to acquire such a KDC: (1) a partially
dissipative merger event, that induced a burst of star formation some 4.5
Gyr ago producing the central stellar structure that we see now as a counter
rotating subpopulation with the high metallicity; (2) dissipative accretion
of the gas from the nearby giant gas-rich NGC~772. In the next section we
demonstrate evidences for ongoing gas accretion on another possible dE
progenitor (NGC~127), which is an indirect argument for the second
possibility of KDC formation in NGC~770.

\begin{figure}
\begin{tabular}{c c c}
(a) & (b) & (c) \\
\includegraphics[width=5.5cm]{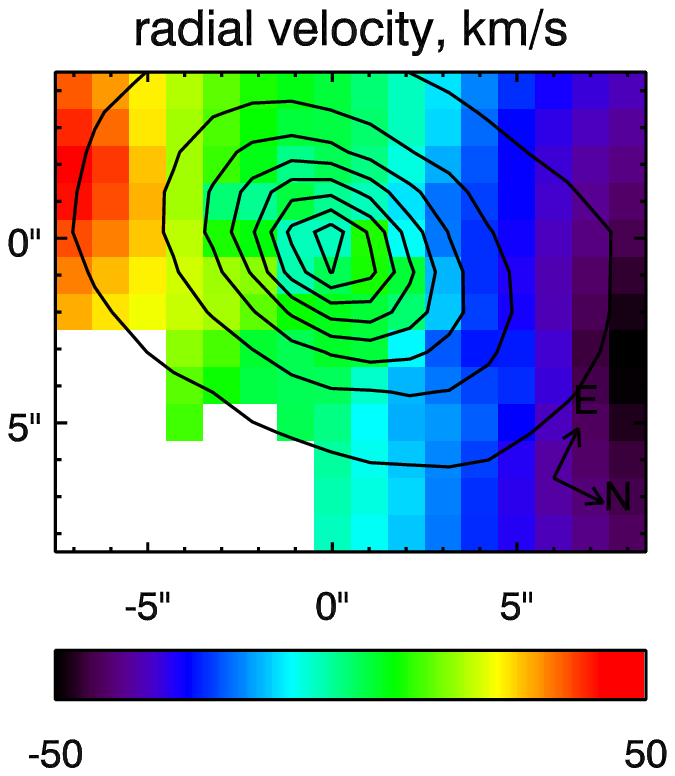} &
\includegraphics[width=5.5cm]{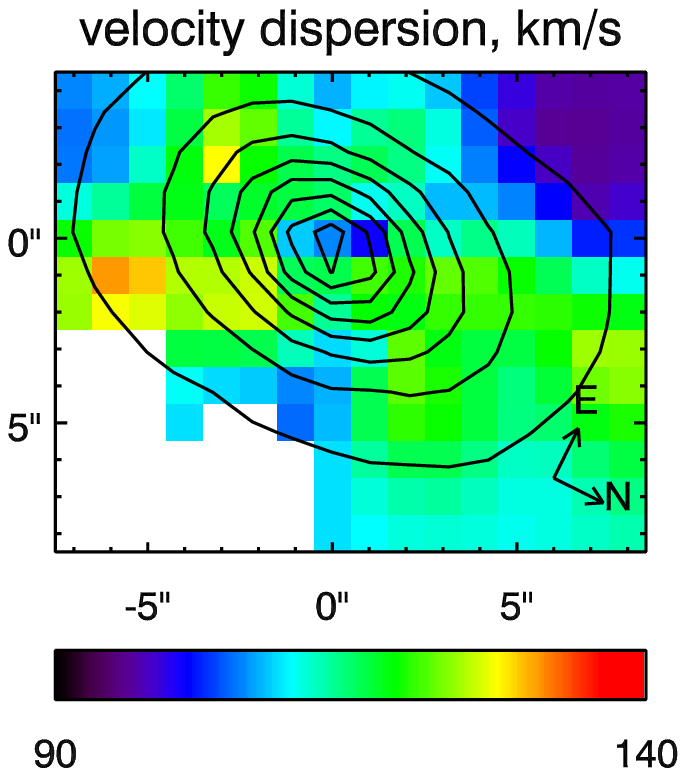} &
\includegraphics[width=5.5cm]{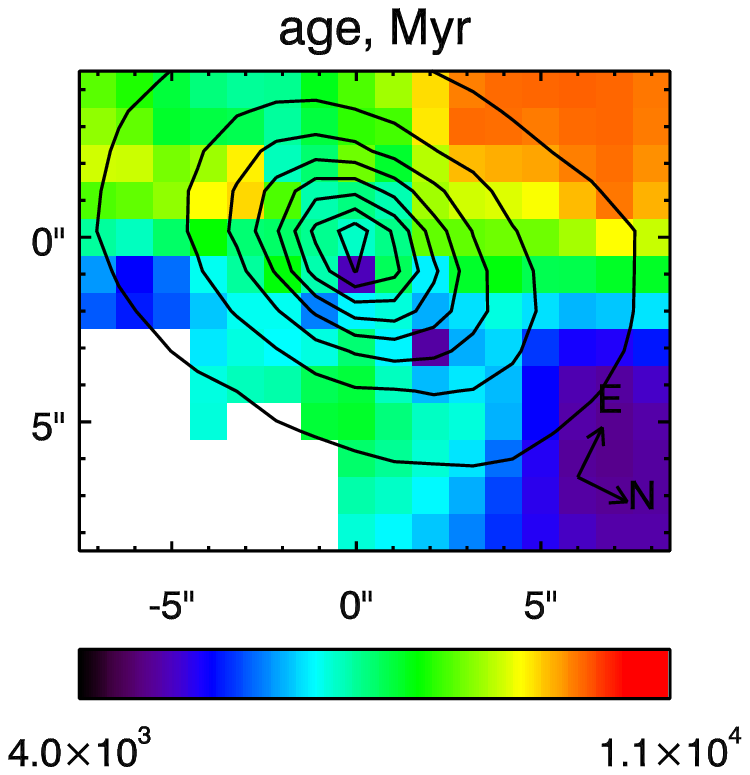} \\
(d) & (e) & (f) \\
\includegraphics[width=5.5cm]{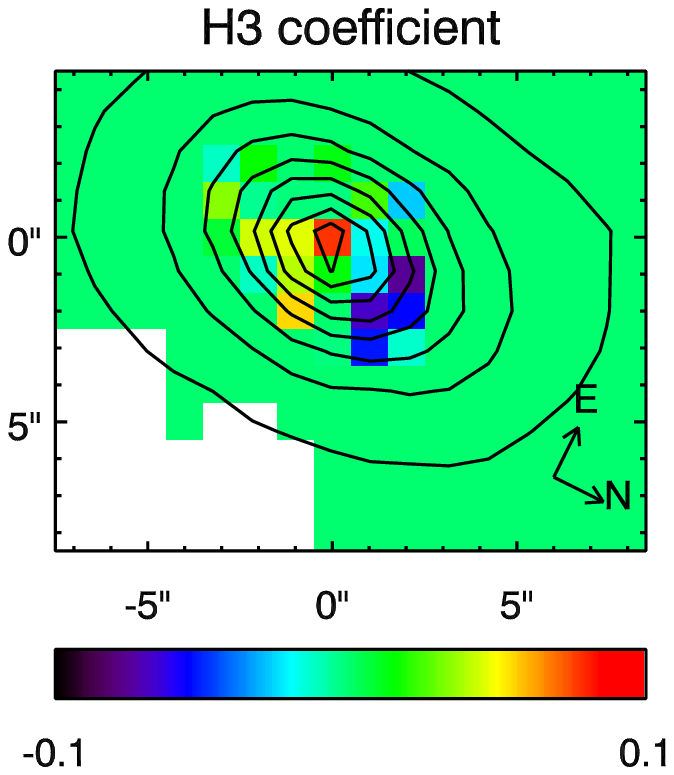} &
\includegraphics[width=5.5cm]{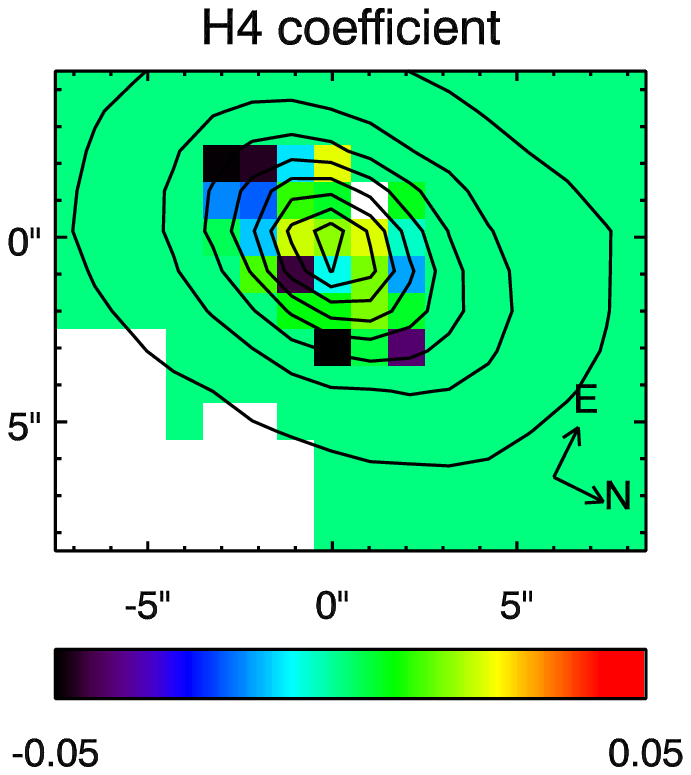} &
\includegraphics[width=5.5cm]{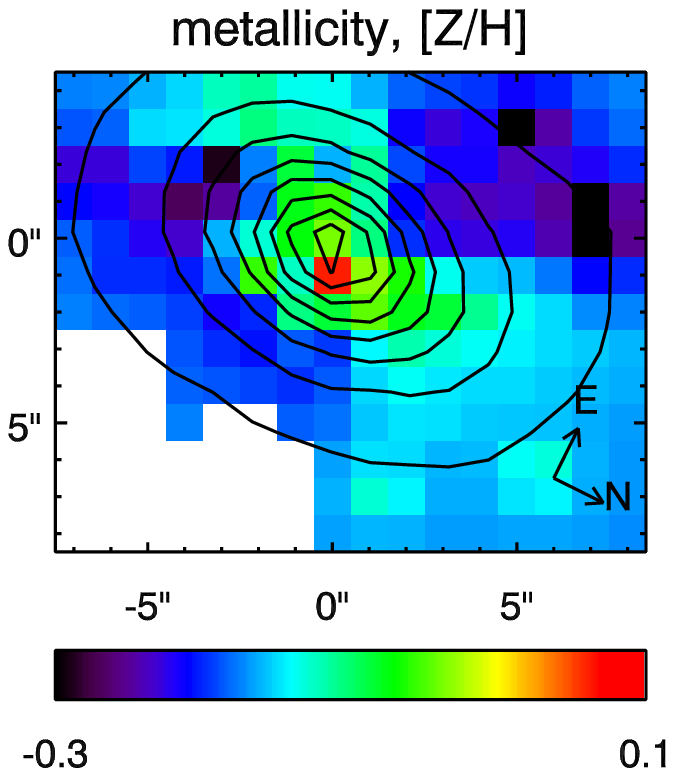} \\
\end{tabular}
\caption{Kinematics and stellar populations of NGC~770. (a), (b), (d), (e)
panels show radial velocity, velocity dispersion, and $h_3$ and $h_4$
coefficients of Gauss-Hermite parametrization. (c) and (f) demonstrate maps
of SSP-equivalent age and metallicity.\label{fign770}}
\end{figure}

\section{NGC~127 (NGC~128 group)}
NGC~128 group is located some 55~Mpc away (assuming $H_0=75$~km~s$^{-1}$).
It includes (spectroscopically confirmed membership): giant gas-rich S0
galaxy NGC~128, giant early-type spiral NGC~125, three low-luminosity
lenticulars in the vicinity of NGC~128 (closer than 100~kpc in the projected
distance): NGC~126 ($M_B=-18.5$), NGC~127 ($M_B=-18.6$), and NGC~130
($M_B=-19.0$), and a handful of late-type spirals located beyond 300~kpc
from the centre of the group.

\begin{figure}
\includegraphics[width=17cm]{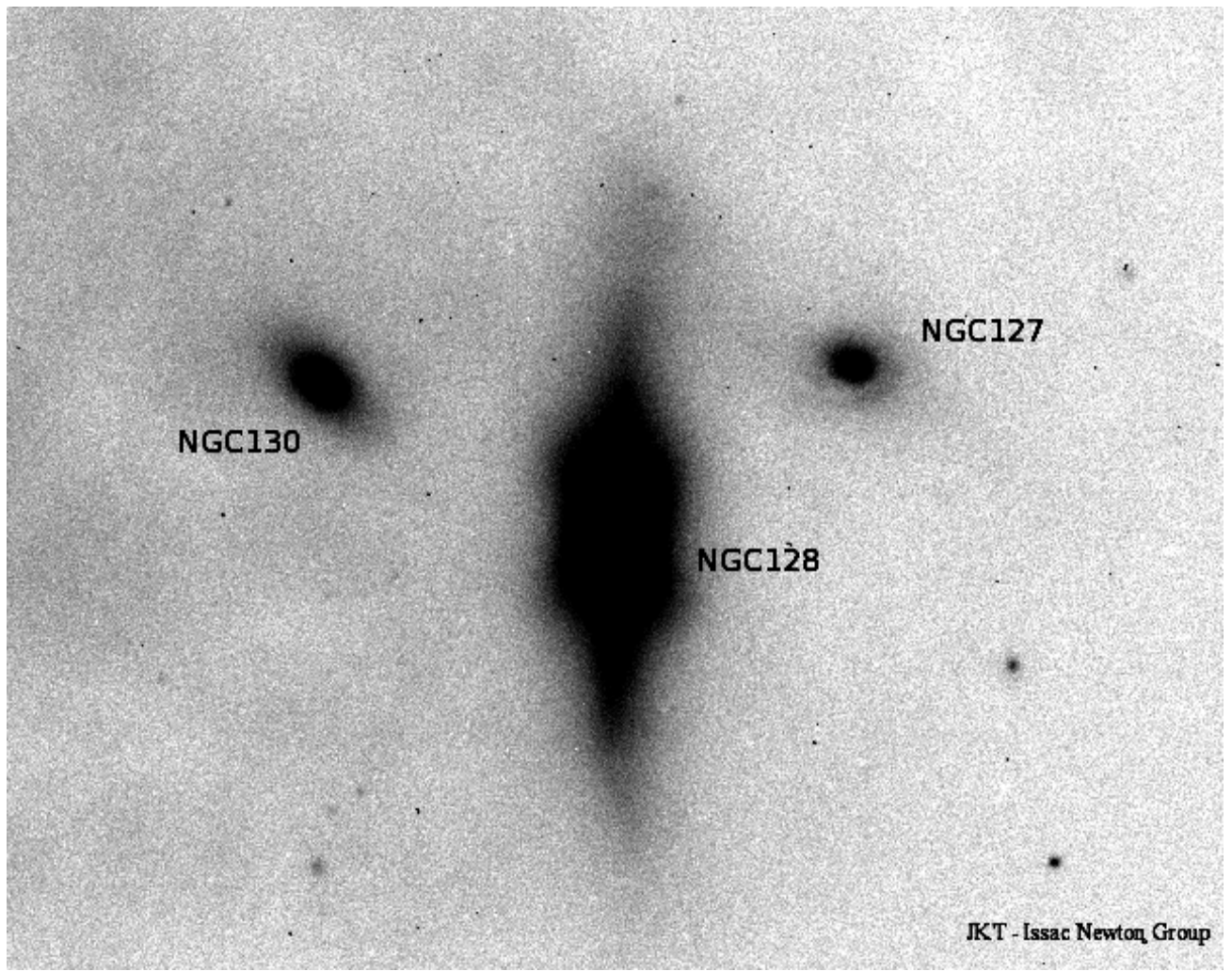}
\caption{Central part of the NGC~128 group in B band. Image obtained with
1.0-m JKT (La Palma) and available through the archive of the Isaac Newton
Group.\label{fign128group}}
\end{figure}

NGC~128 is a giant lenticular galaxy with a peanut-shaped bulge. It is known
to have a gaseous disc counter-rotating to the stars (Emsellem \& Arsenault, 1997).
Faint gaseous tail is seen in the direction of NGC~127 (\ref{fign128group}).

We have obtained MPFS observations for all three low-luminosity early-type
galaxies in the group: NGC~126, NGC~127, and NGC~130 during two observing
runs: in October 2005 and September 2006. 

NGC~126 and NGC~130 show no evidence of emission lines, exhibit relatively
old stellar population (5...7~Gyr) with nearly solar metallicity and quite
regular kinematics. Here we are not discussing these two objects, because
they look quite similarly to what is observed in the clusters, and their
evolution was probably ruled by the same mechanisms.

On the contrary, NGC~127, located about 1~arcmin away from the centre of
NGC~128 (16~kpc in the projected distance) is a gas-rich object, having
early-type morphology (according to NED, HyperLeda). Its spectrum provides
evidence for ongoing star formation event. Using MPFS data we are able to
study both stellar and gas kinematics, and give estimates for stellar
population parameters. A dust lane is seen on the JKT image south of the
NGC~127 core (see Fig.~\ref{fign128group}).

Luminosity-weighted age of NGC~127 appears to be around 1~Gyr, with a slight
gradient from 1.5 in the outskirts to 0.8 in the centre. Metallicity is
supersolar ([Fe/H]=+0.2~dex) everywhere except the very centre of a galaxy,
where it exhibits minimum (-0.1~dex). However, taking into account that we
are not including additive continuum terms in the fitting procedure, this
might be connected to a strong contamination by the nebular continuum,
increasing the continuum level and leading to biased metallicity estimates
(see Chapter~1). Fitting was done after masking regions of emission lines:
$H\gamma$, $H\beta$, [OIII], and [NI]. Residuals contain almost flat
zero-level continuum and emission lines.

Velocity field of gas was computed by fitting a single component Gaussian,
convolved with the instrumental response of MPFS into H$\beta$ emission into
the residuals of stellar population fit. Rotation velocity of gas is higher
($\sim$100~km~s$^{-1}$) than of stars ($\sim$40~km~s$^{-1}$). Velocity field
of gas shows asymmetry on the south-eastern part of NGC~127 (direction of
NGC~128): in that region $v_{gas}-v_{stars}$ is 20~km~s$^{-1}$ higher by
absolute value than in the symmetric north-western part, reaching nearly
-60~km~s$^{-1}$ 6~arcsec south-east of NGC~127 core. This asymmetry might
be caused by perturbed motions of gas near the region, where the flow
reaches NGC~127.

Difference in radial velocities of NGC~127 and NGC~128 is -160~km~s$^{-1}$.
Taking into account the presence of a gaseous bridge between them and kinematical appearance
of NGC~127, we are proposing the following scenario: NGC~127 has recently
passed its pericentre, and now we observe an infall of gas from NGC~128
onto NGC~127. Such a process might have lead to the counter-rotating core as
observed in NGC~770 if the orientation of angular momentum of gas was
different. If one imagines appearance of NGC~127 after gas removal and
several Gyr of passive evolution, it will be indistinctive from ''normal''\ dE
galaxies by morphology and luminosity (its absolute magnitude will become 2~mag
fainter), but only by relatively high metallicity.

\begin{figure} 
\begin{tabular}{c c c}
(a) & (b) & (c) \\
\includegraphics[width=5.5cm]{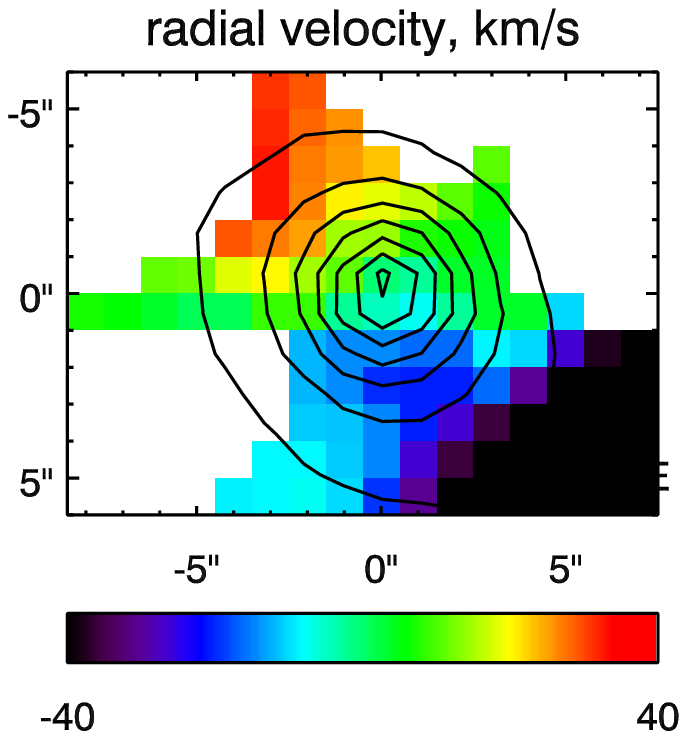} &
\includegraphics[width=5.5cm]{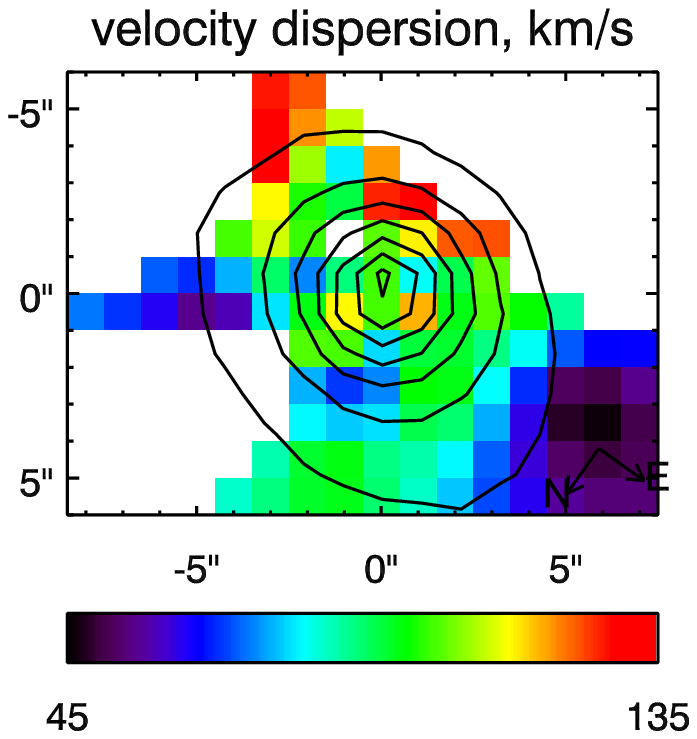} &
\includegraphics[width=5.5cm]{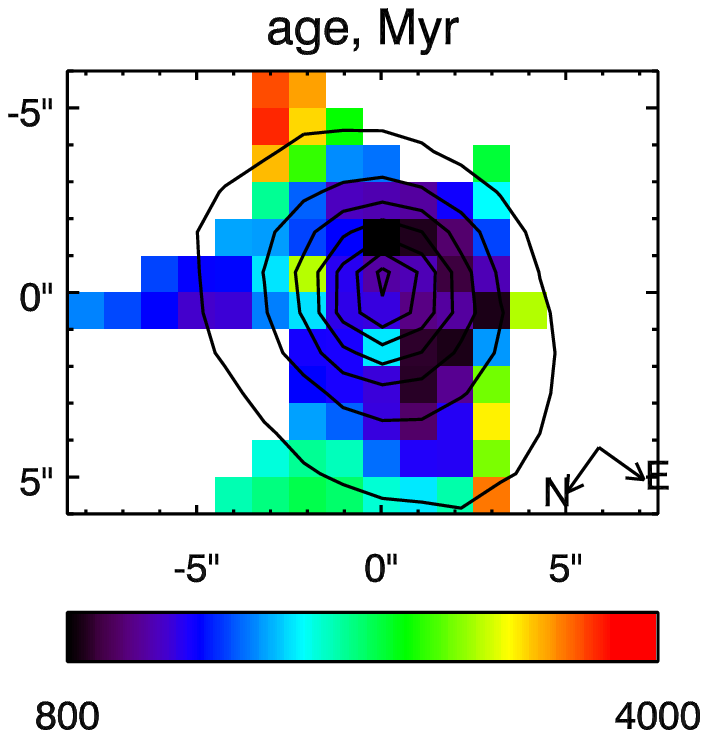} \\
(d) & (e) & (f) \\
\includegraphics[width=5.5cm]{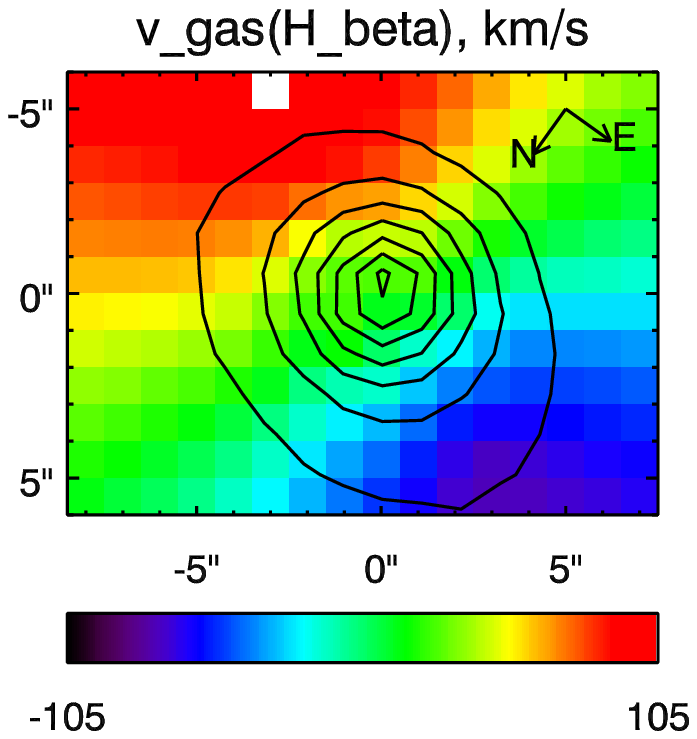} &
\includegraphics[width=5.5cm]{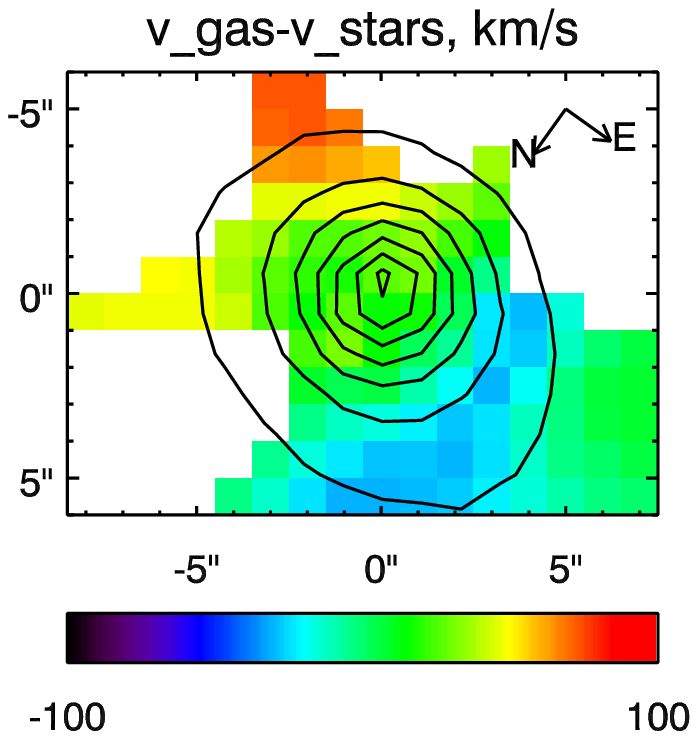} &
\includegraphics[width=5.5cm]{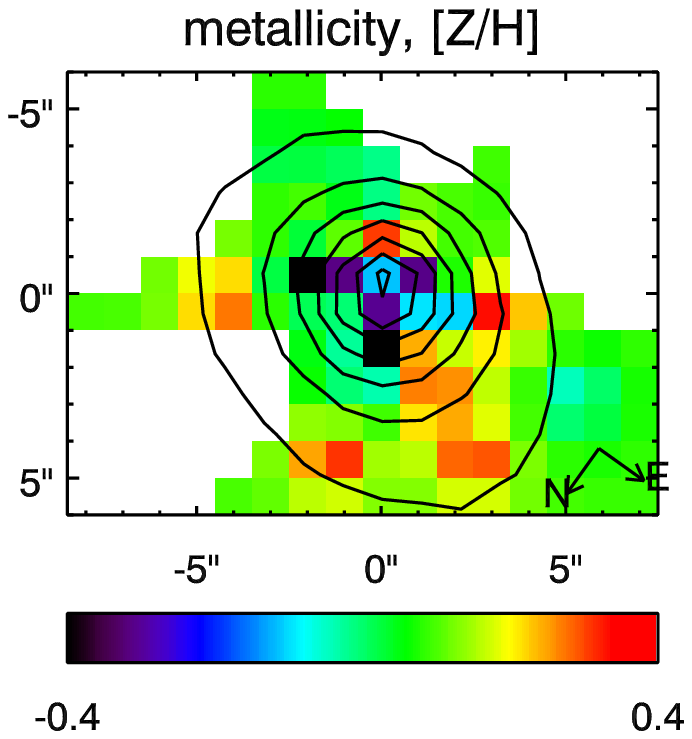} \\
\end{tabular}
\begin{tabular}{c}
(g) \\
\includegraphics[width=17cm,height=6cm]{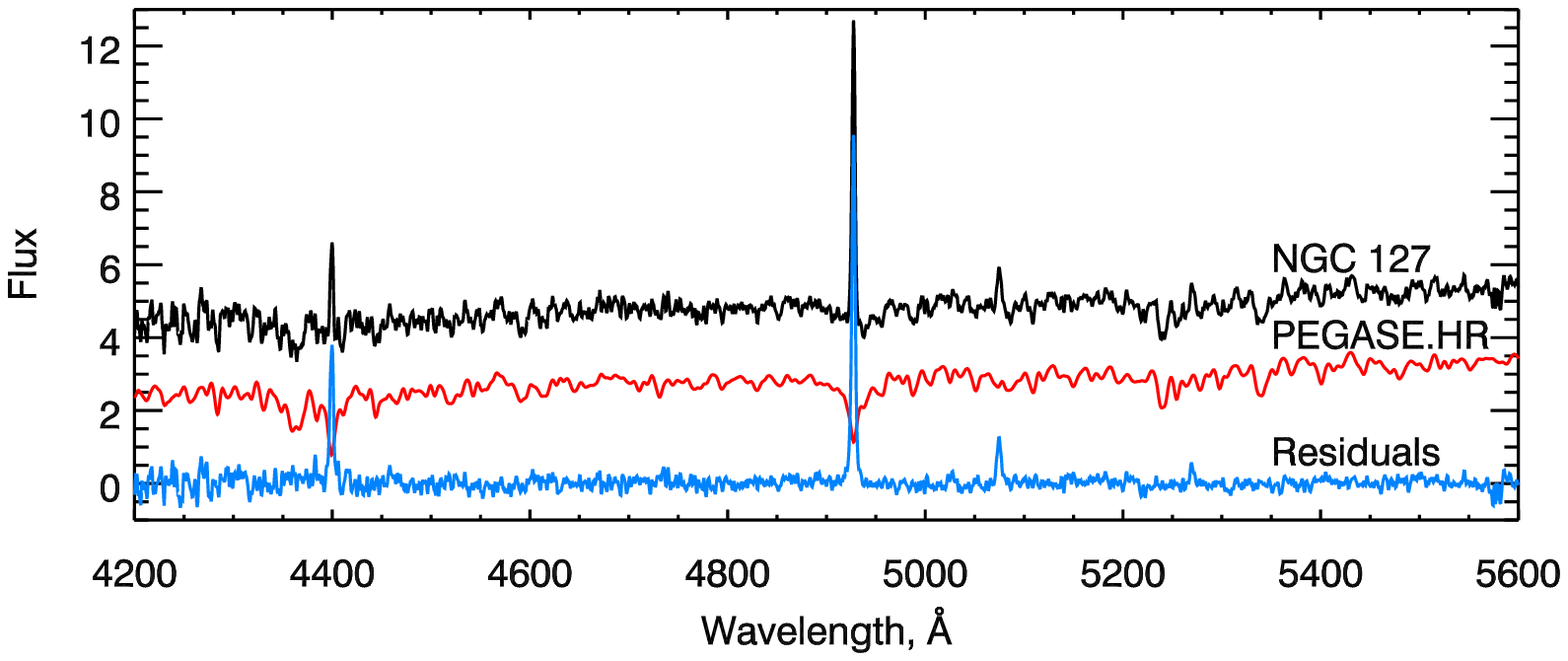} \\
\end{tabular}
\caption{NGC~127: SSP-equivalent stellar population parameters, kinematics
of stars and gas. Panels (a) and (b) show stellar radial velocity and velocity
dispersion; (c) is SSP-equivalent age; (d) is radial velocity of gas
measured on H$\beta$ emission line; (e) is difference between radial
velocities of gas and stars; (f) is SSP-equivalent metallicity; (g) is
a fit for the NGC~127 centre.
\label{fign127}}
\end{figure}

The process of slow accretion which is quite favourable in groups of
galaxies, where relative velocities are rather low, appears to be
improbable scenario for formation of embedded structures in dE's residing in
regions of the Universe with higher density, such as clusters of galaxies.

\chapter{Studies of galaxies in Abell~496}

Abell~496 is a richness class 1 cluster (Abell 1958) of cD type (Struble \&
Rood 1987) at a redshift of 0.0335 (Durret et al. 2000 and references
therein). For a Hubble constant H$_0$=72 km s$^{-1}$ Mpc$^{-1}$ the distance
modulus is 35.71 ($d=139~Mpc$) and the corresponding scale is 40.4
kpc/arcmin.  Abell~496 is a cluster with several hundred measured galaxy
redshifts. The analysis of the distribution of 466 redshifts in the
direction of this cluster has revealed the existence of several structures
along the line of sight; however, the redshift distribution of the 274
galaxies found to belong to the cluster itself implied that Abell~496 has a
regular morphology and a well relaxed structure (Durret et al. 2000). This
is confirmed by X-ray data: the X-ray map obtained from XMM-Newton
observations is indeed quite regular, contrary to most clusters where even
if the X-ray emissivity map appears regular, the temperature map of the
X-ray emitting gas does not (Durret et al. 2005).

\section{Observations and Data Reduction}
\subsection{Imaging observations and reduction}

Images were obtained at the Canada France Hawaii Telescope with the
Megacam camera in the fall of 2003 (program 03BF12,
P.I. V.~Cayatte). Megacam covers a field of 1$^\circ
\times 1^\circ$ on the sky, with a pixel size on the sky of
0.187~arcsec. Deep images were obtained in the u, g, r and i filters.

These images were reduced in a usual way (bias and flat field
corrections, photometric and astrometric calibrations) by the staff of
the Terapix data center at IAP, France. The SExtractor software was
run on the r image (that with the best seeing) to detect objects and
measure their positions and magnitudes. In particular, magnitudes
within a 1.2~arcsec diameter were measured, in order to prepare
Giraffe observations (see below).

Stars were then discarded based on a diagram of aperture minus total
magnitude versus total magnitude for r$<$21. Above this magnitude, all
objects were kept in our galaxy sample. A photometric redshift code
was kindly applied by O.~Ilbert to our catalogue to try eliminating
background galaxies. Finally, the galaxies observed with Giraffe were
taken from this imaging catalogue, with a magnitude within a diameter
of 1.2~arcsec in the r band (r$_{1.2}$) in the [17.5-22] interval.

\subsection{Spectroscopic observations and reduction}

Spectra were obtained at the ESO Very Large Telescope with the Giraffe
instrument with the L682.2 configuration on the nights of 8-9/12/2004. The
Giraffe field of view is 20~arcmin in diameter, with a total number of
fibres of 130; each fiber has an aperture on the sky of 1.3~arcsec in
diameter. The 600 lines/mm grating has been used in the LR4 setup giving a
resolving power about $R=6300$ in the wavelength range 5010-5831\AA.

The first night four exposures have been obtained during 2700, 3300, 2351
and 1699 seconds. The second nights four other exposures have been added
with the same positioner configuration file and effective duration of 2700,
3300 X 2 and 4200 seconds. During the day, the exposures of bias, flats and
comparison lamps for the wavelength calibration have been done in the same
setup and with the two separate sets of MEDUSA fibers. The description of
GIRAFFE instrument can be found in Pasquini, L. et al., 2002 (The Messenger,
110, 1). The spectra were extracted and calibrated using the Python version
of BLDRS - Baseline Data Reduction Software (girbldrs-1.12) available from
http://girbldrs.sourceforge.net and with functions and recipes description
in the BLDRS Software Reference Manual, Doc. No. VLT-SPE-OGL-13730-0040
(Issue 1.12, 20 Septembre 2004). When needed art of the reduction have been
done using IRAF as well as redshift determination. The processing includes
bias subtraction, diffuse light estimation and remove, localisation and
extraction, correction for the fiber transmission variations, wavelength
calibration, division by the continuum lamp spectrum and sky subtraction.

The observed galaxies were taken from the catalogue described above,
with the following priorities: top priority: objects with
17.5$<$r$_{1.2}<20.75$; middle priority: objects with
20.75$<$r$_{1.2}<21.5$; low priority: objects with
21.5$<$r$_{1.2}<22$. 112 galaxy spectra were thus obtained (some fibers
had to be used for guide stars and sky spectra).

Individual 1D spectra were combined with the IRAF software and
redshifts were measured with the rvsao.xcsao package in IRAF, using
various star templates. Redshifts were also measured through the
stellar population synthesis fit described in this work, and both
values agreed within their uncertainties.

Our star-galaxy separation came out to be very good, since no stars
were observed. On the other hand, the rejection of background objects
based on photometric redshifts was not very efficient, since only 52
out of 112 galaxies with measurable redshifts actually belong to the
cluster. 46 of those 52 have sufficient signal-to-noise ratios for
analysis of kinematics and stellar population properties.

Absolute magnitudes is computed using distance modulus mentioned above. All
magnitudes are corrected for intergalactic extinction according to Schlegel
et al. 1998.

\section{Stellar Population Fitting and Results}
To deduce kinematical and stellar populations parameters we have used direct
fitting of the PEGASE.HR (Le Borgne et al. 2004) synthetic spectra into
observed data in the pixel space (see Chapter~1 for details). 

Taking into account high spectral resolution of the Giraffe spectrograph in
the MEDUSA mode (R=7000) PEGASE.HR models based on high-resolution ELODIE.3
stellar library (R=10000) are left as the only alternative to avoid
degradation of the spectral resolution of the observed spectra. In order to
acquire unbiased estimates of the velocity dispersion, one needs to take
into account variations of the spectrograph's line-spread-function (LSF) and
broaden template spectra according to LSF shape (strictly speaking to the
difference between LSF of spectrographs used to obtain the spectra being
analysed and stellar library used for the spectral synthesis purposes).
To achieve this we fitted twilight spectra obtained with Giraffe at the same
setup as Abell~496 galaxies with solar spectra available in the ELODIE.3
library which obviously have exactly the same intrinsic LSF as stars used
for the spectral synthesis. Instrumental response of Giraffe appeared to be
very stable across fibers. Instrumental width ($\sigma_{inst}$) is changing
smoothly from 19~km~s$^{-1}$ at 5000\AA\ to 15~km~s$^{-1}$ at 5800\AA,
H3 remains stable at about -0.01, and H4 at about -0.07. Slightly negative
values of H4 are trivially explained by sizes of fibers 
(1.2~arcsec) 
which are larger than the normal slit width of the
spectrograph (diffraction limit of the collimator) resulting in the 
$\Pi$-shaped LSF.

Fitting method is not very sensitive to the presence of H$\beta$ feature in
the spectral range: though age estimations have higher uncertainties, they
remain unbiased (see Chapter~1 for details).

We have applied elliptically-smoothed unsharp masking technique (Lisker et
al. 2006) to CFHT/Megacam images for revealing embedded structures. Using
different smoothing radii (major axes of ellipses), from 1.5 to 4 arcsec we
subjectively classified all the objects into three categories: no, weak and
strong embedded structures of the following types: bar, disc, spiral arms,
ring (see Tab~\ref{tabresfit}). 6 and 9 objects turned to have weak and
strong embedded structures respectively. To be stressed that we do not
observe embedded structures among faint galaxies, in return, bright ones
often exhibit strong and complex embedded systems.

In Tab.\ref{tabresfit} we present values of line-of-sight radial velocities,
central velocity dispersions, SSP-equivalent ages and metallicities of the
central regions of galaxies from our sample obtained with direct fitting of
PEGASE.HR synthetic spectra. Besides, morphological classification and
indication for embedded structures based on Megacam multicolor imagery 
(see next section) are given.

\begin{table}
\begin{small}
\begin{tabular}{lccccccccl}
IAU Name & M(B) & morph. & $v_r$, km/s & $\sigma_0$, km/s & $t$, Gyr & $Z$, dex & \multicolumn{2}{c}{emb.str.} \\ 
\hline
A496J043333.53-131852.6 &-18.79 & SBa & 11704 $\pm$ 1.5 & 82 $\pm$ 1.5 & 5.6 $\pm$ 0.5 & -0.07 $\pm$ 0.03 & B & s\\ 
A496J043346.71-131756.2 &-19.06 & S0 & 8365 $\pm$ 1.3 & 72 $\pm$ 1.3 & 14.1 $\pm$ 2.2 & -0.43 $\pm$ 0.02 & S & s\\ 
A496J043331.48-131654.6 &-18.76 & SO/SA & 9586 $\pm$ 1.0 & 50 $\pm$ 1.1 & 6.8 $\pm$ 0.8 & -0.26 $\pm$ 0.03 & S/D & s\\ 
A496J043333.17-131712.6 &-18.89 & cE/E & 9870 $\pm$ 2.0 & 123 $\pm$ 2.3 & 15.5 $\pm$ 1.9 & -0.07 $\pm$ 0.03 & \multicolumn{2}{c}{-}\\ 
A496J043342.10-131653.7 &-18.24 & S0/dS0 & 9517 $\pm$ 2.1 & 77 $\pm$ 2.1 & 10.0 $\pm$ 1.5 & -0.33 $\pm$ 0.05 & B/R? & s\\ 
A496J043341.69-131551.8 &-19.11 & cE/E & 9770 $\pm$ 1.8 & 179 $\pm$ 2.0 & 16.3 $\pm$ 1.7 & -0.08 $\pm$ 0.02 & \multicolumn{2}{c}{-}\\ 
A496J043352.77-131523.8 &-18.35 & S0/dS0 & 8926 $\pm$ 1.9 & 52 $\pm$ 2.0 & 9.6 $\pm$ 2.3 & -0.42 $\pm$ 0.08 & D? & w\\ 
A496J043332.07-131518.1 &-18.38 & dE/E & 9954 $\pm$ 1.4 & 79 $\pm$ 1.4 & 13.1 $\pm$ 1.8 & -0.43 $\pm$ 0.03 & \multicolumn{2}{c}{-}\\ 
A496J043337.35-131520.2 & & cE & 9753 $\pm$ 1.3 & 104 $\pm$ 1.5 & 16.4 $\pm$ 1.9 & -0.04 $\pm$ 0.02 & \multicolumn{2}{c}{-}\\ 
A496J043338.22-131500.7 & & cE/E & 10292 $\pm$ 2.9 & 145 $\pm$ 3.2 & 15.3 $\pm$ 2.7 & -0.19 $\pm$ 0.03 & \multicolumn{2}{c}{-}\\ 
A496J043339.72-131424.6 &-16.80 & dE & 10138 $\pm$ 4.3 & 24 $\pm$ 5.6 & 15.7 $\pm$ 15.5 & -0.92 $\pm$ 0.15 & \multicolumn{2}{c}{-}\\ 
A496J043401.57-131359.7 &-18.97 & S0 & 10281 $\pm$ 1.6 & 148 $\pm$ 1.7 & 13.3 $\pm$ 1.1 & -0.25 $\pm$ 0.02 & \multicolumn{2}{c}{-}\\ 
A496J043403.19-131310.6 &-19.39 & SB0/SBa & 8952 $\pm$ 1.8 & 85 $\pm$ 1.9 & 13.8 $\pm$ 2.3 & -0.27 $\pm$ 0.03 & S & s\\ 
A496J043339.07-131319.7 &-16.26 & dE & 10815 $\pm$ 2.8 & 23 $\pm$ 4.0 & 4.6 $\pm$ 3.0 & -0.08 $\pm$ 0.18 & \multicolumn{2}{c}{-}\\ 
A496J043413.08-131231.6 &-17.22 & dE & 10199 $\pm$ 6.0 & 42 $\pm$ 7.0 & 14.3 $\pm$ 13.5 & -0.83 $\pm$ 0.14 & \multicolumn{2}{c}{-}\\ 
A496J043408.50-131152.7 &-17.31 & dE/dS0 & 9751 $\pm$ 17 & 47 $\pm$ 18 & 8.3 $\pm$ 19.2 & -0.50 $\pm$ 0.71 & B/D & w\\ 
A496J043334.54-131137.1 &-16.59 & dE & 8442 $\pm$ 7.0 & 25 $\pm$ 9.3 & 2.6 $\pm$ 1.9 & 0.03 $\pm$ 0.54 & \multicolumn{2}{c}{-}\\ 
A496J043351.54-131135.5 &-17.06 & dS0/dE & 9459 $\pm$ 2.1 & 21 $\pm$ 3.3 & 2.4 $\pm$ 0.9 & -0.62 $\pm$ 0.21 & \multicolumn{2}{c}{-}\\ 
A496J043411.72-131130.2 &-15.68 & dE & 10497 $\pm$ 15 & 26 $\pm$ 21 & 1.0 $\pm$ 1.3 & 0.20 $\pm$ 0.96 & \multicolumn{2}{c}{-}\\ 
A496J043413.00-131003.5 &-18.86 & S0 & 10835 $\pm$ 1.3 & 50 $\pm$ 1.4 & 8.9 $\pm$ 1.3 & -0.21 $\pm$ 0.05 & \multicolumn{2}{c}{-}\\ 
A496J043355.55-131024.9 &-16.75 & dS0/dE & 8430 $\pm$ 3.6 & 20 $\pm$ 4.9 & 14.4 $\pm$ 18.1 & -0.35 $\pm$ 0.17 & \multicolumn{2}{c}{-}\\ 
A496J043342.83-130846.8 &-18.54 & S0/E & 10533 $\pm$ 1.6 & 54 $\pm$ 1.6 & 7.5 $\pm$ 1.4 & -0.56 $\pm$ 0.06 & \multicolumn{2}{c}{-}\\ 
A496J043329.79-130851.7 &-17.12 & dE & 8640 $\pm$ 3.1 & 30 $\pm$ 4.0 & 3.7 $\pm$ 1.8 & -0.59 $\pm$ 0.10 & \multicolumn{2}{c}{-}\\ 
A496J043410.60-130756.7 &-17.39 & dS0 & 8373 $\pm$ 5.3 & 36 $\pm$ 6.2 & 3.7 $\pm$ 2.7 & -0.30 $\pm$ 0.13 & \multicolumn{2}{c}{-}\\ 
A496J043359.03-130626.7 &-18.13 & dS0 & 10552 $\pm$ 2.3 & 25 $\pm$ 3.1 & 4.4 $\pm$ 2.0 & -0.46 $\pm$ 0.12 & \multicolumn{2}{c}{-}\\ 
A496J043348.59-130558.3 &-17.72 & dE & 9777 $\pm$ 1.4 & 46 $\pm$ 1.5 & 11.1 $\pm$ 2.1 & -0.37 $\pm$ 0.06 & \multicolumn{2}{c}{-}\\ 
A496J043349.08-130520.5 &-18.56 & S0/Sa & 9770 $\pm$ 1.8 & 56 $\pm$ 1.8 & 9.5 $\pm$ 2.0 & -0.43 $\pm$ 0.07 & S & w\\ 
A496J043343.04-130514.1 &-18.03 & dS0 & 9681 $\pm$ 1.8 & 56 $\pm$ 1.9 & 12.7 $\pm$ 3.3 & -0.45 $\pm$ 0.05 & \multicolumn{2}{c}{-}\\ 
A496J043345.67-130542.2 &-17.91 & dS0 & 9689 $\pm$ 1.3 & 46 $\pm$ 1.4 & 5.4 $\pm$ 0.8 & -0.25 $\pm$ 0.05 & \multicolumn{2}{c}{-}\\ 
A496J043350.17-125945.4 &-16.65 & dS0 & 10381 $\pm$ 8.3 & 38 $\pm$ 10 & 2.9 $\pm$ 3.4 & -0.65 $\pm$ 0.45 & \multicolumn{2}{c}{-}\\ 
A496J043356.18-125913.1 &-18.58 & dE & 11203 $\pm$ 1.0 & 42 $\pm$ 1.3 & 0.84 $\pm$ 0.04 & -0.28 $\pm$ 0.04 & \multicolumn{2}{c}{-}\\ 
A496J043343.04-125924.4 &-16.92 & dS0/dE & 10221 $\pm$ 4.7 & 23 $\pm$ 6.7 & 4.9 $\pm$ 4.1 & -0.38 $\pm$ 0.33 & D & w\\ 
A496J043326.49-131717.8 &-17.73 & dS0 & 8942 $\pm$ 4.1 & 37 $\pm$ 4.9 & 4.8 $\pm$ 3.2 & -0.24 $\pm$ 0.19 & D & w\\ 
A496J043318.95-131726.9 &-17.47 & dE & 8625 $\pm$ 3.5 & 49 $\pm$ 3.7 & 5.3 $\pm$ 1.8 & -0.32 $\pm$ 0.14 & \multicolumn{2}{c}{-}\\ 
A496J043325.15-131715.9 &-15.92 & dE & 9598 $\pm$ 5.3 & 8 $\pm$ 15 & 17.1 $\pm$ 44.0 & -0.36 $\pm$ 0.27 & \multicolumn{2}{c}{-}\\ 
A496J043317.75-131536.6 &-17.18 & dS0/dE & 11064 $\pm$ 5.5 & 20 $\pm$ 8.5 & 2.4 $\pm$ 1.6 & -0.32 $\pm$ 0.42 & B? & w\\ 
A496J043325.40-131414.6 &-17.27 & dE & 8374 $\pm$ 2.6 & 38 $\pm$ 2.9 & 8.5 $\pm$ 3.6 & -0.32 $\pm$ 0.14 & \multicolumn{2}{c}{-}\\ 
A496J043324.91-131342.6 &-17.82 & dE/E & 9199 $\pm$ 2.2 & 41 $\pm$ 2.4 & 10.6 $\pm$ 3.6 & -0.48 $\pm$ 0.09 & \multicolumn{2}{c}{-}\\ 
A496J043306.97-131238.8 &-18.33 & SB0/SBa & 8915 $\pm$ 2.4 & 47 $\pm$ 2.7 & 4.8 $\pm$ 1.6 & -0.12 $\pm$ 0.09 & B & s\\ 
A496J043324.61-131111.9 &-16.28 & dE & 9594 $\pm$ 3.8 & 18 $\pm$ 6.2 & 4.7 $\pm$ 4.6 & -0.23 $\pm$ 0.26 & \multicolumn{2}{c}{-}\\ 
A496J043325.10-130906.6 &-16.61 & dE & 10404 $\pm$ 9.0 & 32 $\pm$ 11 & 2.0 $\pm$ 1.7 & -0.25 $\pm$ 0.45 & \multicolumn{2}{c}{-}\\ 
A496J043312.08-130449.3 &-17.10 & dE & 9197 $\pm$ 6.0 & 30 $\pm$ 7.4 & 10.0 $\pm$ 13.5 & -0.65 $\pm$ 0.33 & \multicolumn{2}{c}{-}\\ 
A496J043321.37-130416.6 &-17.61 & dS0/dSa & 7735 $\pm$ 1.8 & 26 $\pm$ 2.5 & 1.2 $\pm$ 0.1 & -0.55 $\pm$ 0.15 & B/S/R? & s\\ 
A496J043325.54-130408.0 &-16.51 & dE & 11094 $\pm$ 8.9 & 26 $\pm$ 12 & 1.7 $\pm$ 2.1 & -0.48 $\pm$ 0.46 & \multicolumn{2}{c}{-}\\ 
A496J043320.35-130314.9 &-19.40 & SB0 & 8764 $\pm$ 1.0 & 73 $\pm$ 1.0 & 7.3 $\pm$ 0.6 & -0.11 $\pm$ 0.02 & B/R/S? & s\\ 
A496J043308.85-130235.6 &-19.18 & Sc & 10862 $\pm$ 1.7 & 45 $\pm$ 1.9 & 4.6 $\pm$ 1.0 & -0.24 $\pm$ 0.06 & D & s\\ 
\hline
\multicolumn{10}{l}{$^1$\footnotesize{type of embedded structure as follows B: bar, D: disc, R: ring, S: spiral; and strength: strong (s) or weak (w)~~~~~~~~~ ~~~~~~~~~ ~~~~~}}\\
\end{tabular}
\caption{Absolute magnitudes, radial velocities, velocity dispersions, SSP-equivalent ages
and metallicities of the galaxies in Abell~496.\label{tabresfit}}
\end{small}
\end{table}

The principal limitation of our method which might make impossible to determine
correctly parameters for objects exhibiting non-solar [Mg/Fe] abundance
ratios is due to contents of the ELODIE.3 library including only stars in
the nearest solar neighbourhood known to have [Mg/Fe] correlated with their
metallicities [Fe/H] (see Chen et al. 2003 and references there). Thus
fitting spectra having non-solar [Mg/Fe] results in template mismatch, which
can bias our estimations of stellar population parameters. In Chapter~1 we
have shown that [Mg/Fe] ratios do not bias age estimations.

To obtain [Mg/Fe] abundance ratios for Abell~496 galaxies we used
stellar population models dealing with Lick indices of magnesium and iron
(Thomas et al. 2003). For computing Lick indices we degraded spectral
resolution to match one needed to compute Lick indices (Worthey et al. 1994,
Thomas et al. 2003) by convolving original spectra with a Gaussian having
width equal to the square root of difference between squares of Lick
resolution, LSF ($\sigma_{inst}$) and velocity dispersion values found by
spectral fitting. Spectral range of Giraffe in the setup we used
($5010$\AA$< \lambda < 5800$\AA) and mean redshift $z=0.033$ allow to
compute the following Lick indices: Fe$_{5015}$, Mg$b$, Fe$_{5270}$,
Fe$_{5335}$, and Fe$_{5406}$. Uncertainties of the indices are computed
according to Cardiel et al. (1998). Measurements of indices and derived
values of the [Mg/Fe] abundance ratio are presented in Tab~\ref{tablick}.

\begin{table}
\begin{small}
\begin{tabular}{lcccccc}
IAU Name & Fe$_{5015}$ & Mg$b$ & Fe$_{5270}$ & Fe$_{5335}$ & Fe$_{5406}$ & [Mg/Fe] \\
\hline
A496J043333.53-131852.6 & $4.88 \pm 0.19$ & $3.53 \pm 0.09$ & $2.77 \pm 0.10$ & $2.44 \pm 0.11$ & $1.75 \pm 0.09$ & $ 0.13 \pm  0.10$ \\
A496J043346.71-131756.2 & $3.58 \pm 0.17$ & $3.50 \pm 0.08$ & $2.61 \pm 0.08$ & $2.13 \pm 0.10$ & $1.31 \pm 0.07$ & $ 0.21 \pm  0.08$ \\
A496J043331.48-131654.6 & $4.23 \pm 0.17$ & $3.24 \pm 0.08$ & $2.93 \pm 0.09$ & $2.12 \pm 0.10$ & $1.33 \pm 0.08$ & $ 0.05 \pm  0.09$ \\
A496J043333.17-131712.6 & $4.87 \pm 0.23$ & $4.84 \pm 0.10$ & $3.33 \pm 0.11$ & $2.64 \pm 0.13$ & $1.77 \pm 0.10$ & $ 0.22 \pm  0.09$ \\
A496J043342.10-131653.7 & $4.57 \pm 0.26$ & $3.53 \pm 0.12$ & $2.80 \pm 0.13$ & $2.19 \pm 0.15$ & $1.42 \pm 0.12$ & $ 0.15 \pm  0.12$ \\
A496J043341.69-131551.8 & $4.59 \pm 0.15$ & $5.02 \pm 0.07$ & $2.96 \pm 0.08$ & $2.55 \pm 0.09$ & $1.73 \pm 0.07$ & $ 0.35 \pm  0.07$ \\
A496J043352.77-131523.8 & $3.38 \pm 0.31$ & $3.15 \pm 0.14$ & $2.85 \pm 0.16$ & $2.49 \pm 0.18$ & $1.62 \pm 0.14$ & $ 0.00 \pm  0.14$ \\
A496J043332.07-131518.1 & $4.37 \pm 0.17$ & $3.68 \pm 0.08$ & $2.53 \pm 0.09$ & $2.01 \pm 0.10$ & $1.41 \pm 0.07$ & $ 0.28 \pm  0.08$ \\
A496J043337.35-131520.2 & $5.58 \pm 0.17$ & $4.79 \pm 0.08$ & $3.26 \pm 0.08$ & $2.97 \pm 0.10$ & $1.95 \pm 0.07$ & $ 0.19 \pm  0.07$ \\
A496J043338.22-131500.7 & $4.14 \pm 0.27$ & $4.85 \pm 0.12$ & $2.65 \pm 0.14$ & $2.10 \pm 0.16$ & $1.75 \pm 0.12$ & $ 0.43 \pm  0.09$ \\
A496J043339.72-131424.6 & $4.94 \pm 0.80$ & $2.42 \pm 0.39$ & $1.34 \pm 0.44$ & $0.81 \pm 0.51$ & $0.68 \pm 0.38$ & $ 0.36 \pm  0.24$ \\
A496J043401.57-131359.7 & $3.92 \pm 0.14$ & $4.23 \pm 0.06$ & $2.84 \pm 0.07$ & $2.33 \pm 0.08$ & $1.60 \pm 0.06$ & $ 0.27 \pm  0.07$ \\
A496J043403.19-131310.6 & $3.36 \pm 0.22$ & $3.90 \pm 0.10$ & $2.82 \pm 0.11$ & $2.33 \pm 0.13$ & $1.60 \pm 0.10$ & $ 0.20 \pm  0.10$ \\
A496J043339.07-131319.7 & $4.25 \pm 0.81$ & $3.05 \pm 0.38$ & $3.36 \pm 0.41$ & $1.82 \pm 0.49$ & $1.21 \pm 0.37$ & $-0.11 \pm  0.32$ \\
A496J043413.08-131231.6 & $2.57 \pm 0.93$ & $3.06 \pm 0.43$ & $1.90 \pm 0.49$ & $1.90 \pm 0.56$ & $0.47 \pm 0.42$ & $ 0.41 \pm  0.42$ \\
A496J043408.50-131152.7 & $3.39 \pm 2.54$ & $3.40 \pm 1.19$ & $1.53 \pm 1.40$ & $1.63 \pm 1.61$ & $1.46 \pm 1.18$ & $ 0.45 \pm  0.72$ \\
A496J043334.54-131137.1 & $6.10 \pm 1.71$ & $1.71 \pm 0.86$ & $2.15 \pm 0.94$ & $1.80 \pm 1.08$ & $1.58 \pm 0.79$ & $-0.11 \pm  0.51$ \\
A496J043351.54-131135.5 & $1.55 \pm 0.40$ & $1.96 \pm 0.19$ & $2.80 \pm 0.21$ & $1.65 \pm 0.24$ & $0.93 \pm 0.18$ & $-0.30 \pm  0.23$ \\
A496J043411.72-131130.2 & $3.21 \pm 2.97$ & $1.34 \pm 1.48$ & $2.61 \pm 1.61$ & $0.35 \pm 1.95$ & $-0.12 \pm 1.45$& $-0.12 \pm  0.46$  \\
A496J043413.00-131003.5 & $4.81 \pm 0.24$ & $3.80 \pm 0.11$ & $2.68 \pm 0.12$ & $2.27 \pm 0.14$ & $1.55 \pm 0.11$ & $ 0.23 \pm  0.11$ \\
A496J043355.55-131024.9 & $3.00 \pm 1.02$ & $2.49 \pm 0.49$ & $2.84 \pm 0.52$ & $2.17 \pm 0.61$ & $1.98 \pm 0.44$ & $-0.14 \pm  0.45$ \\
A496J043342.83-130846.8 & $3.51 \pm 0.22$ & $2.89 \pm 0.10$ & $2.15 \pm 0.12$ & $1.99 \pm 0.13$ & $1.52 \pm 0.10$ & $ 0.25 \pm  0.13$ \\
A496J043329.79-130851.7 & $3.60 \pm 0.54$ & $2.40 \pm 0.26$ & $2.33 \pm 0.29$ & $0.96 \pm 0.34$ & $0.82 \pm 0.25$ & $ 0.16 \pm  0.30$ \\
A496J043410.60-130756.7 & $5.49 \pm 0.96$ & $2.17 \pm 0.47$ & $2.30 \pm 0.52$ & $2.26 \pm 0.60$ & $1.20 \pm 0.45$ & $-0.11 \pm  0.49$ \\
A496J043359.03-130626.7 & $3.94 \pm 0.48$ & $2.08 \pm 0.23$ & $1.68 \pm 0.26$ & $2.06 \pm 0.30$ & $1.78 \pm 0.22$ & $ 0.11 \pm  0.23$ \\
A496J043348.59-130558.3 & $3.77 \pm 0.26$ & $3.55 \pm 0.12$ & $2.70 \pm 0.13$ & $2.38 \pm 0.15$ & $1.71 \pm 0.11$ & $ 0.16 \pm  0.12$ \\
A496J043349.08-130520.5 & $3.82 \pm 0.27$ & $3.03 \pm 0.13$ & $2.51 \pm 0.14$ & $2.06 \pm 0.16$ & $1.53 \pm 0.12$ & $ 0.15 \pm  0.14$ \\
A496J043343.04-130514.1 & $4.64 \pm 0.28$ & $3.62 \pm 0.13$ & $2.42 \pm 0.15$ & $1.86 \pm 0.17$ & $1.28 \pm 0.13$ & $ 0.31 \pm  0.12$ \\
A496J043345.67-130542.2 & $4.87 \pm 0.23$ & $2.89 \pm 0.11$ & $2.92 \pm 0.12$ & $2.28 \pm 0.14$ & $1.68 \pm 0.11$ & $-0.06 \pm  0.12$ \\
A496J043350.17-125945.4 & $2.59 \pm 1.23$ & $1.77 \pm 0.61$ & $1.14 \pm 0.68$ & $2.44 \pm 0.76$ & $1.84 \pm 0.56$ & $ 0.13 \pm  0.34$ \\
A496J043356.18-125913.1 & $1.15 \pm 0.10$ & $1.54 \pm 0.05$ & $1.42 \pm 0.06$ & $1.29 \pm 0.07$ & $0.77 \pm 0.05$ & $ 0.11 \pm  0.08$ \\
A496J043343.04-125924.4 & $2.64 \pm 1.10$ & $2.90 \pm 0.52$ & $2.23 \pm 0.58$ & $2.36 \pm 0.67$ & $3.15 \pm 0.46$ & $ 0.17 \pm  0.42$ \\
A496J043326.49-131717.8 & $4.38 \pm 0.82$ & $2.55 \pm 0.39$ & $3.54 \pm 0.42$ & $1.91 \pm 0.49$ & $1.70 \pm 0.36$ & $-0.24 \pm  0.27$ \\
A496J043318.95-131726.9 & $3.97 \pm 0.56$ & $2.78 \pm 0.26$ & $3.02 \pm 0.29$ & $2.18 \pm 0.33$ & $1.42 \pm 0.25$ & $-0.11 \pm  0.24$ \\
A496J043325.15-131715.9 & $2.08 \pm 1.79$ & $4.80 \pm 0.76$ & $2.81 \pm 0.89$ & $2.29 \pm 1.03$ & $1.88 \pm 0.76$ & $ 0.37 \pm  0.51$ \\
A496J043317.75-131536.6 & $4.68 \pm 1.21$ & $2.94 \pm 0.59$ & $1.62 \pm 0.68$ & $1.22 \pm 0.78$ & $0.55 \pm 0.60$ & $ 0.48 \pm  0.49$ \\
A496J043325.40-131414.6 & $4.75 \pm 0.51$ & $2.63 \pm 0.25$ & $2.11 \pm 0.28$ & $2.01 \pm 0.32$ & $1.69 \pm 0.23$ & $ 0.19 \pm  0.28$ \\
A496J043324.91-131342.6 & $2.86 \pm 0.41$ & $3.20 \pm 0.19$ & $3.00 \pm 0.21$ & $1.69 \pm 0.24$ & $1.41 \pm 0.18$ & $ 0.07 \pm  0.18$ \\
A496J043306.97-131238.8 & $3.91 \pm 0.45$ & $3.40 \pm 0.21$ & $2.59 \pm 0.23$ & $2.05 \pm 0.27$ & $1.77 \pm 0.20$ & $ 0.20 \pm  0.19$ \\
A496J043324.61-131111.9 & $4.52 \pm 1.04$ & $3.07 \pm 0.49$ & $2.45 \pm 0.55$ & $2.04 \pm 0.63$ & $0.76 \pm 0.48$ & $ 0.18 \pm  0.40$ \\
A496J043325.10-130906.6 & $2.22 \pm 1.62$ & $2.28 \pm 0.77$ & $1.68 \pm 0.88$ & $1.66 \pm 1.02$ & $0.83 \pm 0.77$ & $ 0.22 \pm  0.47$ \\
A496J043312.08-130449.3 & $3.07 \pm 1.15$ & $2.70 \pm 0.55$ & $2.17 \pm 0.62$ & $1.27 \pm 0.73$ & $0.64 \pm 0.55$ & $ 0.30 \pm  0.59$ \\
A496J043321.37-130416.6 & $2.66 \pm 0.24$ & $1.39 \pm 0.12$ & $1.74 \pm 0.13$ & $1.13 \pm 0.15$ & $1.21 \pm 0.11$ & $-0.03 \pm  0.13$ \\
A496J043325.54-130408.0 & $2.02 \pm 1.52$ & $2.28 \pm 0.73$ & $0.94 \pm 0.85$ & $2.61 \pm 0.94$ & $0.59 \pm 0.75$ & $ 0.30 \pm  0.51$ \\
A496J043320.35-130314.9 & $5.48 \pm 0.13$ & $3.67 \pm 0.07$ & $2.94 \pm 0.07$ & $2.50 \pm 0.08$ & $1.39 \pm 0.06$ & $ 0.10 \pm  0.08$ \\
A496J043308.85-130235.6 & $4.63 \pm 0.29$ & $3.07 \pm 0.14$ & $2.46 \pm 0.15$ & $2.46 \pm 0.18$ & $1.71 \pm 0.13$ & $ 0.12 \pm  0.14$ \\
\hline
\end{tabular}
\caption{Selected Lick indices and values of derived values of [Mg/Fe]
according to Thomas et al. (2003). All values of indices are in \AA, [Mg/Fe]
in dex. \label{tablick}}
\end{small}
\end{table}

For majority of the low-mass galaxies in our sample, exhibiting
$-0.15 < \mbox{[Mg/Fe]} < 0.15$ we believe to have unbiased estimates for
overall metallicities [Fe/H] reported by the spectral fitting procedure.

\begin{figure}
\includegraphics{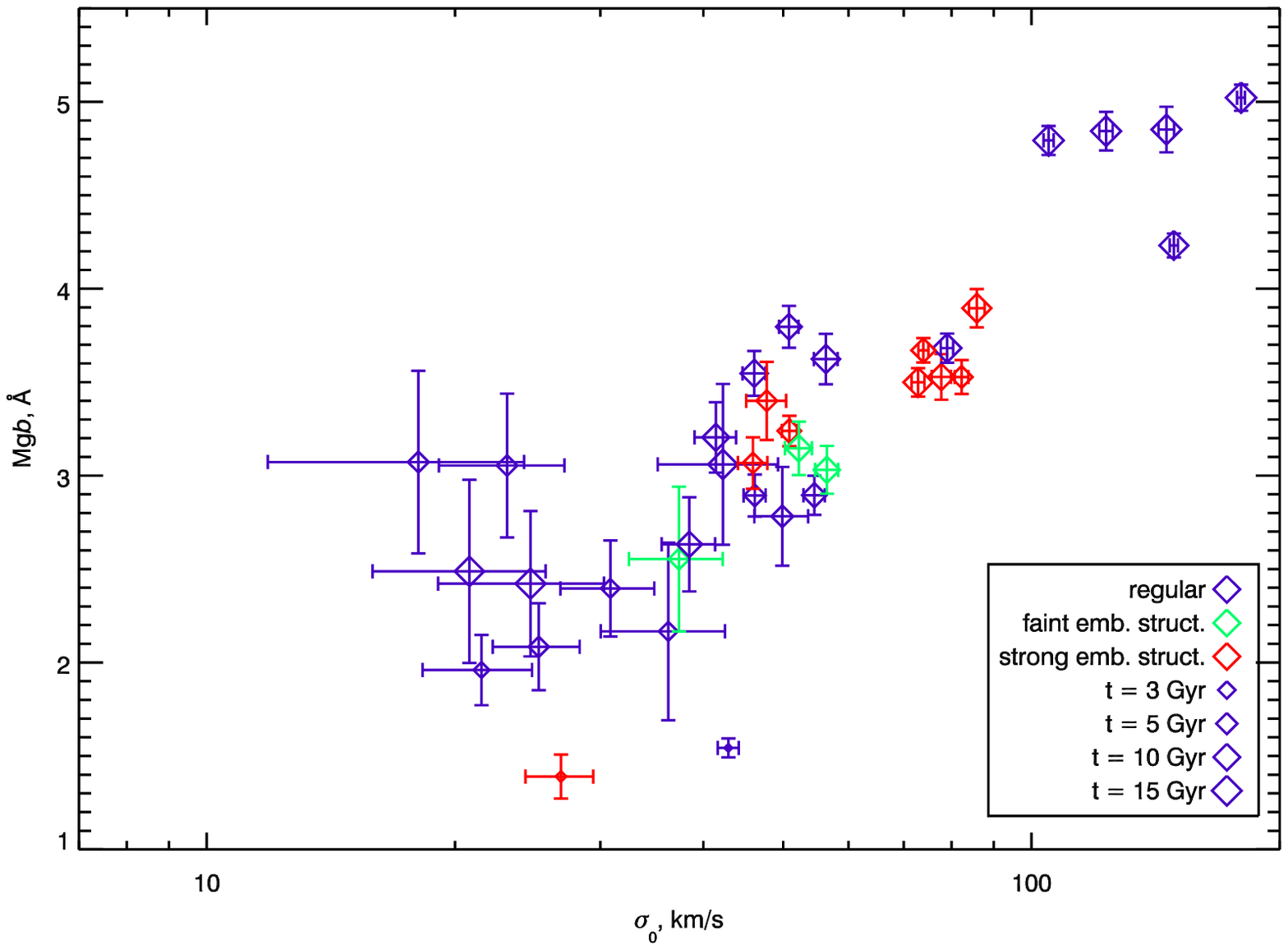}
\caption{Mg$b$ - $\sigma_0$ relation. Only points having
$\Delta(\mbox{Mg}b)<0.5$~\AA\ are shown. \label{figmgsig}}
\end{figure}

\begin{figure}
\includegraphics{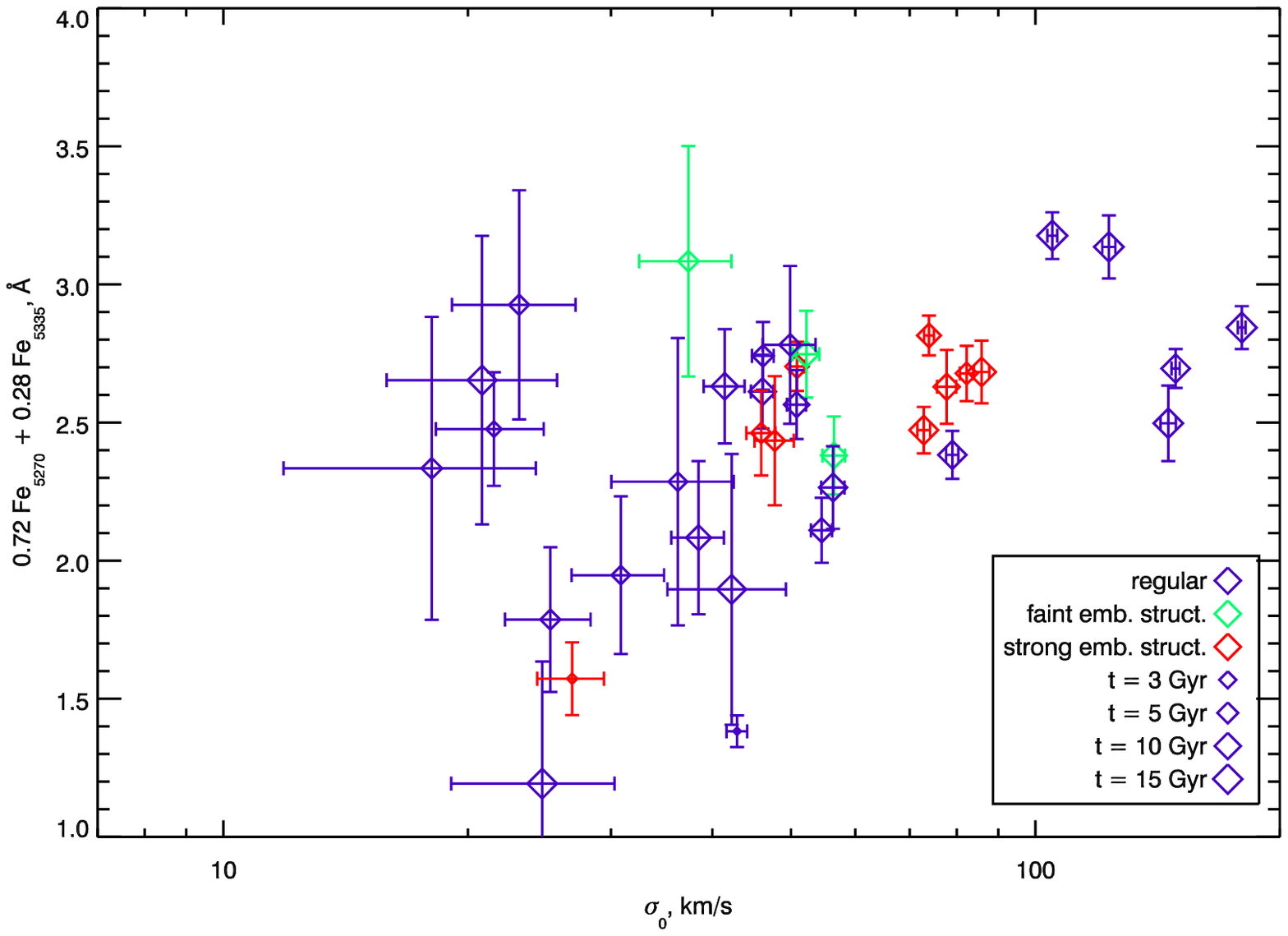}
\caption{$<$Fe$>$ - $\sigma_0$ relation. Only points having
$\Delta(\mbox{Mg}b)<0.5$~\AA\ are shown. \label{figfesig}}
\end{figure}

\begin{figure}
\includegraphics{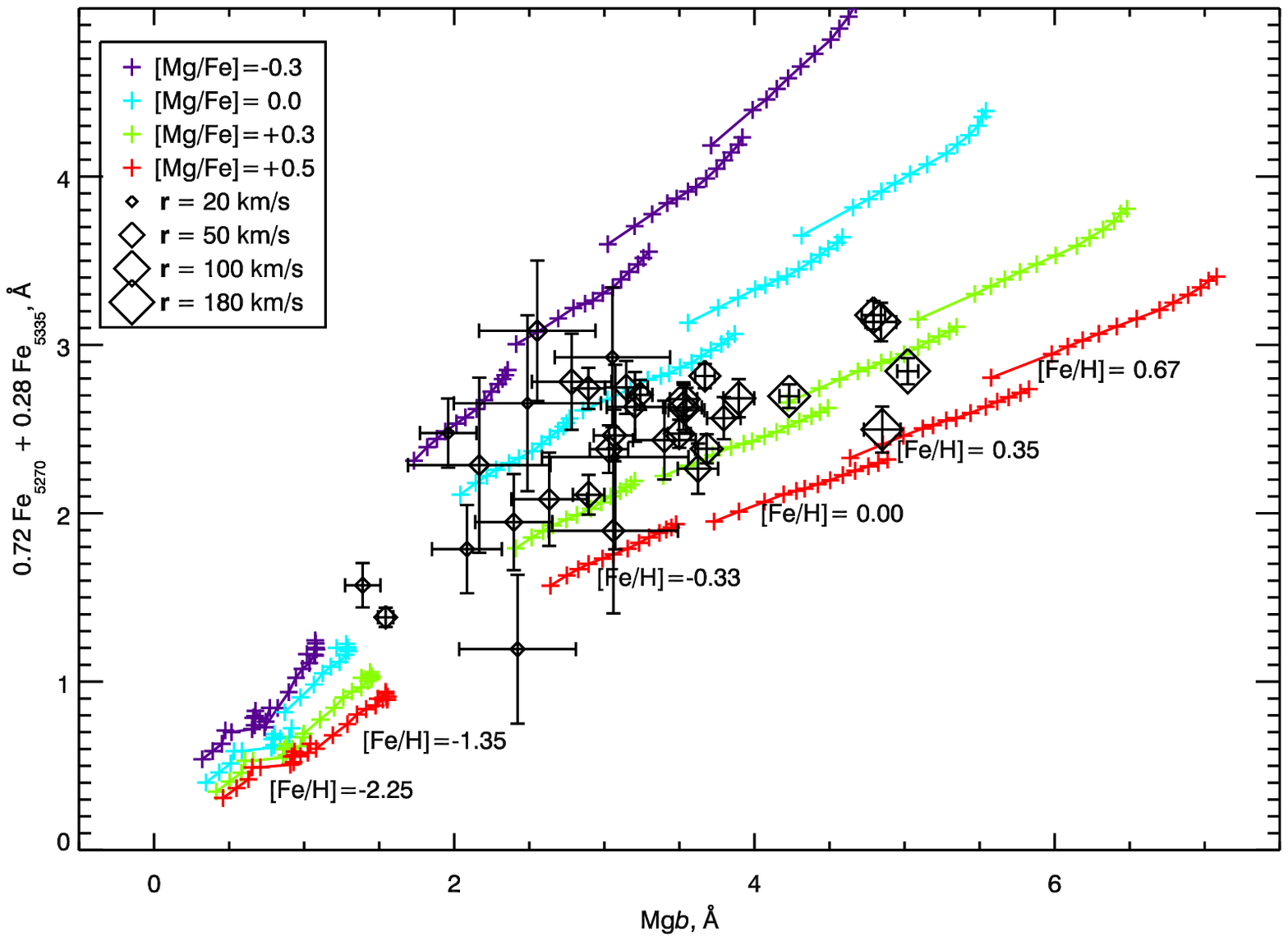}
\caption{Mg$b$ vs $<$Fe$>$ with indication of central velocity dispersions as
sizes of symbols. Models from Thomas et al. (2003) for different values of
[$\alpha$/Fe] enrichment are overplotted. Only points having
$\Delta(\mbox{Mg}b)<0.5$~\AA\ are shown. \label{figmgfe}}
\end{figure}

In Figures \ref{figmgsig},\ref{figfesig}, and \ref{figmgfe} measurements of
Mg$b$, $<\mbox{Fe}> = 0.72 \mbox{Fe}_{5270} + 0.28 \mbox{Fe}_{5335}$, and
central velocity dispersions are shown. One can notice quite tight
correlation between Mg$b$ and $\log \sigma_0$ for $\sigma_0 >
25$~km~s$^{-1}$. Massive galaxies tend toward higher [Mg/Fe] values.

\begin{figure} 
\includegraphics{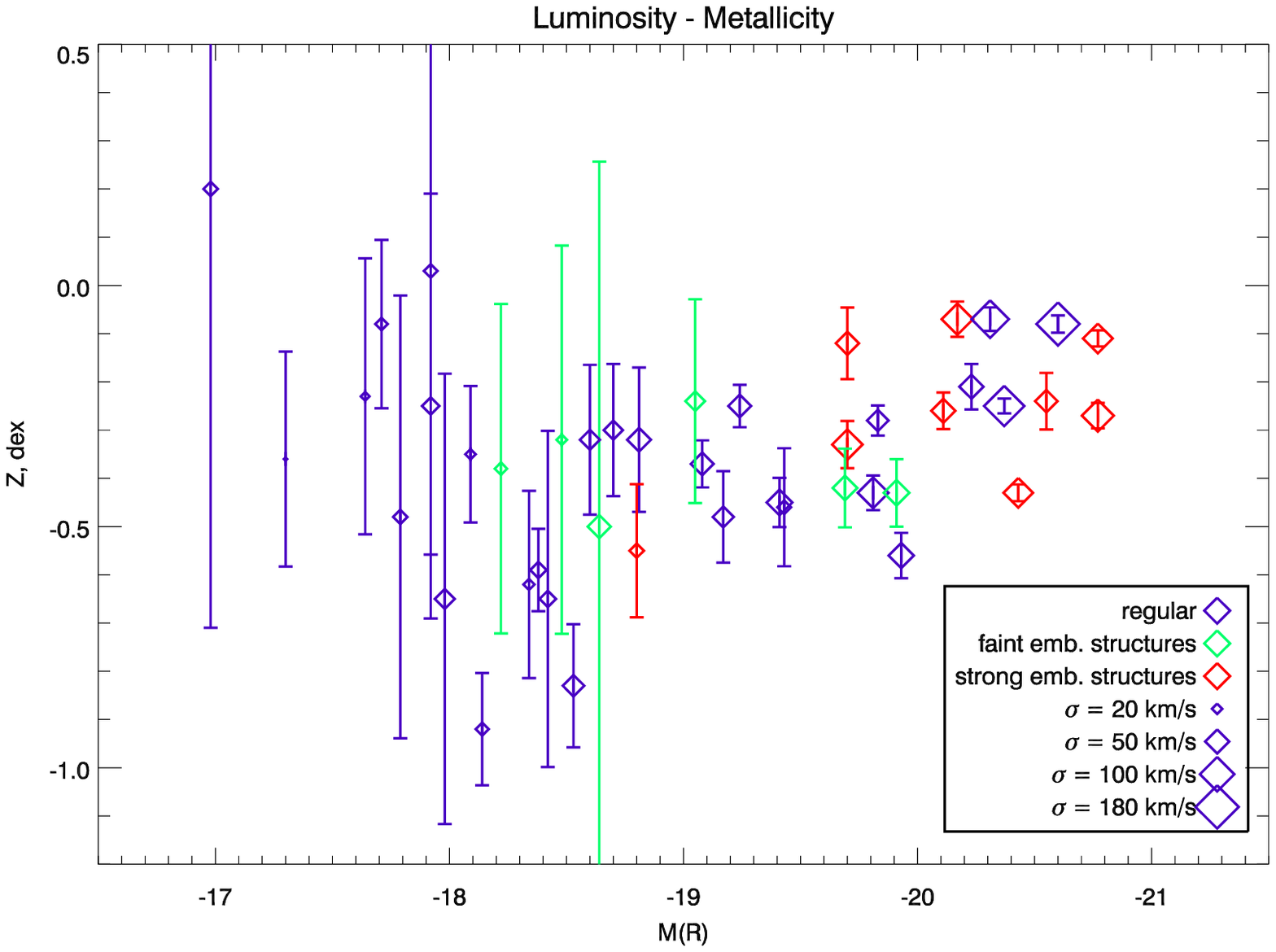}
\caption{R-band luminosity-metallicity relation. Presence of embedded
structures is indicated. \label{figZL}}
\end{figure}

\begin{figure} 
\includegraphics{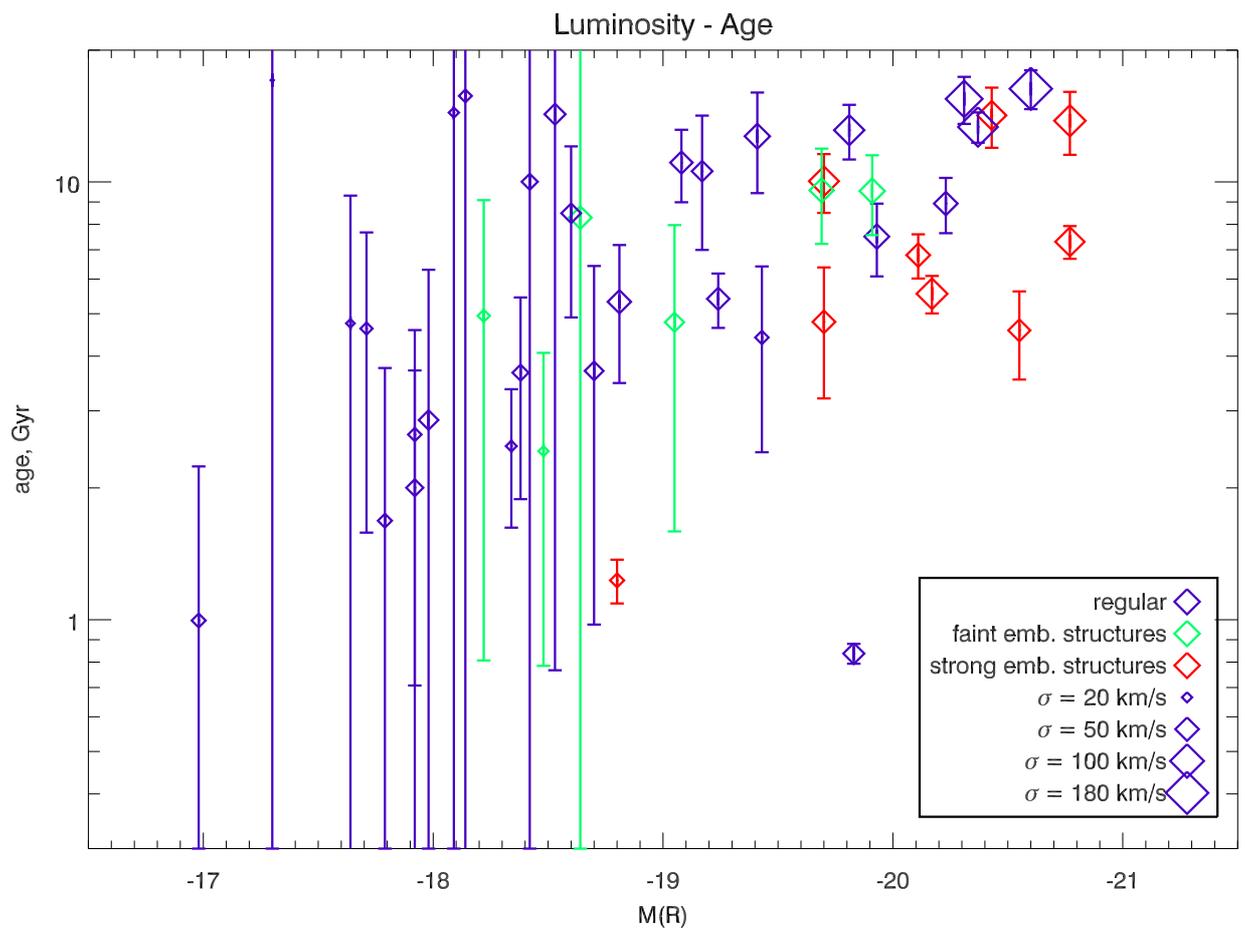}
\caption{R-band luminosity-age relation. Presence of embedded
structures is indicated. \label{figtL}}
\end{figure}

Figures \ref{figZL} and \ref{figtL} demonstrate luminosity -- metallicity
and luminosity -- age relations. It is clearly seen that low-luminosity
galaxies tend toward low metallicities and younger ages, though spread of
age estimations is quite high due to low signal-to-noise values in the
spectra of faint objects.

\section{Scaling Relations}

Combining photometrical and kinematical data provides a possibility to study
fundamental properties of our galaxies and compare them to the literature.

\begin{figure}
\includegraphics{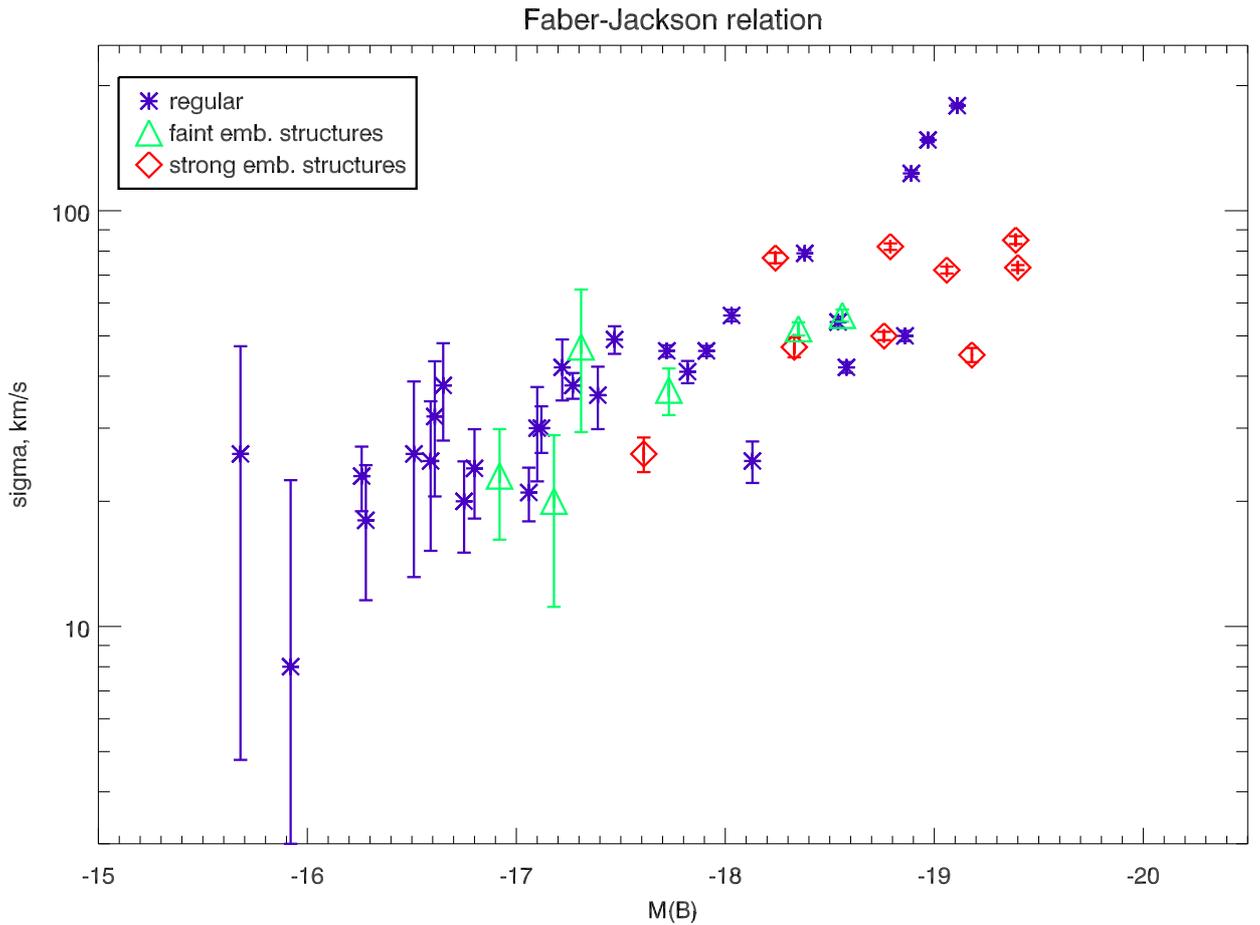}
\caption{Faber-Jackson relation. \label{figFJR}}
\end{figure}

Faber-Jackson relation (Faber \& Jackson 1976), reflecting connection of
dynamical and stellar masses is shown in Fig.~\ref{figFJR}.

We have built the Fundamental Plane (Djorgovski \& Davis 1987) combining
kinematical data with surface photometry. Original
equation by Djorgovski \& Davis (1987) rewritten using $R_e$, $I_e$, and 
$\sigma_0$ (e.g. Guzman et al. 1993) can be transformed into:
\begin{equation}
\log R_e = -8.666 + 0.314 \mu_{B} + 1.14 \log \sigma_0,
\label{eqfp}
\end{equation}
where $\mu_{eff}$ is effective surface brightness in B band in
mag~arcsec$^{-2}$, $R_e$ is effective radius in kpc, and $\sigma_0$ is
central velocity dispersion in km~s$^{-1}$.

\begin{figure}
\includegraphics[width=14cm]{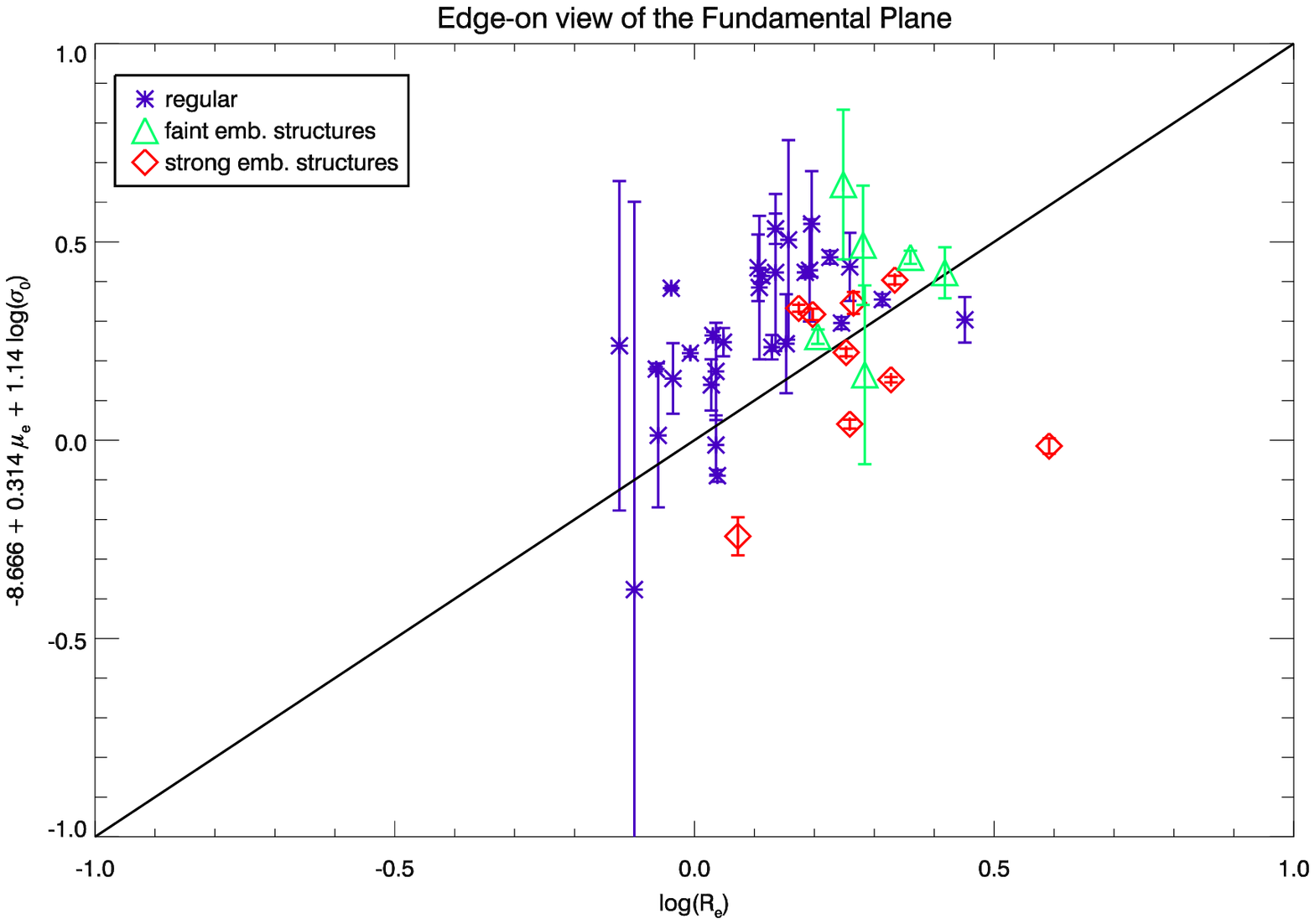}\\
\includegraphics[width=14cm]{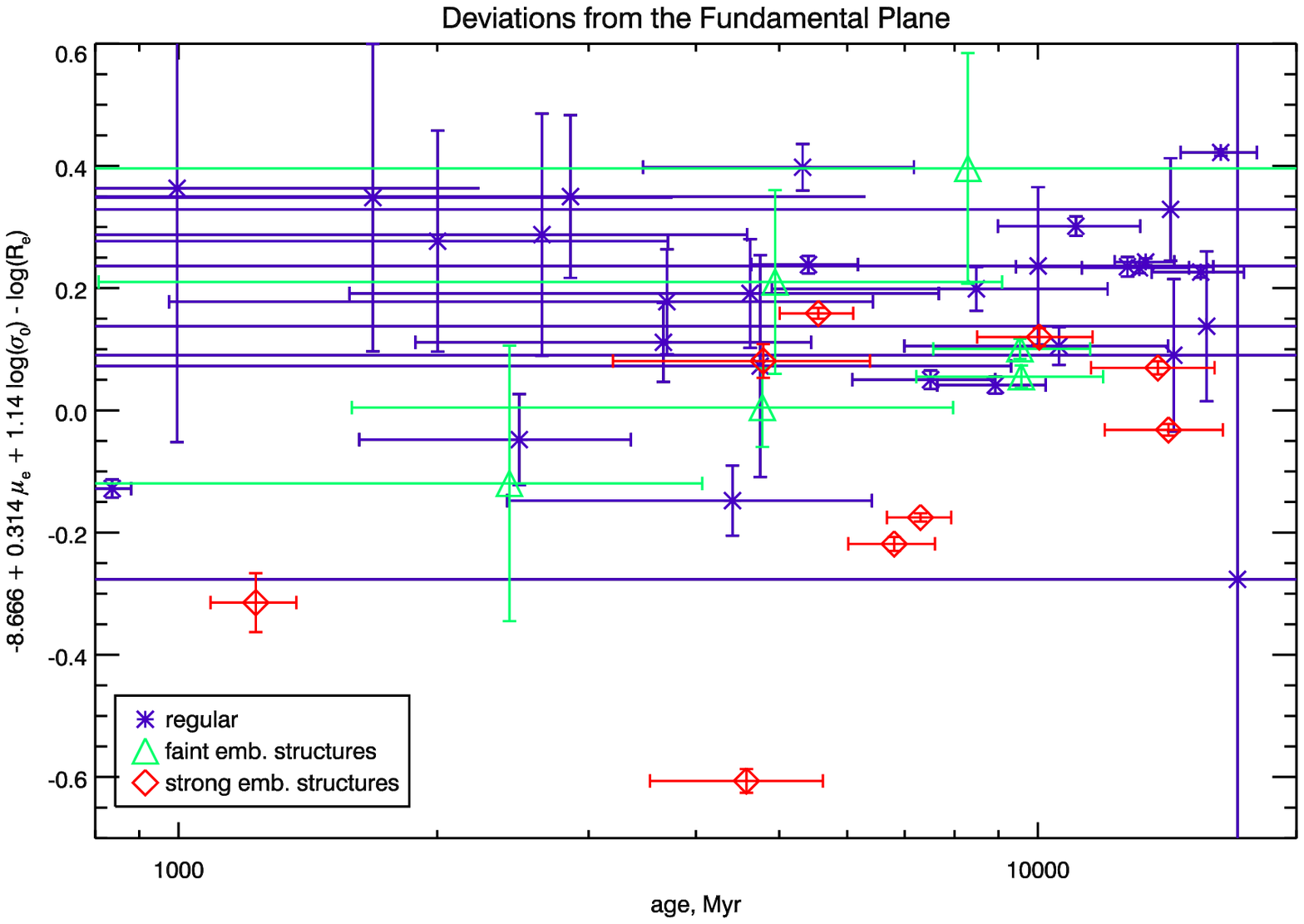}
\caption{Edge-on view of the Fundamental Plane (top) and residuals from FP
versus SSP-equivalent ages (bottom). Presence of embedded structures is indicated.
\label{figfpedge}}
\end{figure}

Edge-on view of FP is shown in Fig.~\ref{figfpedge}. Residuals from the
fundamental plane versus SSP-equivalent age are shown in
Fig.~\ref{figfpedge}(bottom). One could notice an anti-correlation of these
residuals and age, which can be explained by higher surface brightness of
young population, when other parameters remain fixed.

\section{Discussion and Conclusions}

Two possible scenarios of the gas removal usually considered for dE galaxies
are: (1) supernova-driven winds on the early stage of galaxy evolution; and
(2) ram pressure stripping while crossing central regions of the cluster. 
The idea of supernova-driven winds is based on the assumption that the
gravitational field of small galaxies is not sufficiently strong to keep the
interstellar medium from being swept out by SN~II explosions during first
intense star formation episode. 

This scenario leads to abrupt gas loss and interruption of the star
formation episode after short time ($10^8$ years). Consequently, later
explosions of SN~Ia will not contribute to the Fe-enrichment of the stellar
population. Thus, a short star formation episode will lead to overabundance
of $\alpha$-elements over iron ([$\alpha$/Fe] $>$ 0). This phenomenon is observed in
globular clusters and often in giant early-type galaxies (Sil'chenko 2006,
Kuntschner et al. 2006). If gas is removed from dE galaxies by
supernova-driven winds, we would expect to see [Mg/Fe] $>$ 0, and it should
increase when the dynamical mass of galaxies (or sigma) decreases.

However, in case of Abell 496, [Mg/Fe] = 0 for nearly all low-massive objects
($\sigma_0 <$ 60~km~s$^{-1}$). Hence we cannot consider the scenario of
supernova-driven gas removal as the only explanation for the observed
properties of dE galaxies. But we cannot avoid it completely, otherwise
there would have been no observed correlation between metallicity and
luminosity (Fig.\ref{figZL}). The only conclusion is that observed solar
[Mg/Fe] ratios in low-massive objects put the lower limit of at least 1--2
Gyr on the duration of the star formation epoch -- a minimal required time
to complete iron enrichment (Matteucci 1994). On the other hand,
ram-pressure stripping of late-type dwarf galaxies appears to be acceptable.
If we assume that late-type galaxies had formed outside the central region
of the cluster, and later fell down onto it, enough time is left for the
iron enrichment: the typical infall time is several Gyr.

If ram pressure stripping plays the leading role in gas removal, one would
expect large spread of luminosity-weighted ages for low-massive objects
which can be completely stripped during the first cross of the cluster
centre, because it can occur at any moment of galaxy lifetime. Due to low
statistics using our data we cannot give decisive answer whether spread of
age estimations in Fig.~\ref{figtL} is a result of ram-pressure stripping of
late-type progenitors, or just due to low quality of measurements -- deeper
observations are needed.

The presence of faint embedded discs in some galaxies is another strong
argument for an evolutionary connection between early and late type dwarf
galaxies. This result is in agreement with N-body modelling of morphological
evolution of late-type galaxies in clusters (Mastropietro et al. 2005),
suggesting that discs will not be completely destroyed. Our conclusion is
that dE galaxies have late-type progenitors, formed in the peripheral parts
of the cluster, and experienced tidal interactions with the cluster
potential and other cluster galaxies, and ram-pressure stripping while
crossing the cluster centre.

\section{Appendix: M~32 twin in Abell~496}
We discovered a galaxy belonging to a very rare class of compact elliptical
galaxies (cE). Presently, only five galaxies of this class are known: M32,
NGC4486B, NGC5846A, and two objects recently discovered in the Abell 1689
cluster (Mieske et al. 2005). 

Compact elliptical galaxies are very different from diffuse ones. On the
Kormendy diagram they are located on the extension of the sequence formed by
giant ellipticals and bulges of spirals toward small effective radii,
whereas dE galaxies form a distinct branch. We found the sixth object of
this class near the centre of the cluster (projected distance 16 kpc). Like
three of five known cE's it resides in the halo of a large galaxy: the
central CD galaxy of Abell~496 (PGC~15524). The central velocity dispersion,
105~km~s$^{-1}$, is quite high for its luminosity probably showing evidence
for the presence of a central black hole, as implied by the peak of velocity
dispersion in the centre of M~32 (Simien \& Prugniel, 2002). Its stellar
population parameters are: [Fe/H]$=-0.04 \pm 0.04$~dex; [Mg/Fe]$=0.15 \pm 0.07$~dex; 
$t = 16 \pm 3$~Gyr. It is significantly older than M~32 (t=7~Gy, Caldwell et
al. 2003, Rose et al. 2005), though its metallicity is nearly the same.

\begin{figure}
\includegraphics[width=15cm]{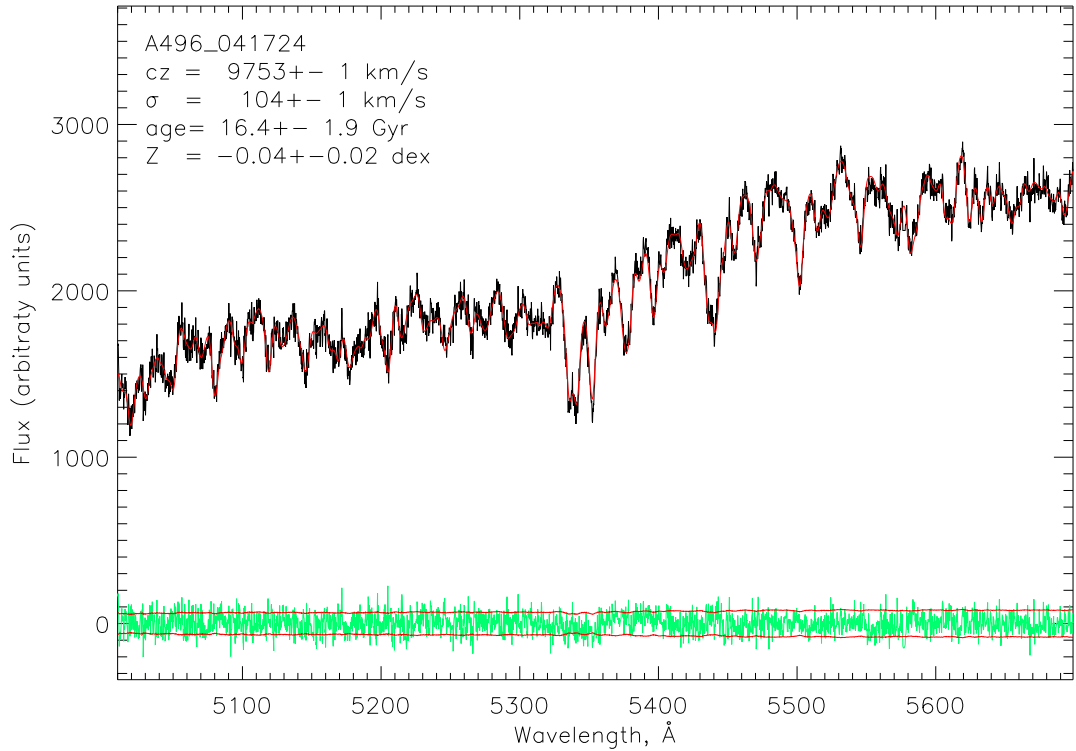}
\caption{Spectrum of the compact elliptical (cE) galaxy in
Abell~496.\label{A496_cE}}
\end{figure}

Spectrum and its best fitting template are shown in Fig.~\ref{A496_cE}.

\conclusion{Summary}
This thesis is devoted to observational studies of evolution of dwarf
elliptical galaxies. Despite these objects are the numerically-dominant type
of galaxies in the Universe, their origin and evolution still remains a
matter of a debate. Based on the similar morphology, most of the existing
studies conclude about evolutionary connection of dwarf elliptical galaxies
with gas-rich dwarf irregulars: secular evolution transforms dIrr's into
dE's. Observations demonstrate that dE galaxies do not exhibit ongoing star
formation and contain no ionised gas. Thus, the fundamental problem for
building the theory of dE evolution is a proper choice of a gas loss
scenario, resulting in an interruption of star formation. Though large
samples of spectroscopic observations of dE's are published, no serious
attempts have been made to establish connection between kinematics and
stellar populations of these objects. In this work we present such an
attempt, based on the high-quality observations obtained with large
telescopes.

Chapter~1 presents a new technique for fitting spectra, integrated along a
line of sight, by the PEGASE.HR evolutionary models. Thanks to its high
sensitivity, our methods opens new horizons in the studies of kinematics
and stellar populations using absorption-line spectra.

Analysis of observations of dE galaxies in clusters and groups is given in
chapters 2, 3, and 4. Data have been obtained using MPFS IFU spectrograph
(Russian 6-m telescope), FLAMES-Giraffe multiobject spectrograph (ESO VLT),
and wide-field Megacam imager (CFHT). Our work is the first example of
studies of a sample of dE galaxies observed with 3D-spectroscopy technique.

Analysis carried out in this work allows to conclude that evolution of dwarf
galaxies is driven by environmental effects. Only ram pressure stripping in
clusters, central parts of groups, containing hot gas, or halos of giant
galaxies does not contradict to any of the appearances of dE's. Another
scenario, which can not be completely excluded is a gravitational harassment
due to numerous encounters with other cluster members. Though this
possibility is predicted by the cosmological simulations, it can be realized
only if at most 1 percent of cluster members is observable directly, but
other 99+ percent are dark matter halos without any presence of visible
matter.

Possibility to form embedded discs (including counter-rotating) in dE
galaxies evolving in a quiescent environment of groups of galaxies is
demonstrated on a real example.

Despite 3D spectroscopy now became a widely-used observational technique, no
commonly accepted standards for storing, accessing and retrieving 3D data
exists. This question became even more important at the era of rapid
evolution of the Virtual Observatory concept. In the appendix we demonstrate
a possibility of building complete and self-sufficient description of 3D
data (data model), essential to provide access to such datasets, and
consequently, for creation of science-ready data archives.

\section*{Major results of the thesis}
\begin{enumerate}

\item New technique for simultaneous fitting of stellar population and
internal kinematics using synthetic template spectra

\item Two-dimensional fields of radial velocities, velocity dispersions, and
SSP-equivalent parameters of stellar populations (age and metallicity) of
dwarf elliptical and lenticular galaxies in the Virgo cluster: IC~783,
IC~3468, IC~3509 и IC~3653; and low-luminosity early-type galaxies in
groups: NGC~127 (NGC~128 group) и NGC~770 (NGC~772 group); two dimensional
velocity field of ionised gas in NGC~127. All the results obtained by
analysing 3D spectroscopic data originating from MPFS IFU spectrograph at
the Russian 6-m telescope

\item Catalogue of parameters, including radial velocities, central velocity
dispersions, measurements of Lick indices, SSP-equivalent ages,
metallicities, and [Mg/Fe] abundance ratios in 46 early type galaxies (28 of
those are dE and dS0) in the Abell~496 cluster. Results obtained by analysis
of the high-resolution (R=7000) multiobject spectroscopy using
FLAMES-Giraffe spectrograph at ESO VLT

\item Discovery of evolutionary-decoupled cores in dwarf elliptical galaxies
in the Virgo cluster. Taking into account this result and complex kinematics
of those galaxies, evident for a presence of embedded stellar discs, a
conclusion about the most probably scenario of gas loss in dE's, ram pressure
stripping, is made.

\item Techniques for universal description, storage, and data access
mechanisms for the 3D spectroscopic data in the Virtual Observatory.
\end{enumerate}

\section*{Acknowledgments}
Author is very grateful to the thesis advisors: Olga Sil'chenko and Philippe
Prugniel. I greatly appreciate the support of all my collaborators, in
particular of: Victor Afanasiev (SAO RAS, Russia), Francois Bonnarel (CDS,
Strasbourg, France), Veronique Cayatte (Observatoire de Paris-Meudon,
France), Sven De Rijcke (University of Ghent, Belgium), Marie-Lise Dubernet
(Observatoire de Paris-Meudon, France), Florence Durret (IAP, France), Mina
Koleva (CRAL Observatoire de Lyon, France / University of Sofia, Bulgaria),
Mireille Louys (CDS, Strasbourg, France), Pierre Le Sidaner (Observatoire de
Paris-Meudon, France), Jonathan McDowell (Harvard-Smithsonian CfA, USA),
Chantal Petit (CRAL Observatoire de Lyon, France), Francois Simien (CRAL
Observatoire de Lyon). We are very grateful to Alexei Moiseev for supporting
the observations of dE galaxies at the 6-m telescope. Visits in France were
supported through a bilateral CNRS grant and EGIDE (through Scienctific
department of the Embassy of France in Russia). This PhD is supported by the
INTAS Young Scientist Fellowship (04-83-3618). The dwarf galaxies
investigation is supported by the bilateral Flemish-Russian collaboration
(project RFBR-05-02-19805-MF\_a). We appreciate support provided by the
organizing committees of the following meetings: JENAM-2004, ADASS, ''Mapping
the Galaxy and Nearby Galaxies'', ''Science Perspectives for 3D Spectroscopy'',
and to the International Astronomical Union for providing support to attend
IAU Colloquium 198. Special thanks to the Large Telescopes Time Allocation
Committee or the Russian Academy of Sciences for providing observing time
with MPFS. We are grateful to the staff of the Terapix data center at IAP,
France, for their efficiency and competence in reducing our Megacam imaging
data.

Special thanks to a PhD student in the Moscow University, Ivan Zolotukhin
for valuable assistance in preparation of the Russian version of the
manuscript.

Author is grateful to his parents: Vladimir and Lyudmila Chilingarian for
their support during 3 years of PhD preparation.

\appendix
\chapter{3D data in the Virtual Observatory}
\section{Introduction to the 3D spectroscopy}
Integral field (or 3D) spectroscopy is a modern technique in 
astrophysical observing that was proposed by Georges Court\'es in the late
60's. The idea is to get a spectrum for every point in the field of
view of a spectrograph.

One of the approaches is to use a scanning Fabry-Perot interferometer. In
this case after reducing the data, one gets a set of narrow-band direct
images with slightly overlapping bands, or a so-called data cube -- a three
dimensional structure, containing spatial and spectral information -- a
short spectrum for every spatial pixel. Resulting data cubes have wide
spatial dimensions and relatively narrow spectral ones. A similar approach has
been used in radio astronomy for a couple of decades, and their datasets look
nearly the same.

Another approach is to slice a field of view using a micro-lens
array or special image slicer device (Integral Field Unit, or IFU)
and feed a ''classical''\ spectrograph (see review in
P\'econtal-Rousset et al., 2004 for a description of different image
slicing techniques).
The resulting datasets are normally smaller in spatial dimensions than
Fabry-Perot datasets, however the spectral dimension is usually one
to two orders longer.

The first implementations of 3D spectrographs came in the 1980s and
immediately demonstrated great benefits of this technique for
studying both extended and point sources. Presently, there are more
than a dozen 3D spectrographs being operated on nearly
every large telescope all over the world. A growing amount of 3D
data is being produced by these instruments, and the question of
dissemination of these data in the Virtual Observatory has become
an important challenge.

Creating the archive implies the following necessary metadata to be defined:
{\bf Data Description}, {\bf Data Storage Format}, and {\bf Query Interface
and Data Retrieval}. While there is a good and interoperable solution for
the data storage format, the Euro3D FITS Format developed within Euro3D research
training network (Kissler-Patig et al. 2004), the two other aspects rely on
the Virtual Observatory community.

\section{Characterisation Data Model of IVOA}

An abstract, self-sufficient and standardised description of the
astronomical data is known as a data model. Such a description is
supposed to be sufficient for any sort of data processing and
analysis. The Data Modeling working group of the International Virtual
Observatory Alliance (IVOA)
is responsible for defining data models for different types of
astronomical data sets, catalogues, and more general concepts e.g.
''quantity''. The most general description of any sort of
observational or theoretical data sets will be given by the
forthcoming Observational DM (McDowell et al. 2004 in prep.). Its
main subclasses are: Observation, DataCollection, Curation,
Provenance and Characterisation. The latter one gives a
physical insight to the dataset, while others provide more
instrument-specific or sociological information. Characterisation
was reorganised as a separate data model (McDowell et al. 2006 in prep.),
which is now being intensively developed.

Characterisation DM is a way to say where, how extended and in which way the
Observational or Simulated dataset can be described in a multidimensional
parameter space, having the following axes: {\bf spatial}, {\bf temporal},
{\bf spectral}, {\bf observed} (e.g. flux), {\bf polarimetric}, as well
as other arbitrary axes. For every
axis there are three characterisation properties: {\bf coverage}, {\bf
resolution}, and {\bf sampling}. Every axis also contains a specific {\bf
axisFrame} subclass used for error assessment and including some general
axis-specific metadata. Four levels of characterisation, reflecting
different levels of details in the description can be given for every axis:
\begin{enumerate}
\item {\bf location} or {\bf reference value}, giving average position of the data on a
given parameter axis
\item {\bf bounds}, providing a bounding box
\item {\bf support}, describing more precisely regions on a parameter axis
as a set of segments
\item {\bf map}, showing a detailed sensitivity map, containing the absolute
transmission factor for every volume element in the parameter space
\end{enumerate}

The first two levels of characterisation provide basic information that
usually already exists in the metadata given by the different data
processing pipelines, or is easy to compute from science-ready
datasets. These levels can be easily provided as searchable criteria
by the data access services (e.g. Simple Spectral Access Protocol
(Tody et al. 2006 in prep.) admits service-specific parameters).

\section{Characterising 3D datasets}
Due to complexity of 3D datasets we propose that data centres give only the
first two levels of characterisation for the whole dataset. Further levels
can be given optionally for every spectral segment (in case of IFU data) or
image plane (in case of Fabry-Perot or radio data cube).

\begin{figure}
\includegraphics[width=17cm]{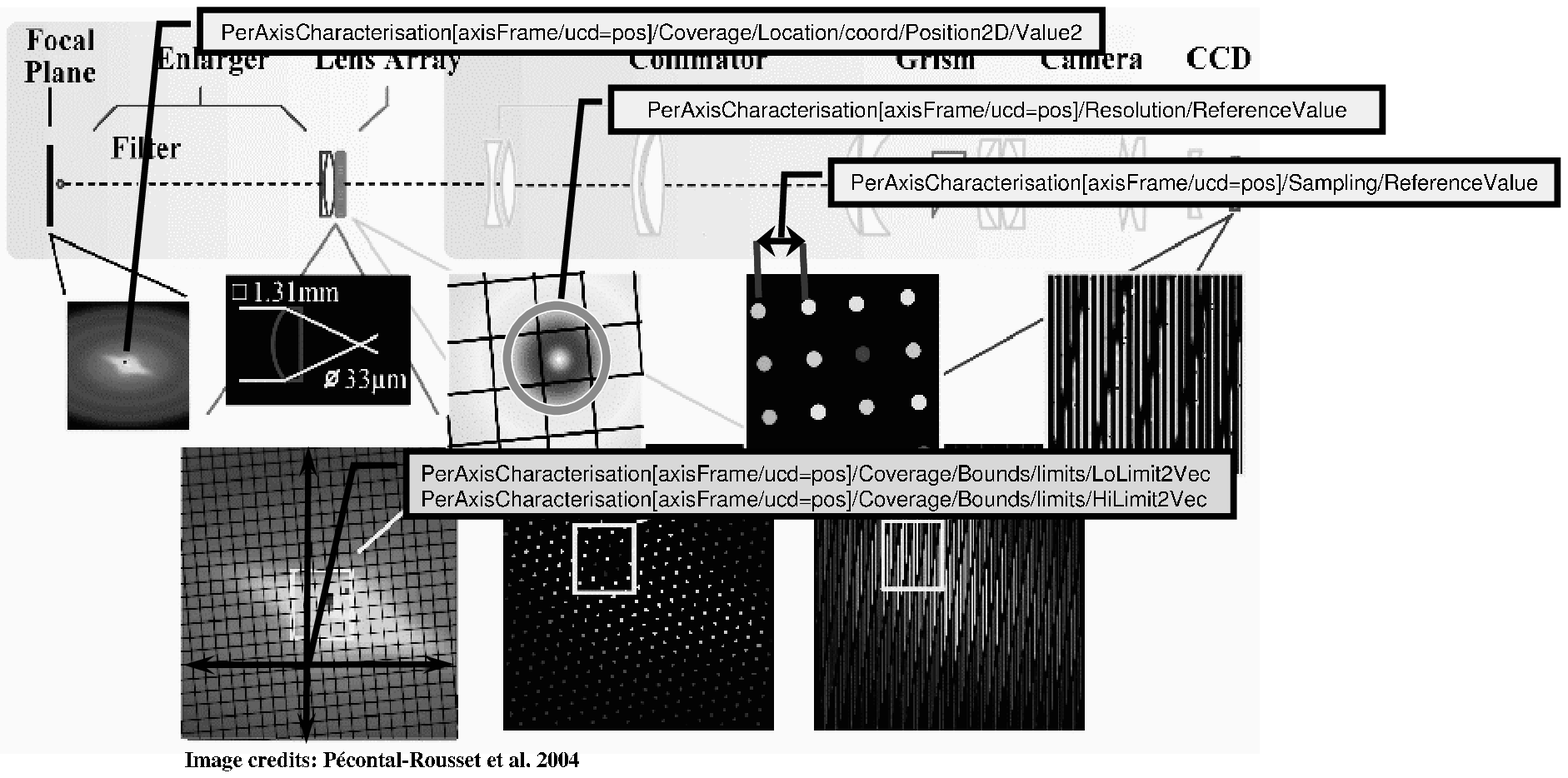}
\caption{Characterisation of the spatial axis for IFU datasets.
\label{O7.8-fig-1}}
\end{figure}

\begin{figure}
\includegraphics[width=17cm]{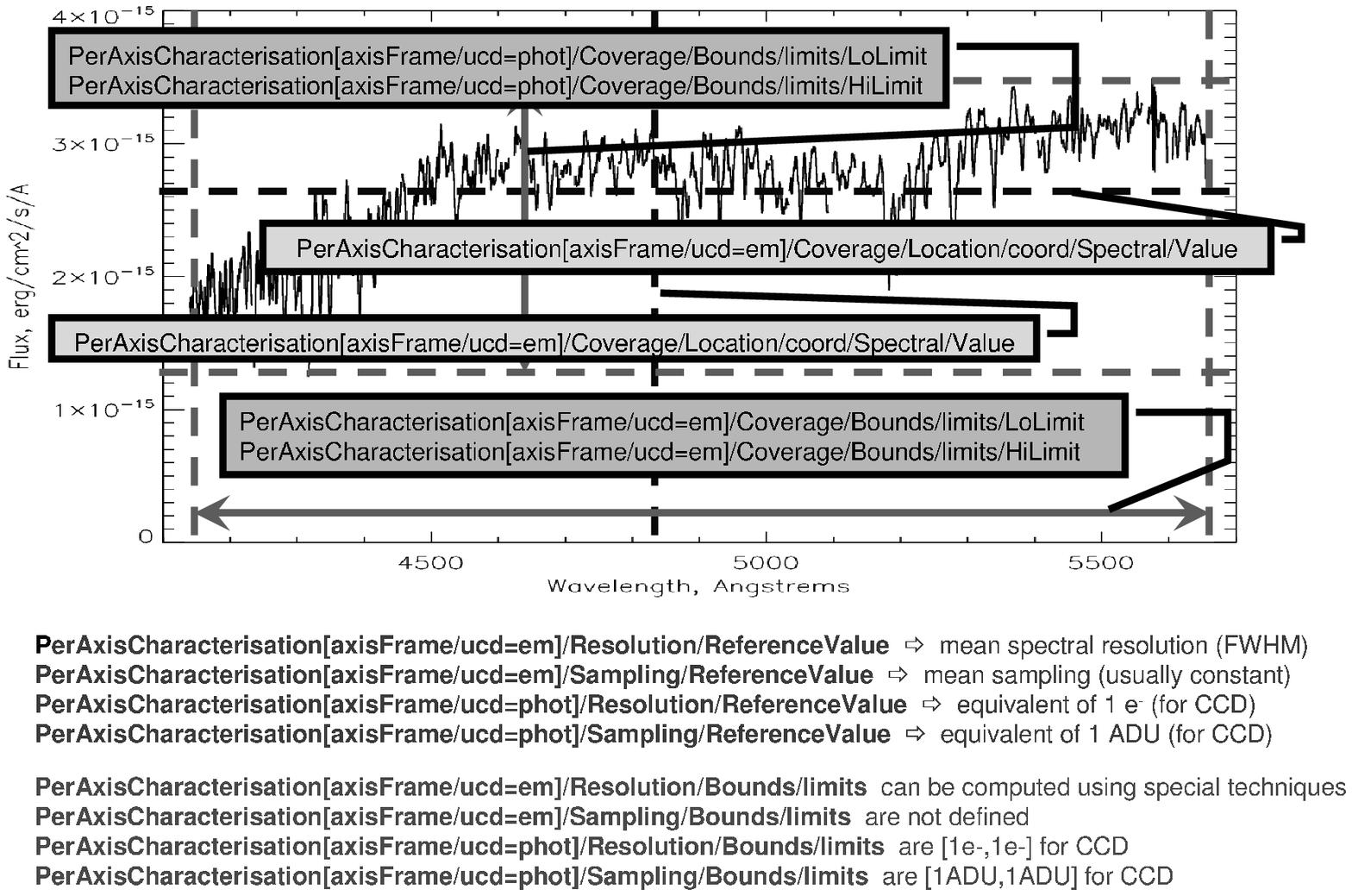}
\caption{Characterisation of the spectral and observed axes for IFU
datasets. The procedure has to be run over all the spectral segments.
\label{O7.8-fig-2}}
\end{figure}

We present a way of characterising IFU datasets (actually other types of 3D
data can be characterised in a similar way). On Figure~\ref{O7.8-fig-1} and
Figure~\ref{O7.8-fig-2} we demonstrate how to compute characterisation
metadata for spectral, spatial, and observable axes from the real dataset.
We have developed a software package for computing characterisation metadata
for IFU datasets stored in the Euro3D format. We also suggest some
modifications to Euro3D FITS format for storing temporal and resolution-specific
information.

A ''live''\ example of the first two levels of the characterisation metadata
for the data, obtained with the Multi-Pupil Fiber Spectrograph based on the
Russian 6m telescope at SAO RAS, in the XML format can be found here:
{\bf http://www.sai.msu.su/\~~chil/VO/CharMPFS.xml} 

\section{Summary}
The data modeling is a crucial point for building VO-compliant data
archives and tools for data processing and analysis.
Characterisation DM has sufficient flexibility
and completeness to be applied for such complex datasets as 3D data.

Considering the Characterisation DM and the extensibility of the Simple
Spectral Access Protocol, we conclude that all the necessary
infrastructural components exist for building VO-compliant archives
of science-ready 3D data and tools for dealing with them. We expect
the first 3D archives to appear in the beginning of 2006.

\end{document}